\documentclass[12pt,a4paper,twoside,openright]{report}
\usepackage[utf8]{inputenc}  
\usepackage[centertags]{amsmath}           
\usepackage{amssymb}
\usepackage{bbm}               
\usepackage{delarray}
\usepackage{hyperref}
\usepackage{enumerate}
\usepackage{pstricks,pst-node,pstcol}
\usepackage{subfigure}
\usepackage{flafter}

\usepackage{psfrag}

\usepackage{fancybox}          
\usepackage{amsfonts,a4wide-adi}
\usepackage{latexsym,fancyhdr}

\fancypagestyle{plain}{\fancyhf{}}

\newcommand{\clearemptydoublepage}{\newpage{\pagestyle{empty}\cleardoublepage}}

\usepackage{graphicx}

\newcommand{\ie}{i.\,e.\ } 

\newcommand{\eg}{e.\,g.\ }

\newcommand{\bra}[1]{\langle #1|}
\newcommand{\ket}[1]{|#1\rangle}
\newcommand{\braket}[2]{\langle #1 | #2 \rangle}

\newcommand{\x}{\textsf{x}}

\begin{document}
\begin{titlepage}
\begin{center}

\vspace*{1cm}

{\huge \bf{Dalitz Plots and Hadron\\[1ex] Spectroscopy}}

\vspace{3cm}

{\Large \bf Master Thesis}

\vspace{0.5cm} Faculty of Science, University of Bern

\vspace{1.5cm} Author:

\vspace{0.5cm} {\bf Adrian W\"uthrich}

\vspace{2cm} May 2005

\vspace{0.5cm} (July 2005: minor changes for publication on
\texttt{hep-ph}) 

\vspace{2cm} Supervisor:

\vspace{0.5cm} {\bf Prof.\ P.~Minkowski}

\vspace{0.5cm}
Institute for Theoretical Physics, University of Bern\\[1ex]
CH-3012 Bern, Switzerland

\end{center}
\end{titlepage}


\clearemptydoublepage
\begin{abstract}

In this master thesis I discuss how the weak three-body decay $B^+\to
\pi^-\pi^+ K^+$ is related to the strong interactions between the final-state
particles.  Current issues in hadron spectroscopy define a region of interest
in a Dalitz plot of this decay. This region is roughly characterized by
energies from 0 to 1.6 GeV of the $\pi^-\pi^+$ and the $\pi^- K^+$ system, and
energies from 3 to 5 GeV of the $\pi^+ K^+$ system. Therefore I propose to use
for the former two systems elastic unitarity (Watson's theorem) as an
approximate ansatz, and a non-vanishing forward-peaked amplitude for the
latter system.  I introduce most basic concepts in detail with the intention
that this be of use to non-experts interested in this topic.

\end{abstract}


\clearemptydoublepage
\setcounter{page}{1}
\pagenumbering{roman}
\tableofcontents
\clearemptydoublepage
\listoffigures \addcontentsline{toc}{chapter}{List of Figures}
\clearemptydoublepage
\chapter*{Acknowledgments}\addcontentsline{toc}{chapter}{Acknowledgments}

It is a pleasure to thank Peter Minkowski for supervising this diploma
thesis. I appreciated very much learning a lot of particle physics from
him. Many of the ideas presented here stem directly from discussions with him
and inputs of his.

I am also grateful to my parents for making it possible for me to dedicate my
time to this work.


\clearemptydoublepage
%
\setcounter{page}{1}
\pagenumbering{arabic}

\chapter{Introduction}

\section{Dalitz plots and hadron spectroscopy}

\emph{Dalitz plots} are representations of a three-body decay,
\begin{equation}
  X\to abc,
\end{equation}
in a two-dimensional plot. The two axis of the plot are nowadays usually the
invariant masses squared of two of the three possible particle pairs, \ie for
instance\footnote{$p^2\equiv E^2-\vec{p}^2$.} 
\begin{equation}
  \begin{split}
    s_{ab} &\equiv (p_a+p_b)^2,\\
    s_{ac} &\equiv (p_a+p_c)^2.
\end{split}
\end{equation}
One can also choose for the axis the invariant masses not squared or, as it was
done mostly in the first Dalitz plots, the kinetic energies of two of the
three decay products, for instance $T_b$ and $T_c$. These coordinates are
equivalent in the sense that they are linearly related, \eg
\begin{equation}
    s_{ab} = m_X^2 + m_c^2 -2m_X(m_c+T_c).
\end{equation}

Dalitz plots owe their name to Richard Dalitz who developed this
representation technique in order to analyze the decay $K^+\to
\pi^+\pi^+\pi^-$ \cite{dalitz53,fabri}.\footnote{Some of the kaons were then
  called ``$\tau$-meson''. See also \cite[p.~141]{weinberg}.} The basic idea
is to assign each decay event its coordinates with respect to the two axis of
the plot. One thus can obtain a ``landscape'' where the ``mountains''
correspond to a lot of events and the ``valleys'' to very few or no events.

\subsection{``Old'' and ``new'' Dalitz plots}

\begin{figure}
\centering
\includegraphics[width=.5\linewidth]{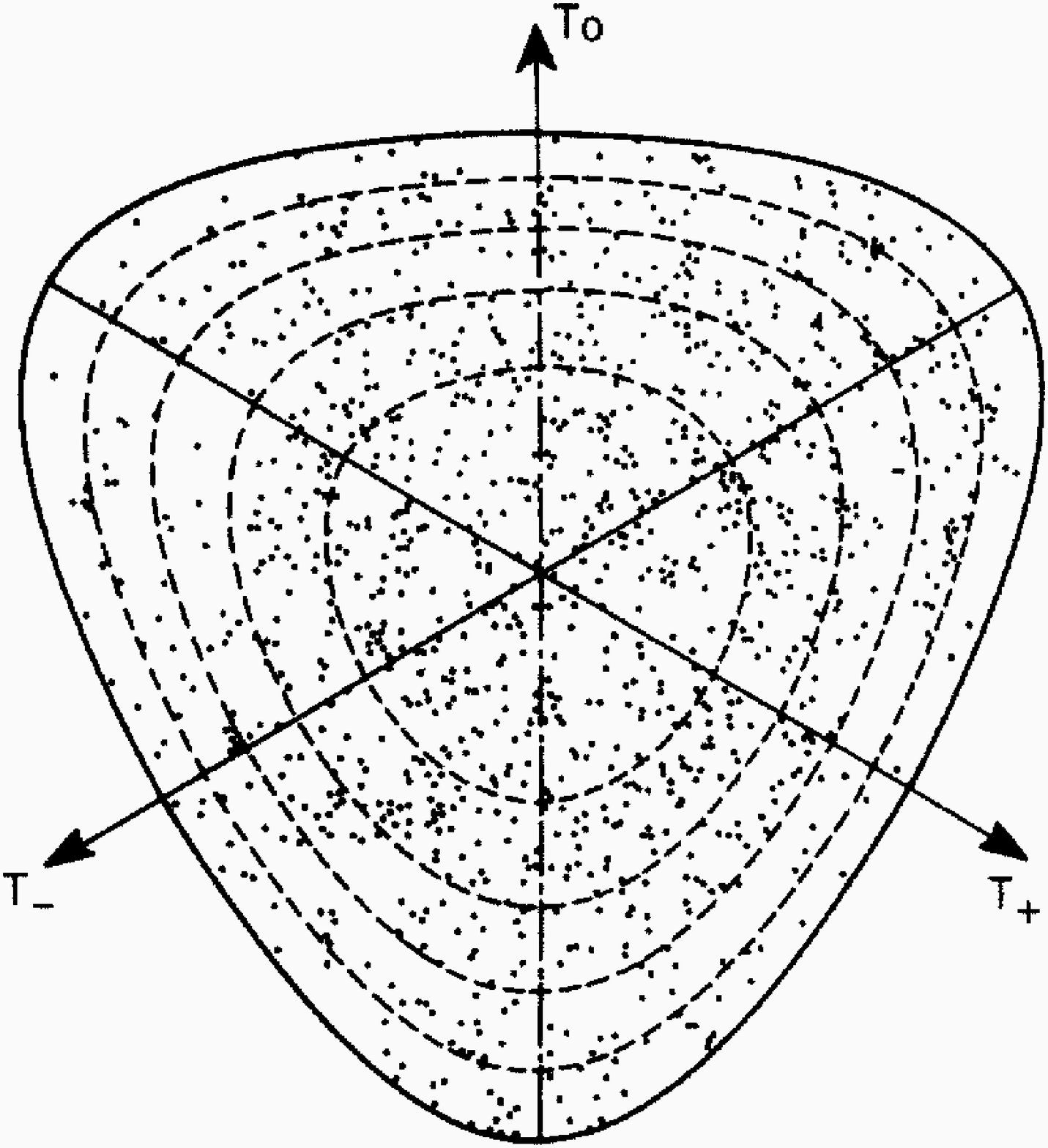}
\caption[Dalitz plot for the decay $\omega^0\to\pi^+\pi^0\pi^-$.]{Dalitz plot
  for the decay $\omega^0\to\pi^+\pi^0\pi^-$, from 
  \cite{alff} cited in \cite[p.~320]{gasio}.}
\label{fig:omega}
\end{figure}

\begin{figure}
  \centering
  \includegraphics[width=.9\linewidth]{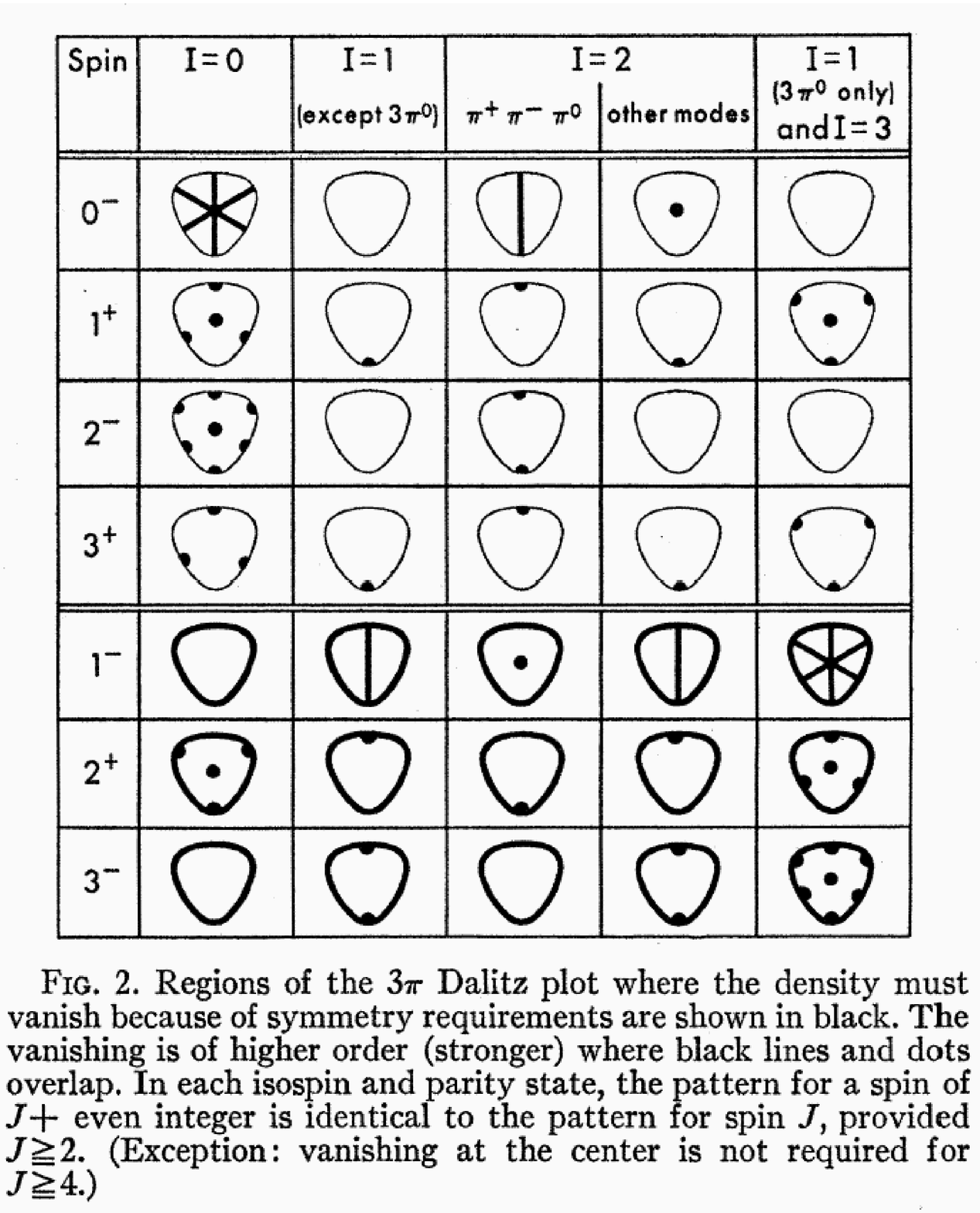}
  \caption[Characteristic patterns in a Dalitz plot.]{Characteristic patterns
    in a Dalitz plot from which spin and parity of the decaying particle can
    be determined; from \cite{zemach}.}
  \label{fig:patterns}
\end{figure}

Originally, the main application of Dalitz plots was to use it in order to
determine spin and parity of the decaying particle. A prominent example is the
decay
\begin{equation}
  \omega^0\to\pi^+\pi^0\pi^-,
\end{equation}
see figure~\ref{fig:omega}.  Spin and parity of the decaying particle can be
read off from characteristic patterns in the Dalitz plot as discussed
prominently in ref.~\cite{zemach}, see figure~\ref{fig:patterns}.

\begin{figure}
  \centering
  \includegraphics[width=.7\linewidth]{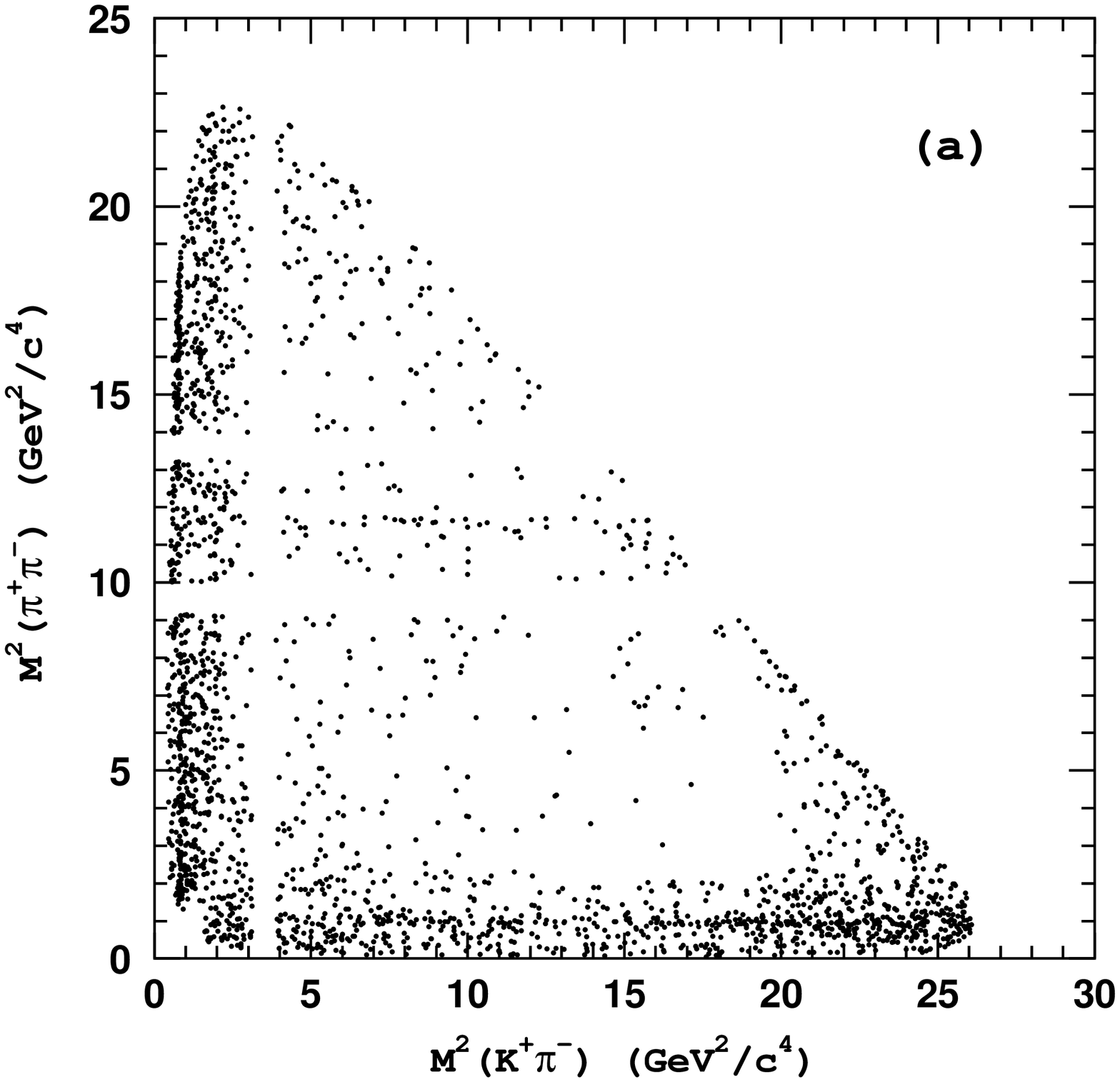}
  \caption[A ``modern'' Dalitz plot.]{A ``modern'' Dalitz plot of
    $B^+\to\pi^-\pi^+K^+$; from \cite{belle04}.}
  \label{fig:belle}
\end{figure}

In recent years Dalitz plots came to the fore again mainly in studies of $D$
and $B$ decays, see for instance \cite{d-cp} and \cite{belle04}. These ``new''
Dalitz plots are distinguished from older Dalitz plots like the one in
figure~\ref{fig:bands} by that they contain a number of events of 1000 to
2000, which reduces statistical fluctuations. Such modern Dalitz plots
(figure~\ref{fig:belle}) are therefore sensitive to the presence or absence of
slight variations in the event distribution. Older Dalitz plots like the one
in figure~\ref{fig:bands} could be used rather for ``resonance hunting''. One
searched for bands in the plot which can be associated to resonances.

Thanks to the sensitivity of the new Dalitz plots, they provide new input for
problems in hadronic spectroscopy, see for instance \cite{Minkowski:2004xf}
and \cite{focus}. 

\subsection{Issues in the spectroscopy of hadrons}

Today's controversial issues in hadron spectroscopy include the isoscalar
scalars, more precisely the states with the quantum numbers of the vacuum,
\begin{equation}
  J^{PC} = 0^{++}.
\end{equation}

One reason why these states are of particular interest is that two-pion
exchange, which has $J^{PC} = 0^{++}$, is the next-to-longest range
contribution for the strong force \cite[p.~1185]{morgan93}.  The longest
range contribution is the exchange of pions, the lightest hadrons. The
isoscalar scalar sector is also relevant for \emph{glueballs}. Various
models (see references in \cite{mink-eur}) predict the lightest glueball to
have the quantum numbers of the vacuum.

To learn more about the scalars one investigates the component with total
angular momentum $J=0$ (\emph{S wave}) and isospin $I=0$ (iso-scalar) mainly
of the $\pi\pi$ scattering amplitude, see figure~\ref{fig:dragon}, top panel.
For the low energy region from zero to about 500 MeV there is Chiral
Perturbation Theory that can state various results. However, the chiral
expansion is hardly valid for energies up to about 1.6 GeV, which are to be
taken into account in discussing the scalars.

The characteristic feature of $\pi\pi$ isoscalar S wave cross section, called
\emph{red dragon} by Minkowski and Ochs, can be explained by the negative
interference of the $f_0(980)$ and the $f_0(1500)$ with the broad glueball
with $J^{PC} = 0^{++}$, \cite{mink-eur}. Negative interference means that the
rapid phase shift of $\pi$ does not start from zero as for a pure Breit-Wigner
resonance but from a background phase of about already $\pi/2$ and then again
from $3\pi/2$, see figure~\ref{fig:phase-maple} on
page~\pageref{fig:phase-maple}.  Alternative explanations postulate the
$\sigma$ (also known as $f_0(600)$) and the $f_0(1370)$. The two alternatives
yield at least to different scalar nonets as described in
\cite{Minkowski:2004xf}:
\begin{itemize}
\item A scalar nonet with $f_0(980)$ and no lighter particles, including
  $a_0(980)$, $K_0^*(1430)$ and $f_0(1500)$, and
\item a scalar nonet with $f_0(980)$ and no heavier particles, including
  $a_0(980)$, $\kappa(850)$ and $\sigma(600)$.
\end{itemize}

\subsection{Weakly coupled decays and strong interactions}

Remarkably enough, $\pi\pi$ scattering cannot be studied experimentally by
scattering a pion beam off a pion target. This is impossible because of the
short lifetime of about $10^{-8}$s for the charged pions and $10^{-16}$s for
the neutral pion. A pion target would therefore have disappeared even before
any scattering can happen. Chew and Low \cite{chew} proposed around 1960 a
surrogate laboratory for unstable targets. Using basically the idea of Chew
and Low a big part of the experimental input for $\pi\pi$ scattering comes
from experiments such as $\pi p\to\pi\pi n$, see also section~\ref{sec:kpi}.

Another surrogate laboratory are three body decays. A prominent example is the
decay $J/\psi \to \phi\pi\pi$. This is a process of the strong interaction but
it is weakly coupled.\footnote{The Particle Data Group reports a branching
  fraction of $8\times 10^{-3}$ for the mode $\phi\pi^+\pi^-$ \cite{PDBook}.}
More recently three-body decays by the weak interaction of $B$ and $D$ mesons
have attracted attention. Dalitz plots from these decays that were produced in
the first place for studies of CP violation serve with their high statistics
as another source of a priori very precise experimental input.  To use such
Dalitz plots for the study of hadronic two-body scattering one has to relate
the weak amplitude of the three-body decay to these hadronic amplitudes. To
profit from the high statistics and to extract information not only from
clearly visible peaks and dips it is important that the resulting ansatz for
fitting the Dalitz plots is sensitive also to slight variations in the event
distribution.

The ansatz that I will try to elaborate is guided by \emph{unitarity} in two
respects. For the amplitudes where we are interested in the energy region
where at least approximately only the elastic channel is open, unitarity leads
to a form of the decay amplitude that has the same phase as the strong
interaction amplitude of the hadrons in the final state. This result is known
as \emph{Watson's theorem} \cite{watson52}.  For the amplitude in higher
energy regions inelastic channels have to be considered and the characteristic
high energy behavior of the elastic amplitude may explain the main features.
I will discuss whether for the high energy amplitude it is again unitarity
that can be used to obtain via the optical theorem a rough idea of its
behavior.

\section{Outline}







In chapter~2 I will introduce basic features of Dalitz plots. I will briefly
discuss some points concerning background suppression. Then I will stress the
possibility of interference between amplitudes and how in the presence of
interference branching fractions are to be understood.

In chapter~3 the $S$ and $T$ matrix are defined. I discuss normalization and
completeness of particle states, give formulations of unitarity conditions and
derive the \emph{optical theorem}. The detailed account of normalization
allows me also to derive an equation defining the \emph{boundary} of a Dalitz
plot.

Chapter~4 deals with the kinematics of a three-body decay. There again one
obtains equations for the boundary of a Dalitz plot. This time not through
phase space considerations but through the definition of two-body subsystems
in the three-body final state. These two-body subsystems are of interest
because they can be related to the strong two-body scattering amplitudes in
the final state and can be used to define alternative coordinate systems in a
Dalitz plot appropriate for a partial-wave analysis.

In chapter~5 I will elaborate on the relation between the decay amplitude and
the scattering amplitudes of the final state particles. For the scattering
amplitudes I will first apply the constraints from unitarity (Watson's theorem
and the optical theorem). Then I try to see what consequences the known
features of the (strong) two-body amplitudes have on an ansatz for the (weak)
amplitude of the three-body decay.


\chapter{Background, resonances and interference}

\section{Background suppression}

Consider as an example the study of the charmless $B^+\to K^+\pi^+\pi^-$ decay
\cite{belle03}. First, one has to detect the three particles with an
appropriate detector. But not all $K$'s and $\pi$'s that are detected during
the experiment may be produced in a charmless $B$ decay, and some particles
may be misidentified as $K$'s and $\pi$'s.

The $K^+\pi^+\pi^-$ final states that come from other reactions than a $B$
decay can be vetoed by requiring that the total invariant mass of the three
particles differ not much from the invariant mass of a $B^+$. By a similar
criterion, then, one can further exclude the final states which are produced
through a charmed (instead of charmless) $B$ decay (\eg $B^+\to
\bar{D}^0\pi^+$ followed by $\bar{D}^0\to K^+\pi^-$)
\cite[p.~7]{belle03}.

\subsection{Sidebands}

The final states where the invariant masses are in a certain neighborhood of
the $B$ invariant mass belong to the \emph{signal region}; other
$K^+\pi^+\pi^-$ final states belong to the so-called \emph{sidebands}. The
events in the sidebands are background events. But not all events in the
signal region are \emph{signal events}, \ie they do not all come from a
charmless $B^+$ decay. In other words, not all background events are excluded
by a sideband criterion. In particular, the background that comes from
misidentification is still present. This sort of background is usually
estimated by an extrapolation from the sidebands: When one has excluded all
events from the sidebands region, the question remains which events in the
signal region do still not come from a charmless $B^+\to K^+\pi^+\pi^-$ decay.
To answer this question one may try the following procedure: Assume that the
background that is not excluded by a restriction to the signal region comes
only from misidentification. Try to estimate how much misidentified events
there are in the sidebands. Assume that the same number of misidentified
events is also present in the signal region. Subtract this number from the
total events in the signal region and you will be left with events that indeed
come from a charmless decay $B^+\to K^+\pi^+\pi^-$.

\subsection{Background and correlations}\label{sec:bg-pw}

A Dalitz plot analysis consists essentially in extracting information from
deviations from homogeneous event distribution over the Dalitz plot. A band of
higher than average density of events parallel to one of the axes, for
instance, is the \emph{correlation} between the four-momenta of two out of the
three final state particles. The information one can extract under certain
circumstances is that the two particles in question form a resonance.  Another
type of information extraction from correlations in this context is
\emph{partial wave analysis}.

In order to extract information from correlations one has to subtract the
background in a way that does not give rise to correlations that are not due
to the physics of the process. To avoid such spurious correlations one has to
take care of two points in particular:

\paragraph{Partial wave decomposition of the background.}

Also the background amplitudes should be decomposed into partial waves.  This
is often not done. As a positive example in this respect I can cite
ref.~\cite{aston}, where the function that characterizes the acceptance of the
detector is expanded in spherical harmonics.

\paragraph{Forming triplets of signal events.}

A point in a Dalitz plot represents a triplet of signal pions and kaons that
come from the same $B$. Assume that constraints like energy and momentum
conservation or detection time do not determine which pions and kaons make up
a Dalitz plot event. Assume as a simple example that there is a set of 8
$\pi^-$'s, a set of 7 $\pi^+$'s and a set of 9 $K^+$'s that are candidates for
coming from the three-body decay of one of 5 $B^+$'s, see
table~\ref{tab:spur}.  This means that the background estimation gives a ratio
of signal to background of $5/8$, $5/7$ and $5/9$, respectively. Due to the
possible quantum mechanical superposition of background and signal amplitudes
it is impossible to say of an individual event whether it is a background or
signal event and---if a signal event---from which of the 5 $B$'s it has come.
But in order to avoid spurious correlations it is not sufficient just to fix
the signal to background ratio of candidates for forming a Dalitz plot
triplet.  One must not form, for instance, pairs of pions for the $\pi^-\pi^+$
mass projection and $\pi^-K^+$ pairs for the $\pi^-K^+$ mass projection where
the $K^+$ and the $\pi^+$ are not assigned to belong to the same Dalitz plot
triplet. As an illustration see table~\ref{tab:spur}.

\begin{table}
  \centering
\begin{tabular}{ccc}
$\pi^-$ & $\pi^+$ & $K^+$ \\ \hline
\rnode{2}{\psframebox{\x}}& \rnode{3}{\psframebox{\x}} \ncline{2}{3} & \x \\
\rnode{2a}{\psframebox{\x}}& \rnode{3a}{\psframebox{\x}} \ncline{2a}{3a} & \x \\
\rnode{2b}{\psframebox{\x}}& \rnode{3b}{\psframebox{\x}} \ncline{2b}{3b} & \x \\
\rnode{2c}{\psframebox{\x}}& \rnode{3c}{\psframebox{\x}} \ncline{2c}{3c} & \x \\
\rnode{2d}{\psframebox{\x}}& \rnode{3d}{\psframebox{\x}} \ncline{1d}{2d} \ncline{2d}{3d} & \x \\
\x & \x & \x \\
\x & \x & \x \\
\x &    & \x \\
   &    & \x \\
  \end{tabular}
\hspace{2cm}
  \begin{tabular}{ccc}
$\pi^-$ & $\pi^+$ & $K^+$ \\ \hline
\x & \x & \x \\
\x & \x & \x \\
\x & \rnode{2}{\psframebox{\x}}& \x \\
\x & \rnode{2a}{\psframebox{\x}}& \x \\
\x & \rnode{2b}{\psframebox{\x}}&
\rnode{3}{\psframebox{\x}} \nccurve[angleB=180]{2}{3} \\
\x & \rnode{2c}{\psframebox{\x}}&
\rnode{3a}{\psframebox{\x}} \nccurve[angleB=180]{2a}{3a} \\
\x & \rnode{2d}{\psframebox{\x}}&
\rnode{3b}{\psframebox{\x}} \nccurve[angleB=180]{2b}{3b} \\
\x &    & \rnode{3c}{\psframebox{\x}} \nccurve[angleB=180]{2c}{3c} \\
   &    & \rnode{3d}{\psframebox{\x}} \nccurve[angleB=180]{2d}{3d} \\
  \end{tabular}\hspace{2cm}
  \begin{tabular}{ccc}
$\pi^-$ & $\pi^+$ & $K^+$ \\ \hline
\rnode{1}{\psframebox{\x}} & \rnode{2}{\psframebox{\x}} &
\rnode{3}{\psframebox{\x}} \ncline{1}{2} \ncline{2}{3} \\
\rnode{1}{\psframebox{\x}} & \rnode{2}{\psframebox{\x}} &
\rnode{3}{\psframebox{\x}} \ncline{1}{2} \ncline{2}{3} \\
\rnode{1}{\psframebox{\x}} & \rnode{2}{\psframebox{\x}} &
\rnode{3}{\psframebox{\x}} \ncline{1}{2} \ncline{2}{3} \\
\rnode{1}{\psframebox{\x}} & \rnode{2}{\psframebox{\x}} &
\rnode{3}{\psframebox{\x}} \ncline{1}{2} \ncline{2}{3} \\
\rnode{1}{\psframebox{\x}} & \rnode{2}{\psframebox{\x}} &
\rnode{3}{\psframebox{\x}} \ncline{1}{2} \ncline{2}{3} \\
\x & \x & \x \\
\x & \x & \x \\
\x &    & \x \\
   &    & \x \\
  \end{tabular}


  \caption{An example of forming pairs for mass projections and Dalitz plot triplets that is consistent with a fixed signal to background ratio but can lead to spurious correlations.}
  \label{tab:spur}
\end{table}

\subsection{Experimental difficulties}

Particular experimental difficulties are encountered in obtaining a Dalitz
plot in decays where one of the three decay products is a photon or where some
decay products can further decay into photons, as in
\begin{equation}
  J/\psi \to \gamma\pi\pi \qquad \text{and} \qquad B \to K\eta\pi.
\end{equation}
The $J/\psi$ can also decay via a $\rho$ and a $\pi^0$ into a four-particle
final state $\pi\pi\gamma\gamma$,
\begin{equation}
  J/\psi \to \rho\pi^0 \to (\pi\pi)(\gamma\gamma).
\end{equation}
Photons escape quite easily any detection. If this happens the reaction
$J/\psi \to \rho\pi^0 \to (\pi\pi)(\gamma\gamma)$ is erroneously identified as
a Dalitz plot event $J/\psi \to \gamma\pi\pi$; cf.\ \cite[p.~1193]{morgan93},
where these experimental difficulties are briefly recalled, and the work of
the Mark~III collaboration (\eg ref.~\cite{mark}), one of the most successful
in overcoming these difficulties.

In the case of $B \to K\eta\pi$, on the other hand, the loss of a $\gamma$ may
have as a consequence that a Dalitz plot event is not identified as such. This
is because one way to detect the final state $\eta$ is via its subsequent
decay into two photons.

\section{Dalitz plots}

\subsection{Basic features}

Dalitz plots have been introduced as a convenient method of analyzing
reactions with three-body final-states in ref.~\cite{dalitz53}, see also
ref.~\cite{fabri}. Consider a three-particle final-state consisting of the
three particles $abc$. Out of these three particles one has three
possibilities to form a pair of two particles: $ab$, $ac$ and $bc$. For a
Dalitz plot the invariant masses squared of two of the three pairs are
represented as the two coordinate axes.\footnote{Originally, see \eg
  ref.~\cite{dalitz53}, each coordinate axis represented the kinetic energies
  of one of the three particles.} Each of the three-body final-states produced
in an experiment can be represented as a dot in the plane defined by these
coordinate axes. Theoretically such dots are confined to a certain boundary
depending on the kinematics of the reaction because outside this boundary the
phase space volume is zero. For the interior of this boundary the phase space
volume is constant, see section~\ref{sec:phase-space}.  Therefore, the density
of the points inside the boundary is a constant multiple of the reaction
matrix element squared.  Thus, every structure in the density of the plots is
due to dynamical characteristics of the reactions and not of kinematical
origin.

\begin{figure}
  \centering
  \includegraphics[width=.9\linewidth]{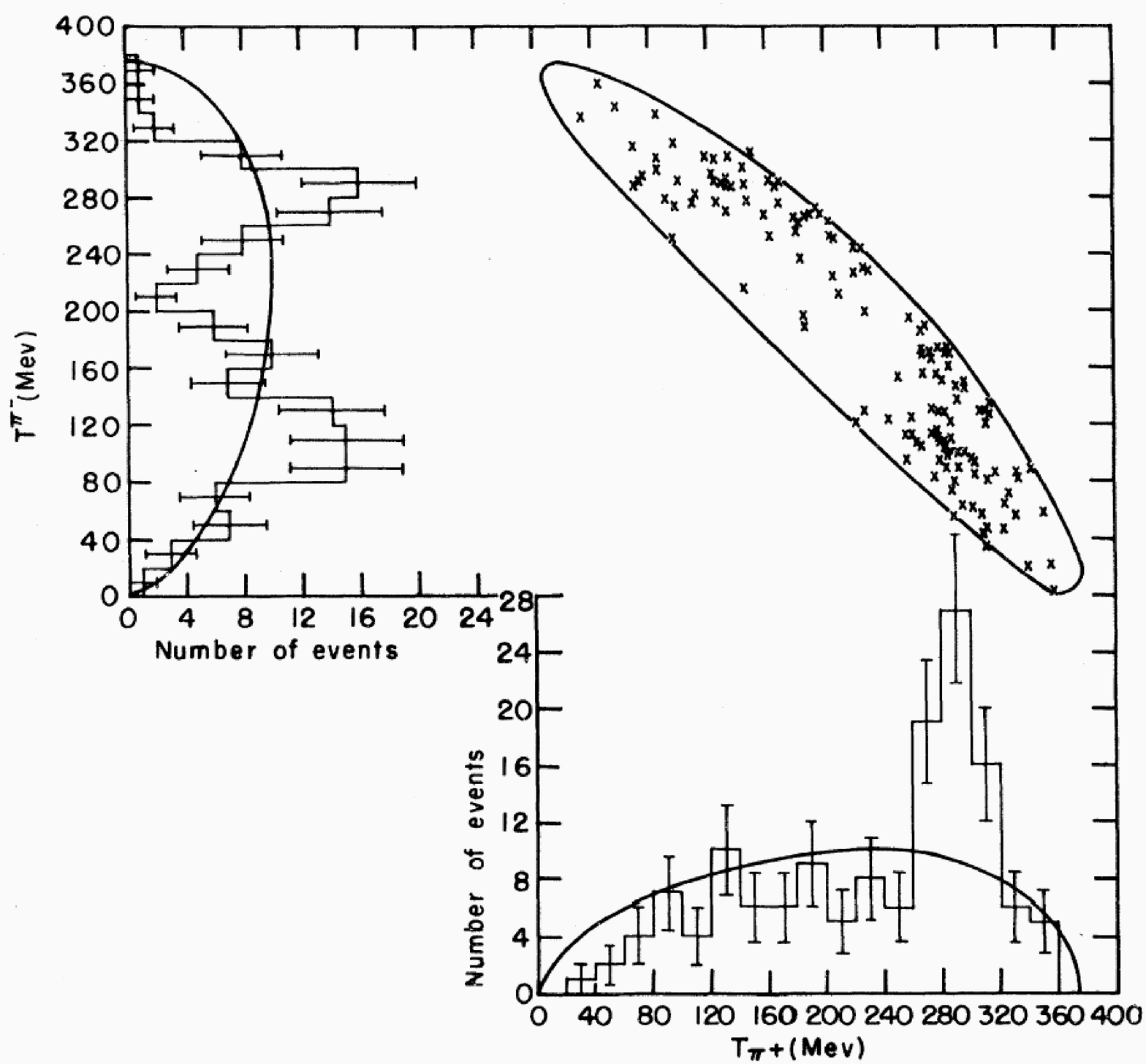}
  \caption[Bands in a Dalitz plot due to resonances.]{Bands in a Dalitz plot
    of $K^-p\to \pi^\pm \Lambda\pi^\mp$ due to the pion-lambda resonance
    $\Sigma^\pm(1385)$; from \cite{alston}.  Also shown: Histograms of
    projections on the axis representing kinetic energies of two out of the
    three final-state particles.}
  \label{fig:bands}
\end{figure}

If the reaction proceeds via an almost stable two-body intermediate state,
\begin{equation}
  X\to (ab)c\to abc,
\end{equation}
the three-body final states will be such that the distribution of the
invariant mass of the pair $ab$ is centered around the invariant mass of the
two-body intermediate state. This manifests itself in the Dalitz plot as a
band of higher than average density of points, see figure~\ref{fig:bands}.

A three-body decay $X\to abc$ has three two-body channels: $ab$, $ac$, and
$bc$. If \eg the pair $ab$ can form a quasi-stable state, a resonance, this
appears in the Dalitz plot as a band along the region of the plot where the
the two-particle invariant mass of the pair $ab$ is equal to the mass of the
resonance.\footnote{This statement needs qualification, see \eg
  ref.~\cite{against-bw}.} If \eg the pair $ab$ may form a resonance, and one
of the axes of the Dalitz plot is the invariant mass of the pair $ab$, then
the band is perpendicular to this axis.

The same type of particle, say $ab$, may form more than one resonances. The
various resonances then appear in the simplest case as parallel bands of a
certain width in the Dalitz plot at respective positions given by the mass of
the resonance. In general, interference effects can generate different
structures.

\subsection{Interference of a resonance with itself}\label{sec:w-itself}

Consider as an example of an interference effect a three-body decay where two
of the three final-state particles are of the same type,
\begin{equation}
  X \to abb. 
\end{equation}
Suppose further that a particle of type $a$ and a particle of type $b$ may
form an almost stable intermediate state, a resonance $R^*$, say. Moreover,
the channel via the resonance $R^*$ be the only process by which the particle
of type $X$ decays into a particle of type $a$ and in two particles of type
$b$. Then the decay amplitude is a superposition of two amplitudes: The
intermediate state can be made up of the particle of type $a$ and one or the
other particle of type $b$, \bigskip

\begin{equation}
X\to
  \begin{array}\{{ccc}\}
    \rnode{A}{\psframebox{a}} & \rnode{B}{\psframebox{b}} & b\ncbar[angleA=90]{A}{B}\\ 
    \rnode{A}{\psframebox{a}} & b & \rnode{B}{\psframebox{b}}\ncbar[angleA=-90]{A}{B}
  \end{array}
\to abb.
\end{equation}
\bigskip

\begin{figure}
  \centering
  \includegraphics[width=\linewidth]{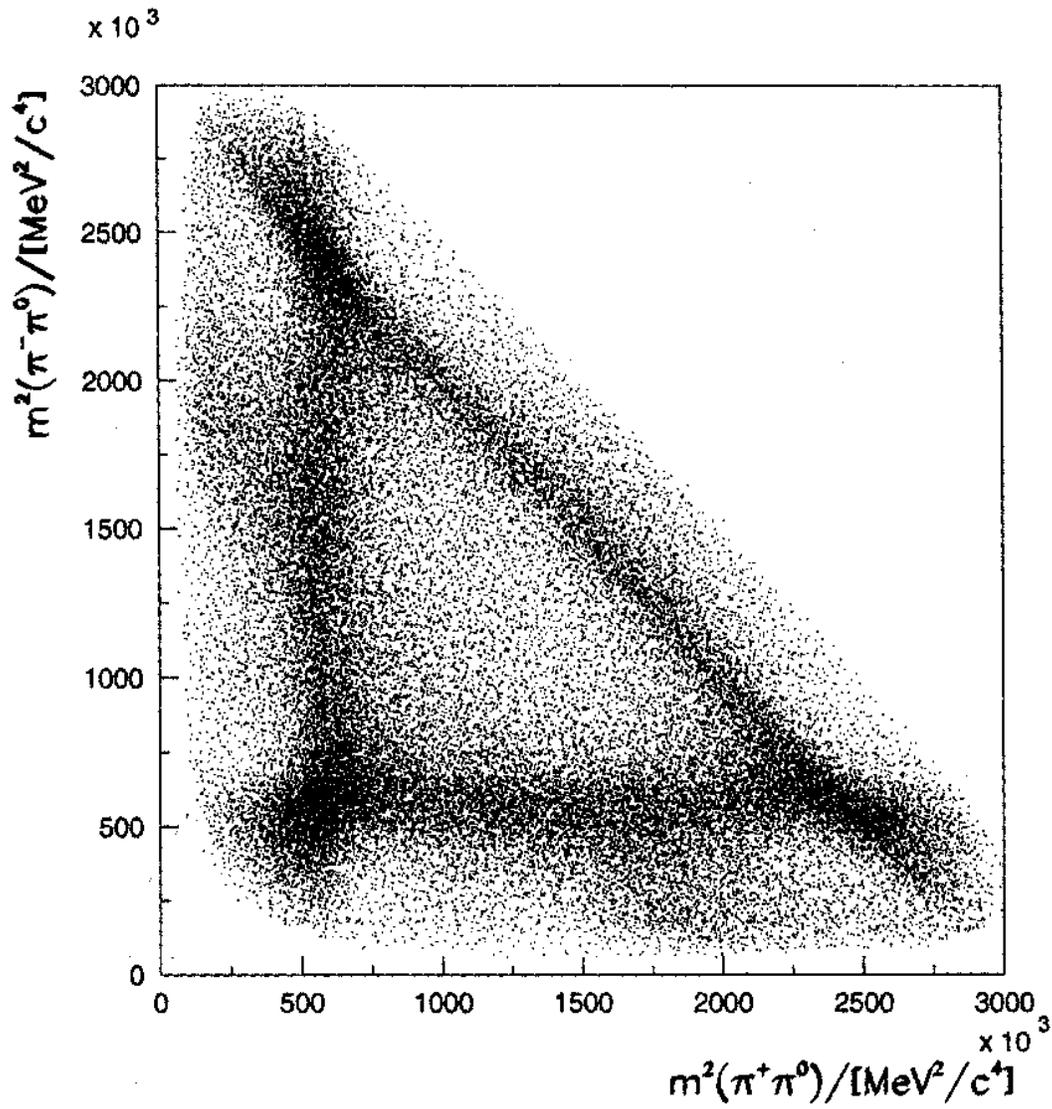}
  \caption[Overlapping bands.]{Overlapping bands showing the interference of
    the $\rho$ resonance with itself (different charge states) in
    $p\bar{p}\to\pi^+\pi^-\pi^0$; from \cite{abele} cited in
    \cite[p.~36]{klempt}.}
  \label{fig:overlap}
\end{figure}

\begin{figure}
  \centering
  \includegraphics[width=.8\linewidth]{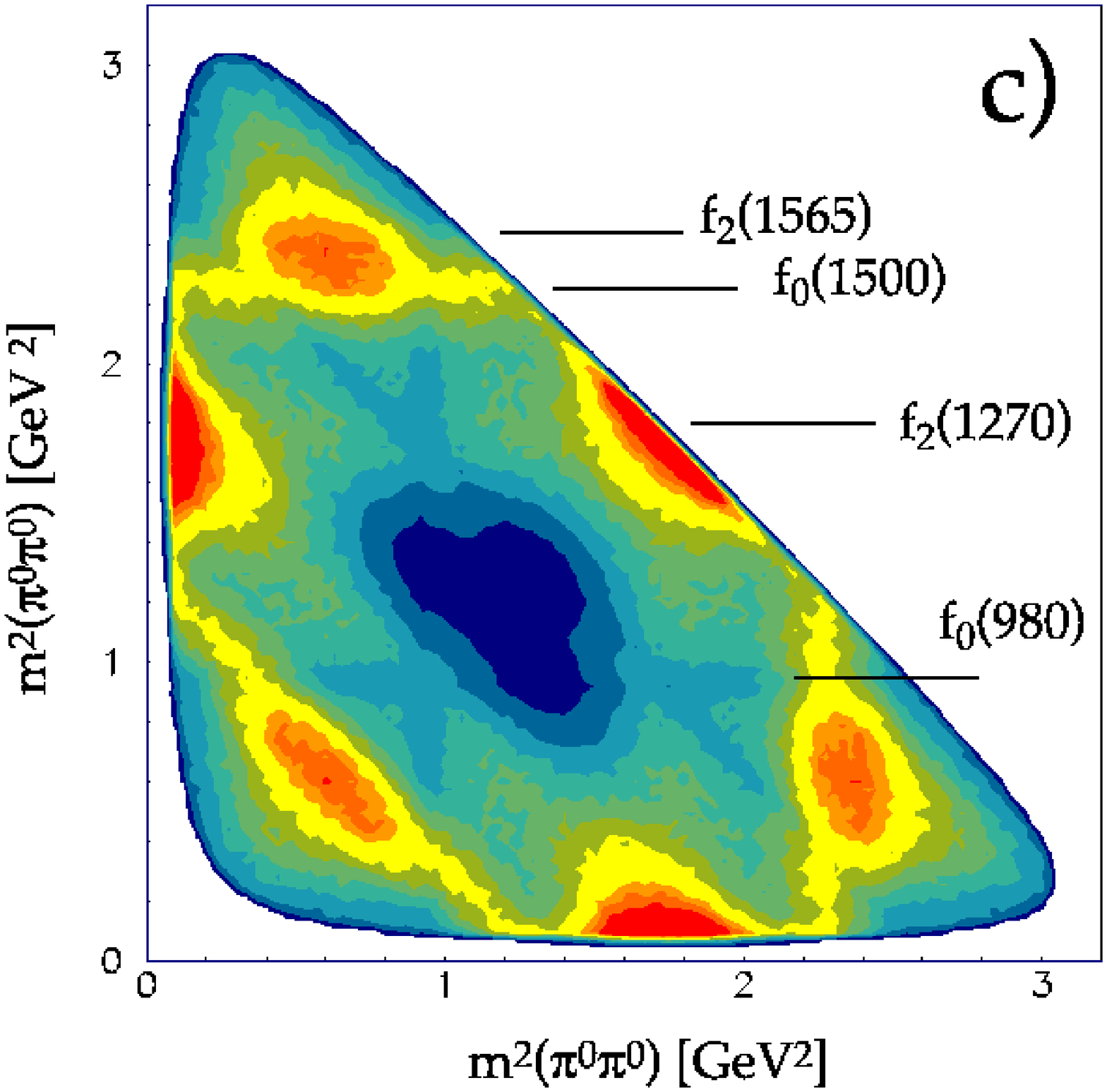} 
  \caption[Overlapping of valleys due to the resonance
  $f_0(980)$.]{Overlapping of valleys due to the resonance $f_0(980)$ in
    $p\bar{p}\to 3\pi^0$; from \cite{Amsler:1997up}. Red and blue regions
    correspond respectively to high and low event densities.}
  \label{fig:overlap-neg}
\end{figure}

In the Dalitz plot of such a three-body decay, with $m_{ab}$ of one
combination $ab$ on one axis and $m_{ab}$ of the other combination on the
other axis, there are two resonance bands perpendicular to one of the axes,
respectively, at the position given by the mass of the resonance $R^*$.
Depending on the kinematics of the process the two bands may or may not
overlap. In case they do, there is constructive interference of the resonance
amplitude in one of the two resonant two-body channels with the amplitude in
the other. Thus in the region of the overlap the total intensity is not twice
the intensity of an isolated band but four times because it is the amplitudes
and not the intensities (\ie the amplitudes squared) that are added. Two
examples with even three overlapping bands are shown in
figures~\ref{fig:overlap} and \ref{fig:overlap-neg}. In
figure~\ref{fig:overlap} the bands are regions of higher than average density,
\ie \emph{peaks}. In figure~\ref{fig:overlap-neg} the light blue bands due to
the resonance $f_0(980)$ are regions with lower than average density, \ie
\emph{dips} or \emph{valleys}.  Besides the overlap of bands this figure thus
shows remarkably that resonances (here the $f_0(980)$) do not necessarily show
up as peaks in a cross section.

\section{Branching fractions}\label{sec:bf}

For a lot of unstable particles there is more than one possible decay
product. A particle $X$ may \eg decay in two particles $ab$, or $cd$, or in
three particles $efg$. The probability for the particle $X$ to decay in one of
the possible products is the respective \emph{branching fraction}. The most
likely decay could for example be $X \to ab$ with a probability of 80\%. With
15\% probability the particle $X$ may decay in the particle pair $cd$, and
with 5\% probability into $efg$. If there is only one possibility for the
particle to decay, the branching fraction for this decay is 100\% and for all
other conceivable decays 0\%---or one may prefer not to speak of branching
fractions at all in such a case.

Dalitz plots can be used to determine branching fractions for three-body
decays with almost stable two-body intermediate states. Consider as an example
the following three-body decay that can proceed via three different
resonances:
\begin{equation}
  X\to
  \begin{array}\{{c}\}
    (ab)c\\
    a(bc)\\
    (ac)b
  \end{array}
\to abc.
\end{equation}
One may then ask what the respective probabilities are for these three
possibilities. These probabilities are the (exclusive) branching fractions for
the three processes respectively. The probability for the three-body decay
$X\to abc$ to occur at all, \ie via any intermediate state, is the
\emph{inclusive} branching fraction for the process $X\to abc$.

The task of determining branching fractions from a Dalitz plot may be
seriously complicated by interference effects:
\begin{quote}
  [\dots] we determine the exclusive branching fractions neglecting the
  effects of interference. The uncertainty due to possible interference
  between different intermediate states is included in the final result as a
  model-dependent error. [\dots] We find that the model-dependent errors
  associated with the wide resonances introduce significant uncertainties into
  the branching fraction determination. \cite{garmash02}
\end{quote}

\subsection{Branching fractions, superposition and
  interference}\label{sec:branch-fract-superp} 

The basic idea of branching fractions is to give a quantitative answer to the
following type of questions. ``If various processes are possible under given
circumstances: How likely are the respective processes to occur?'' Or: ``If
various processes could have occurred: How likely is it that one particular
process has occurred?'' 

The quantum mechanical principle of superposition, however, implies that such
types of questions do not always have an answer---except for the reply that
the question is meaningless. This is for example the case in the notorious
double slit experiment. If both slits are open as a matter of principle one
cannot tell for a particular particle whether it has arrived at the screen via
the upper or the lower slit. But still it seems that one can say something
more about the intermediate state of a double slit experiment than just: ``The
particles arrived somehow at the screen.'' In some sense the two possibilities
(passing through the upper or lower slit) are on an equal footing. So one
would like to say that it is equally probable that the particle has taken the
upper slit or the lower slit, but at the same time, because the particle was
in a state of superposition of the two possibilities, it \emph{has not} taken
one or the other route nor both. How can one solve the dilemma between
superposition and the basic idea of a branching fraction? How can one define
branching fractions in a sensible way even in the presence of superposition?

A possible solution is via an \emph{amplitude analysis}. One can identify the
intensity pattern as the square of a total amplitude which in turn decomposes
into a sum of two amplitudes in the double slit case. One of these amplitudes
represents the process via the upper slit the other amplitude represents the
process via the lower slit. The respective squares of these two amplitudes are
equal. It is in this sense that the two processes are on an equal footing. In
this sense both processes (passing through the upper or lower slit) are
equally probable although it is not the case that one or the other process has
occurred but rather a superposition of both. Because the amplitudes of both
processes contribute equally to the total amplitude the two processes should
be assigned equal branching fractions, without interpreting this assignment as
``in one half of a certain number of repetitions of the experiment the
particle passes through the upper slit and in the other half of the
repetitions through the lower slit''. Such an interpretation would be at odds
with the superposition principle of (standard) quantum mechanics.

Let $\psi$ be the total amplitude of the process of a
particle reaching the screen in the double slit setup, and $\psi_1$ and
$\psi_2$ the amplitudes for passing through the upper or the lower slit,
respectively,
\begin{equation}
  \psi = \psi_1 + \psi_2.
\end{equation}
$|\psi|^2$ is the total intensity, which we set as unity of the branching
fractions,
\begin{equation}
  1 = |\psi|^2 = |\psi_1 + \psi_2|^2 = |\psi_1|^2 + |\psi_2|^2 +
  2\text{Re}(\psi_1\psi_2^*). 
\end{equation}
If as in the double slit experiment the two amplitudes \emph{interfere}, we
have $\text{Re}(\psi_1\psi_2^*)\neq 0$, hence $|\psi_1|^2 + |\psi_2|^2 \neq
1$.  This shows that when there is interference, branching fractions do in
general not add up to unity.

The basic idea of branching fractions is to assign probabilities to various
possibilities under given circumstances. Due to quantum mechanical
superposition the totality of possibilities is not a complete set of
alternatives in the sense of ordinary language, where exactly one of the
alternatives is realized. If, moreover, the superposed amplitudes interfere
the sum of all branching fractions is not one, which shows again, that it is
not complete alternatives in the ordinary sense that branching fractions are
assigned to.

\subsection{Fit fractions in Dalitz plots}

In Dalitz plots interference between resonant amplitudes show up as
overlapping bands. As discussed in section~\ref{sec:branch-fract-superp},
in the presence of interference the definition of a branching fraction is less
obvious than it may seem at first sight. A very early determination of
branching fractions from a Dalitz plot with two overlapping bands is
ref.~\cite{yybar63}.

The method applied there is roughly the following. The mass projections of the
Dalitz plot are fitted with a Breit-Wigner distribution \emph{without} taking
into consideration the events that fall in the overlap regions. The density of
events in the resonance bands thus obtained is then extrapolated into the
region of the overlap, assuming constant density of events in the resonance
bands. The observed number of events in the overlap region is in good
agreement with this extrapolation. This shows that the two overlapping
resonances do not interfere. Therefore, the first fit can be corrected
for the overlap regions assuming \emph{incoherent} addition of the two
amplitudes representing the overlapping bands. 

But what about cases \emph{with} interference? As mentioned at the beginning
of section~\ref{sec:bf} the effect of interference on branching fractions is
not always negligible. I cite again from ref.~\cite{garmash02}:
\begin{quote}
  We find that effects of interference between different two-body intermediate
  states can have significant influence on the observed two-particle mass
  spectra and a full amplitude analysis of the three-body $B$ meson decays is
  required for a more complete understanding. This will be possible with
  increased statistics.
\end{quote}
It seems that recent experiments provide sufficient statistics to allow for
such a full amplitude analysis \cite{belle03,babar04}.


\chapter{Scattering amplitudes and particle states}

\section{$S$- and $T$-matrix}

Let $\ket{\alpha}_\text{in}$ and $\ket{\alpha}_\text{out}$ be some
appropriately defined states that are identical to the free-particle states
$\ket{\alpha}$ (without subscript) in the respective limit $t\to -\infty$
$t\to +\infty$. The amplitude for the initial state $\ket{\alpha}_\text{in}$
to be found as the final state $\ket{\alpha}_\text{out}$ is given by the
scalar products between these states. All possible scalar products can be
collected in to rows and columns of a matrix, usually called $S$ (for
\textbf{s}cattering, I suppose).
\begin{equation}
  S_{\alpha\beta} = {_\text{out}\braket{\alpha}{\beta}}_\text{in}.
\end{equation}
We can define an operator $S$ such that sandwiched between free-particle
states it yields the corresponding element of the $S$-matrix:
\begin{equation}
  S_{\alpha\beta}=\bra{\alpha}S\ket{\beta}.
\end{equation}
The amplitude for the transition of state $\ket{\beta}$ to the state
$\ket{\alpha}$ is the superposition of the amplitude $\braket{\alpha}{\beta}$,
which represents the quantum mechanical collapse of the state $\ket{\beta}$ in
its component $\ket{\alpha}$, and an amplitude that represents a transition
caused by some interaction. This latter amplitude is described in terms of the
elements of the $T$-matrix
\begin{equation}
  T_{\alpha\beta}=\bra{\alpha}T\ket{\beta},
\end{equation}
such that
\begin{equation}
  S_{\alpha\beta}=\bra{\alpha}S\ket{\beta}=\braket{\alpha}{\beta}+i\bra{\alpha}T\ket{\beta}, 
\end{equation}
\ie we define the $T$-matrix through the relation
\begin{equation}
  S=\mathbbm{1}+iT.
\end{equation}

\section{Cross sections and decay rates}\label{sec:event-distr}

In this section I try to establish---in some detail but without going into the
very foundations---the relation between the $T$ matrix elements and cross
sections and decay rates. In particular I will focus on a decay of type $B^+
\to abc$.  The final result will be expressed in terms of matrix elements
between state vectors characterized by their particle content and the
four-momentum each particle has. Such states are improper vectors in an
appropriately defined Hilbert space, and they owe their name to the fact that
they lead in some context to undefined mathematical expressions. To avoid
problems with improper vectors one can consider wave-packets instead of states
with definite momentum. A slightly less expensive way is by normalizing the
vectors with definite momentum with respect to a box of finite volume $V$; and
by making explicit that the interaction that causes the decay is not present
for an infinitely long time but only for a finite time $\tau$.

Both wave-packet and box approach have their respective advantages. In the box
normalization the phase space factors are more clearly accounted for. To use
the box normalization for the decay distribution of an instable particle
faces, however, the inconsistency that the energy of the particle is assumed
to be definite whereas a particle can only decay if it is \emph{not} in an
eigenstate of energy. In that respect it is more adequate to represent the
initial state as a superposition of momentum eigenstates with nearly but not
exactly the same eigenvalues, \ie as a wave-packet.

\subsection{Volume normalization}\label{sec:box}

The state vectors $\ket{B^+,\vec{p}_B}$ are considered as plane wave functions
\begin{equation}
  \Psi(t,\vec{x}) = \mathcal{N} e^{iEt}e^{-i\vec{p}\vec{x}}.
\end{equation}
The absolute squares of these wave functions represent a probability density
and have therefore to be normalized to unity. I will do this with respect to a
volume $V=L^3$. Therefore the method shown here is also known as \emph{box
  normalization.} Choosing the volume to be a box makes the calculations
easier. The idea of normalizing states to a finite volume, however, does not
only apply to this special choice of volume. The normalization constant
$\mathcal{N}$ is determined by
\begin{equation}
  1 = \int_V d^3x\, |\Psi(t,\vec{x})|^2 = \mathcal{N}^2 \int_V d^3x = V
\end{equation}
to be 
\begin{equation}
  \mathcal{N} = \frac{1}{\sqrt{V}}.
\end{equation}
In a box of finite volume $V$ the three-momenta are discrete if one imposes
periodic boundary conditions
\begin{align}
  \Psi_p(t;(0,x_2,x_3)) &= \Psi_p(t;(L,x_2,x_3)),\\
  \Psi_p(t;(x_1,0,x_2)) &= \Psi_p(t;(x_1,L,x_3)),\\
  \Psi_p(t;(x_1,x_2,0)) &= \Psi_p(t;(x_1,x_2,L)).
\end{align}
Then one has, for instance,
\begin{multline}
  \Psi_p(t;(0,x_2,x_3)) = \frac{1}{\sqrt{V}}e^{iEt}e^{-i(p_2x_2+p_3x_3)}
  = \frac{1}{\sqrt{V}}e^{iEt}e^{-i(p_1L+p_2x_2+p_3x_3)} \\
  = e^{-ip_1L}
    \Psi_p(t;(0,x_2,x_3)), 
\end{multline}
which requires the eigenvalues for $\vec{p}$ to be quantized, \ie labeled by
a triplet of integers $\vec{n}\in \mathbb{Z}^3$,
\begin{equation}\label{eq:29b}
  \vec{p} = \frac{2\pi}{L} \vec{n}.
\end{equation}
Because the plane wave functions are normalized to unity, the scalar product
between the momentum states in the box is a Kronecker delta,
\begin{multline}
  \braket{\vec{p}(\vec{n}')}{\vec{p}(\vec{n})} = \int_V d^3x\,
  \Psi_{\vec{n}'}(t,\vec{x})\Psi_{\vec{n}}^*(t,\vec{x}) \\
  =  \frac{1}{V} \int_V
  d^3x\, \exp(-i(\vec{p}(\vec{n}') - \vec{p}(\vec{n}))\vec{x}) =
  \delta^3_{\vec{n}'\vec{n}}.
\end{multline}

Be $\ket{f}_{\text{box}}$ and $\ket{i}_{\text{box}}$ a state of $m_f$ ($m_i$)
particles each with three-momenta $\vec{p}_{m_f}$ ($\vec{p}_{m_i}$),
normalized to the box of volume $V$. (Multi-particle states are direct products
of one-particle states.) The transition probability from
$\ket{i}_{\text{box}}$ to $\ket{f}_{\text{box}}$ is given by
\begin{equation}
  \Delta w(f,i) = |_{\text{box}}\bra{f}S\ket{i}_{\text{box}}|^2 =
  |_{\text{box}}\braket{f}{i}_{\text{box}} + i 
  _{\text{box}}\bra{f}T\ket{i}_{\text{box}}|^2 
  = |_{\text{box}}\bra{f}T\ket{i}_{\text{box}}|^2.
\end{equation}
The probability for a transition of $\ket{i}_{\text{box}}$ into \emph{any}
momentum configuration is
\begin{multline}\label{eq:37}
  w(f,i) = \prod_{m_f} \sum_{\vec{n}_{m_f}}
  |_{\text{box}}\bra{f}T\ket{i}_{\text{box}}|^2 \\
  = \prod_{m_f} \frac{V}{(2\pi)^3}
  \sum_{\vec{n}_{m_f}} \frac{(2\pi)^3}{V}
  |_{\text{box}}\bra{f}T\ket{i}_{\text{box}}|^2 \\ 
  = \left(\frac{V}{(2\pi)^3}\right)^{m_f} \left(\prod_{m_f} \int d^3p_{m_f}\,\right)
  |_{\text{box}}\bra{f}T\ket{i}_{\text{box}}|^2,
\end{multline}
where in the last step I performed the limit $L\to\infty$, using Riemann's
definition of the integral and equation~\eqref{eq:29b}. What is the appropriate
expression for $_{\text{box}}\bra{f}T\ket{i}_{\text{box}}$ in the limit of
large volume (\ie $L\to\infty$)? We want the continuous momentum eigenstates
to be normalized as
\begin{equation}\label{eq:36}
  \begin{split}
    \braket{f'}{f} &= \prod_{m_f} (2\pi)^3 2E_{m_f} \delta^3(\vec{p}'_{m_f}-\vec{p}_{m_f})\\
    &= \prod_{m_f}  2E_{m_f} \int d^3x\, \exp(-i(\vec{p}'_{m_f}-\vec{p}_{m_f})\vec{x}), 
\end{split}
\end{equation}
and similarly for $\braket{i'}{i}$.  In the limit of large volume the
right-hand side of equation~\eqref{eq:36} is
\begin{multline}
  \prod_{m_f} 2E_{m_f} \int d^3x\, \exp(-i(\vec{p}'_{m_f}-\vec{p}_{m_f})\vec{x}) \\
  = \prod_{m_f} 2E_{m_f} \int_V d^3x\,
  \exp(-i(\vec{p}(\vec{n}'_{m_f})-\vec{p}(\vec{n}_{m_f}))\vec{x})
  \\
  = \prod_{m_f} 2E_{m_f} V \delta^3_{\vec{n}'\vec{n}} = \prod_{m_f} 2E_{m_f} V
  \braket{\vec{p}_{m_f}(\vec{n}'_{m_f})}{\vec{p}_{m_f}(\vec{n}_{m_f})} \\
  = V^m \left(\prod_{m_f} 2E_{m_f}\right) {_{\text{box}}}\braket{f'}{f}_{\text{box}}.
\end{multline}
It follows that
\begin{equation}
  \ket{f}_{\text{box}} = \left(\prod_{m_f}
  \sqrt{\frac{(2\pi)^3}{2E_{m_f} V}}\right)\ket{f},
\end{equation}
and similarly for $\ket{i}$ such that
\begin{equation}
  _{\text{box}}\bra{f}T\ket{i}_{\text{box}} = \left(\prod_{m_f}
  \sqrt{\frac{(2\pi)^3}{2E_{m_f} V}}\right) \left(\prod_{m_i}
  \sqrt{\frac{(2\pi)^3}{2E_{m_i} V}}\right) \bra{f}T\ket{i},
\end{equation}
and therefore from equation~\eqref{eq:37}
\begin{equation}
  w(f,i) = \left(\prod_{m_i} \frac{(2\pi)^3}{2E_{m_i} V}\right)
\left(\prod_{m_f} \int \frac{d^3p_{m_f}}{2E_{m_f}}\right)
  |\bra{f}T\ket{i}|^2. 
\end{equation}
$T$ conserves energy and momentum. $T$ is zero whenever the energy and
momentum of the final and initial state are different. It is therefore
suitable to consider the multi-particle states $\ket{i}$ (and $\ket{f}$) as
direct products of eigenstates of the total four-momentum and a vector
$\ket{\psi}$ which contains all the remaining characteristics of the
multi-particle state, such as the masses, three-momenta and quantum numbers of
each particle in the state\footnote{The vectors $\ket{\psi}$ span the
  \emph{little Hilbert space}, see \cite[p.~111ff.]{martin-spearman}},
\begin{equation}
  \ket{i} = \ket{p_{\text{tot}}^{(i)}} \otimes \ket{\psi_i}.
\end{equation}
Since $T$ conserves the total four-momentum it acts as the unit operator in
the space of eigenstates of total four-momentum and thus takes the following
form as a direct product,
\begin{equation}
  T = \mathbbm{1} \otimes \mathcal{M}.
\end{equation}
The $T$ elements between final and initial states then look like
\begin{equation}
  \bra{f}T\ket{i} =
  \bra{p_{\text{tot}}^{(f)}}\mathbbm{1}\ket{p_{\text{tot}}^{(i)}}
  \bra{\psi_f} \mathcal{M} \ket{\psi_i} = \delta^4(p_i-p_f) \bra{\psi_f}
  \mathcal{M}\ket{\psi_i} \equiv \delta^4(p_i-p_f) \mathcal{M}_{fi}.
\end{equation}

In terms of $\mathcal{M}$ we have
\begin{equation}\label{eq:38}
  w(f,i) = \left(\prod_{m_i} \frac{(2\pi)^3}{2E_{m_i} V}\right)
  \left(\prod_{m_f} \int \frac{d^3p_{m_f}}{2E_{m_f}}\right) 
  [\delta^4(p_B-p_{\text{tot}})]^2
  |\mathcal{M}_{fi}|^2. 
\end{equation}
To make sense of the two delta-functions we write for one of it its integral
representation in the volume and time interval that is large but explicitly
accounted for. To be explicit in this subtle point I consider the
delta-functions explicitly as distributions in a space of test-functions
$f(p_f)$. 
\begin{multline}
  \int d^4p_f\, [\delta^4(p_i-p_f)]^2 f(p_f) \\
  = \int d^4p_f\, \delta^4(p_i-p_f) \frac{1}{(2\pi)^4} \int_{V\tau} d^4x\,
  \exp(-i(p_i-p_f)x) f(p_f) \\
  = \frac{1}{(2\pi)^4} \int_{V\tau} d^4x\, f(p_i) = \frac{V\tau}{(2\pi)^4}
  f(p_i) \\
  =  \int d^4p_f\, \frac{V\tau}{(2\pi)^4} \delta^4(p_i-p_f) f(p_f). 
\end{multline}
In the usual short-hand notation this reads
\begin{equation}
  [\delta^4(p_i-p_f)]^2 = \frac{V\tau}{(2\pi)^4} \delta^4(p_i-p_f).
\end{equation}
Equation~\eqref{eq:38} then yields
\begin{equation}\label{eq:40}
  w(f,i) = \left(\prod_{m_i} \frac{1}{2E_{m_i}}\right)
  \frac{V^{1-m_i}}{(2\pi)^{3m_i-4}}\, \tau \int dQ\, 
  |\mathcal{M}_{fi}|^2, 
\end{equation}
with
\begin{equation}
  dQ \equiv \delta^4(p_i-p_f) \prod_{m_f} \frac{d^3p_{m_f}}{2E_{m_f}}
\end{equation}
the \emph{phase space} volume for a final state containing $m_f$ particles.
Usually the quantity of interest is the \emph{rate}, \ie the transition
probability per interaction time $\tau$
\begin{equation}\label{eq:39}
  \Gamma(f,i) = \frac{w(f,i)}{\tau} = \left(\prod_{m_i}
  \frac{1}{2E_{m_i}}\right) 
  \frac{V^{1-m_i}}{(2\pi)^{3m_i-4}}\, \int dQ\, 
  |\mathcal{M}_{fi}|^2.
\end{equation}

\subsubsection{$\boldsymbol{m_i = 1}$ (decay)}

In the case of $m_i = 1$ equation~\eqref{eq:40} reads
\begin{equation}\label{eq:41}
   w(f,i) = \frac{1}{2E}
  2\pi \tau \int dQ\, 
  |\mathcal{M}_{fi}|^2.
\end{equation}
According to this equation the transition probability for decay should
increase proportional to the interaction time $\tau$. However, when the
particle has decayed the probability of decay should be zero. So
equation~\eqref{eq:41} is only valid for $\tau$ much less than the mean
lifetime of the decaying particle. This conflicts with the assumption of very
large $\tau$ when representing the $\delta^4$ functions as an integral over
space-time. Also formula~\eqref{eq:41} contains a factor $2E$, where $E$ is
the energy of the decaying particle. However, as mentioned at the beginning of
this section, a particle only decays if it has no sharp energy value. These
problems can be avoided when the initial particle is represented by a wave
packet, see section~\ref{sec:wave-packets}.

\subsubsection{$\boldsymbol{m_i = 2}$ (two-body scattering)}

With $m_i = 2$ equation~\eqref{eq:39} is
\begin{equation}
    \Gamma(f,i) = \frac{w(f,i)}{\tau} = \frac{1}{2E_12E_2} 
  \frac{1}{V(2\pi)^2}\, \int dQ\, 
  |\mathcal{M}_{fi}|^2.
\end{equation}
The \emph{cross section} is defined as rate per \emph{flux}. In the cms
($\vec{p}_1 = - \vec{p}_2 \equiv p \vec{p}_1/|\vec{p}_1|$) the
flux is given by ($\rho$: particle density, $v$: relative velocity between
beam and target particle)
\begin{equation}
  \Phi = \rho v = \frac{v}{V} = \frac{|\vec{v}_1 - \vec{v}_2|}{V}
  = \frac{|\vec{p}_1 / E_1 - \vec{p}_2 / E_2 |}{V} =  \frac{p (E_1 + E_2)}{E_1
  E_2 V},
\end{equation}
and 
\begin{equation}
 s\equiv (p_1 + p_2)^2 = (E_1+E_2)^2 - (\vec{p_1}+\vec{p_2})^2 = (E_1 +
 E_2)^2,  
\end{equation}
such that the cross section for scattering of two spinless particles into a
final state containing $m_f$ particles characterized by a set of quantum
numbers $\Lambda$ reads\footnote{This formula for the cross section is valid
  in all reference frames where the momenta of the beam and target particles
  are parallel or antiparallel and also in the laboratory frame, see
  \cite[p.~154]{martin-spearman}.}
\begin{equation}\label{eq:46}
  \sigma_\Lambda(2\to m_f) = \frac{\Gamma(f,i)}{\Phi} = \frac{1}{16\pi^2
  p\sqrt{s}} 
  \int dQ_{m_f}\, |\mathcal{M}_{(f_\Lambda)i}|^2.  
\end{equation}
The \emph{total} cross section is the sum over 
\begin{itemize}
\item all numbers of particles $M(\sqrt{s})$ that can be produced at a given
  cms energy $\sqrt{s}$, and
\item for each number of particles all (accessible) set of quantum numbers
  $\Lambda$,
\end{itemize}
\begin{equation}
  \sigma_{\text{tot}} = \sigma(2\to \text{anything}) =
  \sum_{m_f=2}^{M} 
  \sum_\Lambda \sigma_\Lambda(2\to m_f), 
\end{equation}
where $M(s)$ is maximal number of particles that can be produced at a given
center of mass energy $\sqrt{s}$. Using ``anything'' in this sense, I adopted
the following normalization and completeness relation (cf.\ 
eq.~\eqref{eq:36}),
\begin{gather}\label{eq:43}
  \braket{p_1,\ldots,p_{m_f};\Lambda_{m_f}}{p_1',\ldots,p_{m_f}';\Lambda_{m_f}'}
  = \delta_{{m_f}{m_f'}}\prod_{m_f} (2\pi)^3 2E_{m_f}
  \delta^3(\vec{p}'_{m_f}-\vec{p}_{m_f}) 
  \delta_{\Lambda_{m_f},\Lambda_{m_f}'},\\
  \begin{split}
    \mathbbm{1} &= \sum_{m_f} \prod_{m_f} \int \frac{d^3p_{m_f}}{2E_{m_f}}
    \sum_{\Lambda_{m_f}}
    \ket{p_1,\ldots,p_{m_f};\Lambda_{m_f}}\bra{p_1,\ldots,p_{m_f};\Lambda_{m_f}}\\
    &= \sum_{m_f} \int d^4p_{\text{tot},m_f}\int dQ_{m_f}\, \sum_{\Lambda_{m_f}}
    \ket{p_{\text{tot}}}\ket{\psi}\bra{p_{\text{tot}}}\bra{\psi}\\
&\equiv \sum_{m_f} \left[ \int d^4p_{\text{tot},m_f}
    \ket{p_{\text{tot}}}\bra{p_{\text{tot}}} \otimes \int d\psi\,
    \ket{\psi}\bra{\psi}\right]\\
    &\equiv \int d\beta\, \ket{\beta}\bra{\beta}.
\end{split}
\end{gather}
The total cross section then can be written as
\begin{equation}\label{eq:44}
  \sigma_{\text{tot}} = \frac{1}{16\pi^2
  p\sqrt{s}} \sum_{m_f=2}^{M} 
  \sum_\Lambda \int dQ_{m_f}\, |\mathcal{M}_{(f_\Lambda)i}|^2.
\end{equation}
We further have with $p_f$ and $p_i$ the total four-momentum of the final and
initial state respectively (cf.\ equation~\eqref{eq:42} and following)
\begin{equation}
  \sum_{m_f=2}^{M} 
  \sum_\Lambda \int dQ_{m_f}\, = \int d^4p_f\,
  \delta^4(p_i-p_f) \sum_{m_f=2}^{M} 
  \sum_\Lambda \int dQ_{m_f}.
\end{equation}
Therefore,
\begin{equation}
  \sigma_{\text{tot}} = \frac{1}{16\pi^2 p\sqrt{s}}
   \int d\beta\, |\bra{\beta}T\ket{i}|^2.
\end{equation}

\subsection{Wave packets}\label{sec:wave-packets}

To obtain the relation between the respective $T$-matrix elements and the
decay distribution of $B^+\to \pi^-\pi^+K^+$ I follow the lines of
\cite[p.~140ff.]{martin-spearman}, which they use to define the scattering
cross section of two particles.

Let $\ket{\alpha}$ be the state vector that represents a collection of $B^+$'s
with a four-momentum distribution $\phi(p_B)$. 
\begin{equation}
  \ket{\alpha} = \int d^4p_B
  \delta(p_B^2-m_B^2)\theta(p_B^0)\phi(p_B)\ket{B^+,p_B}.
\end{equation}
If we consider a collection of $B^+$'s at rest, $\phi(p_B)$ will be narrowly
centered around $p_B=(m_B,0,0,0)$, where $m_B$ is the $B$ mass of around 5.279
GeV. The four-momentum eigenvectors $\ket{B^+,p_B}$ are normalized as
\begin{equation}\label{eq:29}
  \braket{B^+,p'}{B^+,p} = 2p_B^0\delta^3(\vec{p}-\vec{p}'),
\end{equation}
and the norm of the state $\ket{\alpha}$ is the time derivative of the
particle number density
\begin{equation}\label{eq:33}
  N_B = \int dt \braket{\alpha}{\alpha},
\end{equation}
where $N_B$ is the number of $B^+$'s produced in the experiment.

$S$-, $T$- and $\mathcal{M}$-matrices are defined by
\begin{equation}
  S = \mathbbm{1} + iT = \mathbbm{1} + i\delta^4(p_B-p_{\text{tot}})\mathcal{M},
\end{equation}
where $p_{\text{tot}}=\sum_i p_i$.

The transition amplitude for the $B$ to decay in a three-body state
$\ket{\beta}=\ket{\pi^-,p_1;\pi^+,p_2;K^+,p_3}$ of definite momenta of the
three decay products is given by
\begin{equation}
    \bra{\beta}S\ket{\alpha} = i\bra{\beta}T\ket{\alpha}. 
\end{equation}
The probability for a decay into a volume of momentum space is the square of
this amplitude integrated with the appropriate measure (see
\cite[p.~141]{martin-spearman})
\begin{equation}
  \begin{split}
    P(B;p_1,p_2,p_3) &= \int_{\vec{p}_1}^{\vec{p}_1+\vec{k}_1}
    \frac{d^3p_1}{2p_1^0} \int_{\vec{p}_2}^{\vec{p}_2+\vec{k}_2}
    \frac{d^3p_2}{2p_2^0} \int_{\vec{p}_3}^{\vec{p}_3+\vec{k}_3}
    \frac{d^3p_3}{2p_3^0} \\
    &\quad\times \int d^4p_B\,
    d^4p_B' \delta(p_B^2-m_B^2)\theta(p_B^0)\phi(p_B)\\
    &\quad\times \delta({p_B'}^2-m_B^2)\theta({p_B^0}')\phi^*(p'_B)\\
    &\quad\times \bra{B^+,p'_B}T^\dagger\ket{\alpha}
    \bra{\alpha}T\ket{B^+,p_B}.
  \end{split}
\end{equation}
The measure has to do with the normalization of equation~\eqref{eq:29}. To
simplify this expression by an approximation it is useful to express it in
term of elements of $\mathcal{M}$. We have
\begin{equation}
  \bra{\beta}T\ket{\alpha} = \int d^4p_B
  \delta(p_B^2-m_B^2)\theta(p_B^0)\phi(p_B)\delta^4(p_B-p_{\text{tot}})\bra{\beta}\mathcal{M}\ket{B^+,p_B}.
\end{equation}
If the momentum distribution $\phi(p_B)$ is sufficiently narrow, we can
approximate the above expression by assuming that the $\mathcal{M}$ element is
independent of $p_B$ and that it can therefore be pulled outside the
integral. For the decay probability we then obtain
\begin{equation}
  \begin{split}\label{eq:30}
    P(B;p_1,p_2,p_3) &= \int_{\vec{p}_1}^{\vec{p}_1+\vec{k}_1}
    \frac{d^3p_1}{2p_1^0} \int_{\vec{p}_2}^{\vec{p}_2+\vec{k}_2}
    \frac{d^3p_2}{2p_2^0} \int_{\vec{p}_3}^{\vec{p}_3+\vec{k}_3}
    \frac{d^3p_3}{2p_3^0} \\
    &\quad\times |\bra{\beta}\mathcal{M}\ket{B^+,p_B}|^2 \int d^4p_B\,
    d^4p_B' \delta(p_B^2-m_B^2)\theta(p_B^0)\phi(p_B)\\
    &\quad\times \delta({p_B'}^2-m_B^2)\theta({p_B^0}')\phi^*(p'_B)
    \delta^4(p_B-p_{\text{tot}}) \delta^4(p'_B-p_{\text{tot}}).
  \end{split}
\end{equation}
For further simplification I introduce the Fourier-transforms of the momentum
distributions,
\begin{equation}
  \Phi(x) = \int d^4p_B \delta(p_B^2-m_B^2)\theta(p_B^0)\phi(p_B) e^{ip_Bx}. 
\end{equation}
After some algebra one arrives then at
\begin{equation}
  \begin{split}\label{eq:32}
    P(B;p_1,p_2,p_3) &= \int_{\vec{p}_1}^{\vec{p}_1+\vec{k}_1}
    \frac{d^3p_1}{2p_1^0} \int_{\vec{p}_2}^{\vec{p}_2+\vec{k}_2}
    \frac{d^3p_2}{2p_2^0} \int_{\vec{p}_3}^{\vec{p}_3+\vec{k}_3}
    \frac{d^3p_3}{2p_3^0}\\
    &\quad\times |\bra{\beta}\mathcal{M}\ket{B^+,p_B}|^2 (2\pi)^{-8}
    \int d^4x \int d^4y \Phi(x) \Phi^*(y) e^{-ip_{\text{tot}}(x-y)}.
\end{split}
\end{equation}
The integrations in the space of final state momenta can be written as
\begin{equation}\label{eq:42}
  \prod_{i=1}^3 \int_{\vec{p}_i}^{\vec{p}_i+\vec{k}_i}
    \frac{d^3p_i}{2E_i} = \int d^4p_{\text{tot}}\, dQ_3
\end{equation}
with
\begin{equation}\label{eq:48}
  dQ_3 = \prod_{i=1}^3 \int_{\vec{p}_i}^{\vec{p}_i+\vec{k}_i}
    \frac{d^3p_i}{2E_i} \delta^4(p_1+p_2+p_3-p_{\text{tot}}).
\end{equation}
This can be checked using a test functions
\begin{equation}
f(\sum_i E_i,\sum_i\vec{p}_i,\vec{p}_1,\vec{p}_2,\vec{p}_3))\equiv\tilde{f}(\vec{p}_1,\vec{p}_2,\vec{p}_3)   
\end{equation}
as follows ($p_{\text{tot}}=\sum_{i=1}^3 p_i$,
$E_i=\sqrt{m_i^2+\vec{p}_i^2}$): 
\begin{multline}\label{eq:31}
\int d^4p_{\text{tot}}\, dQ_3
f(p^0_{\text{tot}},\vec{p_{\text{tot}}},\vec{p}_1,\vec{p}_2,\vec{p}_3) = \\
\prod_{i=1}^3 \int_{\vec{p}_i}^{\vec{p}_i+\vec{k}_i}
    \frac{d^3p_i}{2E_i}
    f(\sum_i E_i,\sum_i\vec{p}_i,\vec{p}_2,\vec{p}_3) = \\
\prod_{i=1}^3 \int_{\vec{p}_i}^{\vec{p}_i+\vec{k}_i}
    \frac{d^3p_i}{2E_i}\tilde{f}(\vec{p}_1,\vec{p}_2,\vec{p}_3).
\end{multline}
If we integrate over all values of $p_{\text{tot}}$ the above equality is not
exact. However, we will use this equality to evaluate expressions such as that
in equation~\eqref{eq:30}. There we assume the momentum distribution function
$\phi(p_B)$ to be narrowly centered around a certain value ($(m_B,0,0,0)$ in
the $B$ rest system). Because of the factors $\delta^4(p_B-p_{\text{tot}})
\delta^4(p'_B-p_{\text{tot}})$ the integrand in equation~\eqref{eq:30}
vanishes unless $p_{\text{tot}}=p_1+p_2+p_3$ takes on values near the central
value of $p_B$. So the mistake we make by integrating over all values of
$p_{\text{tot}}$ can be made arbitrarily small by making the momentum
distributions arbitrarily narrow. 

We can now make the substitution of equation~\eqref{eq:31} in
equation~\eqref{eq:32} and perform the integration over $p_{\text{tot}}$. This
gives
\begin{equation}\label{eq:34}
  P(B;p_1,p_2,p_3) = \int dQ_3 |\bra{\beta}\mathcal{M}\ket{B^+,p_B}|^2
    (2\pi)^{-4} 
    \int d^4x |\Phi(x)|^2.
\end{equation}
By Fourier-transforming back $\Phi(x)$, using
\begin{equation}
  \begin{split}
    d^4p \delta(p^2 - m^2) \theta(p^0) &= d^3pdp^0 \delta({p^0}^2 -
    (\vec{p}^2 + m^2) ) \\
    &= d^3pdp^0 \frac{1}{2\sqrt{\vec{p}^2 + m^2}}\theta(p^0)\\
    &\quad\times \left[\delta\left(p^0 +
    \sqrt{\vec{p}^2 + m^2} \right) + \delta\left(p^0 - \sqrt{\vec{p}^2 +
    m^2}\right) 
    \right]\theta(p^0)\\ 
    &= \frac{d^3p}{2p^0}
\end{split}
\end{equation}
and assuming once more that the momentum distributions are narrow one can show
that
\begin{equation}
  \int d^3x\, |\Phi(x)|^2 = \frac{(2\pi)^3}{p^0_B} \braket{\alpha}{\alpha} =
  \frac{(2\pi)^3}{p^0_B} \frac{dN_B}{dt}, 
\end{equation}
where the last equality is equivalent to equation~\eqref{eq:33}. The particle
density $\rho(x)$ is then 
\begin{equation}
  \rho(x) = \frac{p^0_B}{(2\pi)^3}|\Phi(x)|^2,
\end{equation}
such that
\begin{equation}
  N_B = \int d^4x\, \rho(x)
\end{equation}
is the total number of $B$'s produced in the experiment.

Thus equation~\eqref{eq:34} reads
\begin{equation}\label{eq:35}
    P(B;p_1,p_2,p_3) = \frac{N_B}{2\pi p^0_B} \int dQ_3
    |\bra{\beta}\mathcal{M}\ket{B^+,p_B}|^2. 
\end{equation}
The differential number of decay events that yield decay products in the
momentum volume $dQ_3$ can be read off to be
\begin{equation}
  \frac{dP(B;p_1,p_2,p_3)}{dQ_3} = \frac{N_B}{2\pi p^0_B}
  |\bra{\beta}\mathcal{M}\ket{B^+,p_B}|^2.   
\end{equation}
In this way the $\mathcal{M}$-elements between momentum eigenstates determine
the decay distribution of the $B$ meson that is characterized by $\phi(p_B)$
as a wave packet.

\section{Three-body phase space in a Dalitz plot}\label{sec:phase-space}


In this section I try to show, that the phase space volume $\int dQ_3$ is
constant over the Dalitz plot. There are alternative proofs in the literature,
\eg \cite[p.~140f.]{weinberg} and \cite[p.~159ff.]{martin-spearman}. However,
to me there are some rather involved steps necessary. Therefore I try here my
own derivation which is still similar to the one of
ref.~\cite{martin-spearman}.

The factors $\int\frac{d^3p_i}{2E_i}$ can be related to an integration over
the four-momenta with a $\delta$- and a $\theta$ function that restrict the
integration to the respective mass shells and to positive energy values
($f(p_i)$: test function, cf.\ \cite[p.~67]{weinberg} and
\cite[p.~497]{martin-spearman}):
\begin{equation}\label{eq:49}
  \begin{split}
    \int d^4p_i \delta(p_i^2 - m_i^2) \theta(p_i^0) f(p_i) &= \int d^3p_idp_i^0
    \delta({p_i^0}^2 - 
    (\vec{p}_i^2 + m_i^2) ) \theta(p_i^0) f(p^0_i,\vec{p}_i)\\
    &= \int d^3p_i\,dp_i^0 \frac{1}{2\sqrt{\vec{p}_i^2 + m_i^2}}\\
    &\quad\times \Bigg[\delta\left(p_i^0 +
        \sqrt{\vec{p}_i^2 + m_i^2} \right) \\
      &\quad + \delta\left(p_i^0 - \sqrt{\vec{p}_i^2 +
    m_i^2}\right)\Bigg]\theta(p_i^0) f(p^0_i,\vec{p}_i)\\
&= \int d^3p_i \frac{1}{2\sqrt{\vec{p}_i^2 + m_i^2}} f(\sqrt{\vec{p}_i^2 +
    m_i^2},\vec{p}) \\
    &= \int \frac{d^3p_i }{2E_i} f(E_i,\vec{p}_i),
\end{split} 
\end{equation}
where $E_i=\sqrt{\vec{p}_i^2 + m_i^2}$.

A decay of type $B^+\to\pi^-\pi^+K^+$ is completely characterized by the four
particle masses, the three-momenta and the energies. All these quantities can
be expressed in just two scalar variables. In chapter~\ref{cha:kin} I did this
explicitly with the variables $s_{12}$ and $s_{13}$. Since these two variables
are related to $E_3$ and $E_2$ by
\begin{equation}\label{eq:61}
  \begin{split}
    s_{12} &= m_B^2 + m_{3}^2 - 2m_B E_{3} \quad\text{and}\\
    s_{13} &= m_B^2 + m_{2}^2 - 2m_B E_{2}
\end{split}
\end{equation}
the latter suit equally well. Whether one chooses either pair to define the
axis of the Dalitz plot is mere convention. To show that phase space is
constant over the Dalitz plot it is, however, more convenient to work with
$E_2$ and $E_3$. 

\begin{figure}
  \centering
  \input{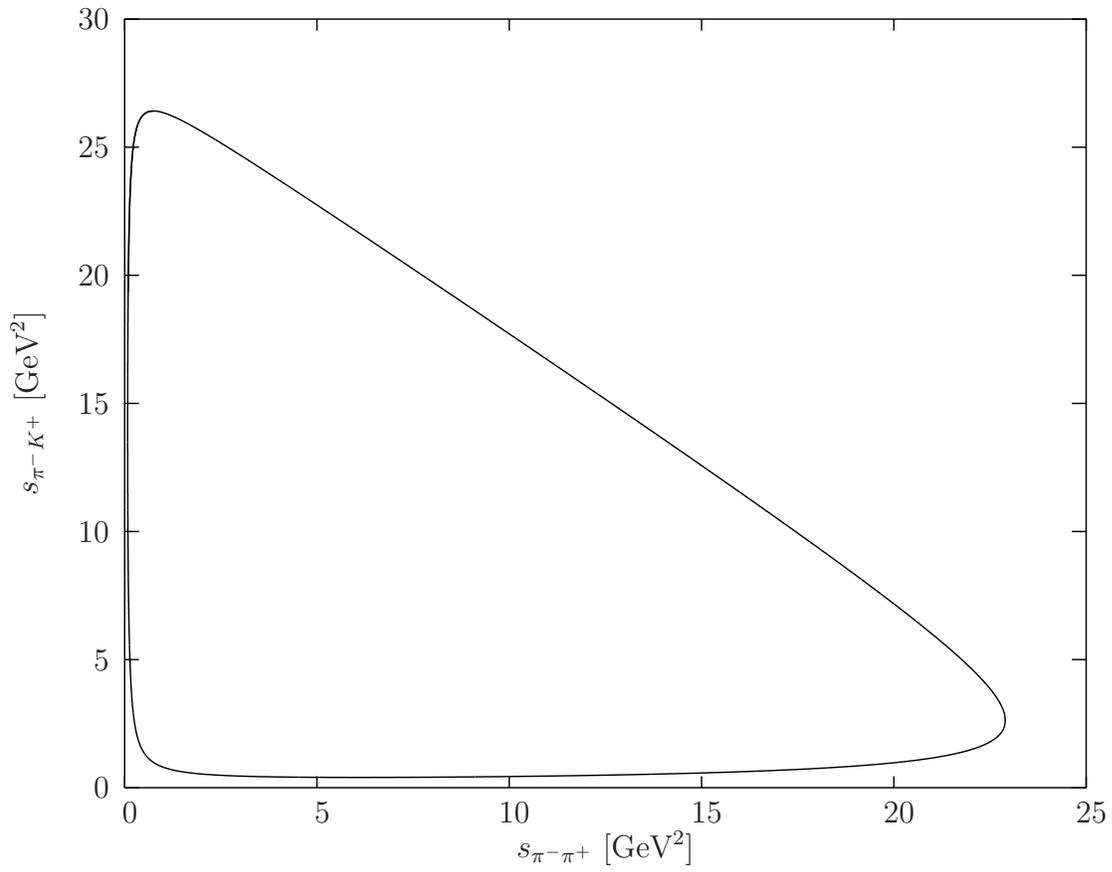}
  \caption[Boundary of the Dalitz plot.]{Boundary of the Dalitz plot according
    to equation~\eqref{eq:58}, with the labeling $1\leftrightarrow\pi^-$,
    $2\leftrightarrow\pi^+$ and $3\leftrightarrow K^+$. The $s$'s are given by
    equation~\eqref{eq:61}.}
   \label{fig:boundary}
\end{figure}

Since the decay is completely specified by just two variables, also the matrix
elements of $\mathcal{M}$ can be expressed in terms of these, 
\begin{equation}
  \bra{\beta}\mathcal{M}\ket{B^+,p_B} = \mathcal{M}(E_2,E_3).
\end{equation}
In the expression for the number of events in a certain region of phase space,
equation~\eqref{eq:35}, we can therefore pull $\mathcal{M}(E_2,E_3)$ outside
the integration over the remaining three variables $dq$, with
$dQ_3=dE_2\,dE_3\,dq$. So let us evaluate the integrals over the remaining
variables. I will do this with respect to the rest frame of the $B$. We start
with (see equation~\eqref{eq:48})
\begin{equation}\label{eq:50}
  \int dQ_3\, = \prod_{i=1}^3 \int \frac{d^3p_i}{2E_i}
  \delta^4(p_1+p_2+p_3-p_{B}).  
\end{equation}
The factor $\frac{d^3p_1}{2E_1}$ we replace with
\begin{equation}
  d^4p_1 \delta(p_1^2 - m_1^2) \theta(p_1^0),
\end{equation}
see equation~\eqref{eq:49}. For the factors $\frac{d^3p_3}{2E_3}$ and
$\frac{d^3p_2}{2E_2}$ we make the substitution
\begin{alignat}{2}
    \frac{d^3p_3}{2E_3} &= \frac{\frac{1}{2}|\vec{p_3}|d(|\vec{p_3}|^2)\,
      d(\cos\theta)\, d\phi}{2E_3} =  & \frac{1}{2} |\vec{p_3}|
      d(\cos\theta)\, d\phi\, 
      dE_3,\\
    \frac{d^3p_2}{2E_2} &= \frac{\frac{1}{2}|\vec{p_2}|d(|\vec{p_2}|^2)\,
      d(\cos\theta')\, d\phi'}{2E_2} = & \frac{1}{2} |\vec{p_2}|
      d(\cos\theta')\, d\phi'\, 
      dE_2,  
\end{alignat}
where without loss of generality we have chosen a coordinate system such that
the polar angle $\theta$ coincides with the angle between $\vec{p_2}$ and
$\vec{p_3}$. This will be of use later. Equation~\eqref{eq:50} now reads
\begin{multline}
  \int dQ_3\, = \int d^4p_1 \delta(p_1^2 - m_1^2) \theta(p_1^0)
  \delta^4(p_1+p_2+p_3-p_{B}) \\
  \times \frac{1}{2} d(\cos\theta)\, d\phi\, |\vec{p_3}| dE_3\,
  \frac{1}{2} d(\cos\theta')\, d\phi'\,
      |\vec{p_2}| dE_2. 
\end{multline}
The integration over $\phi$, $\phi'$ and $\theta'$ yield a factor $8\pi^2$
since the integrand does not depend on either of these variables. Performing
then the $p_1$ integration and the trivial $\phi$ integration yields
\begin{multline}
  \int dQ_3\, = 2\pi^2 \int \delta[m_B^2 + (E_2 + E_3)^2 - 2m_B(E_2 + E_3 )-
  |\vec{p_2}|^2 - |\vec{p_3}|^2 - 2 |\vec{p_2}||\vec{p_3}|\cos\theta
  - m_1^2] \\
  \times \theta(m_B - E_2 - E_3) |\vec{p_2}| |\vec{p_3}| d(\cos\theta)\,
  dE_2\, dE_3.
\end{multline}
Now the integral over the variables other than $E_2$ and $E_3$, from which only
$\mathcal{M}$ depends, is reduced to an integral over $\cos\theta$,
\begin{multline}\label{eq:51}
  \int_{-1}^{1} \delta[m_B^2 + (E_2 + E_3)^2 - 2m_B(E_2 + E_3 )-
  |\vec{p_2}|^2 - |\vec{p_3}|^2 - 2 |\vec{p_2}||\vec{p_3}|\cos\theta
  - m_1^2] \\
  \times |\vec{p_2}| |\vec{p_3}| d(\cos\theta)\\
  = \frac{1}{2} \int_{-1}^{1} \delta[\cos\theta - z_0] d(\cos\theta),
\end{multline}
with
\begin{equation}
  z_0 = \frac{m_B^2 + (E_2 + E_3)^2 - 2m_B(E_2 + E_3 )-
  |\vec{p_2}|^2 - |\vec{p_3}|^2 -m_1^2}{2
  |\vec{p_2}||\vec{p_3}|}.  
\end{equation}
The integral then gives
\begin{multline}\label{eq:52}
  \frac{1}{2} \theta(1 - z_0) = \frac{1}{2} \theta\left(1 - \frac{m_B^2 + (E_2
      + E_3)^2 - 2m_B(E_2 + E_3 )- |\vec{p_2}|^2 - |\vec{p_3}|^2 -m_1^2}{2
      |\vec{p_2}||\vec{p_3}|}\right)\\
  = \frac{1}{2} \theta\left[2 |\vec{p_2}||\vec{p_3}| - m_B^2 - (E_2 + E_3)^2 +
    2m_B(E_2 + E_3 ) +
    |\vec{p_2}|^2 + |\vec{p_3}|^2  + m_1^2 \right]\\
  = \frac{1}{2} \theta\left[2 \sqrt{(E_2^2 - m_2^2)(E_3^2 - m_3^2)} + m_1^2 -
    m_B^2 - m_2^2 - m_3^2 - 2E_2 E_3 + 2m_B(E_2 + E_3 )\right].
\end{multline}
Altogether, we have found that
\begin{multline}\label{eq:53}
  \int dQ_3\, = \pi^2 \int \theta(m_B - E_2 - E_3) \\
  \times \theta\left[2 \sqrt{(E_2^2 - m_2^2)(E_3^2 - m_3^2)} + m_1^2 -
    m_B^2 - m_2^2 - m_3^2 - 2E_2 E_3 + 2m_B(E_2 + E_3 )\right] dE_2\, dE_3. 
\end{multline}
The theta functions define a boundary in the $E_2$-$E_3$ plane outside which
the phase space is zero and inside which it is constant $\pi^2$. The first
theta function gives a straight line in the $E_2$-$E_3$ plane. The more
interesting characteristics of the Dalitz plot boundaries are given in the
second theta function. The condition that its argument is zero corresponds to
the energy-momentum conservation in the special configuration of $\cos\theta =
1$, as can be seen as follows.\footnote{Also $\cos\theta = -1$ corresponds to
  events on the boundary of the Dalitz plot, see
  figures~\ref{fig:z12}--\ref{fig:z23}.}
\begin{align}
  \vec{p_1}+\vec{p_2}+\vec{p_3} &= 0,\\
  E_1 + E_2 + E_3 &= m_B.
\end{align}
The first equation is more useful in the form ($\cos\theta=1$)
\begin{equation}
  \vec{p_2}^2+\vec{p_3}^2 +2|\vec{p_2}||\vec{p_3}| =  \vec{p_1}^2.
\end{equation}
Expressing all momenta in terms of $E_2$ and $E_3$ and the masses, using the
energy conservation condition yields
\begin{equation}
  E_2^2 - m_2^2 + E_3^2 - m_3^2 +2\sqrt{(E_2^2
    - m_2^2)(E_3^2 - m_3^2)} =  (m_B - E_2 -E_3)^2 -m_1^2,
\end{equation}
which is equivalent to
\begin{equation}\label{eq:58}
 2\sqrt{(E_2^2
    - m_2^2)(E_3^2 - m_3^2)} + m_1^2- m_2^2 - m_3^2 - m_B^2 + 2m_B(E_2+E_3) -
 2 E_2E_3 = 0 
\end{equation}
where the left-hand side is the argument of the theta function in
equation~\eqref{eq:53}. The points where this equation is satisfied lie on the
closed curve shown in figure~\ref{fig:boundary}.

That the configuration with $\cos\theta = 1$ gives the boundary of the Dalitz
plot was reflected in the above calculation in the transition from the delta
to the theta function, \ie from equation~\eqref{eq:51} to
equation~\eqref{eq:52}. 


\section{Elastic scattering of two spinless particles}

Scattering of two particles, say $a$ and $b$, where the final state consists
also of two particles $a$ and $b$ is called \emph{elastic (two-body)
  scattering}. In this case the $\mathcal{M}$-matrix elements depend at most
on the two independent scalar variables $s$ (total energy squared in the
center-of-mass system (cms)) and $t$ (4-momentum transfer),
\begin{equation}
  \begin{split}
    s &= (p_1 + p_2)^2\\
    t &= (p_1 - p_3)^2,
\end{split}
\end{equation}
where the particle $a$ ($b$) is labeled by ``1'' (``2'') before the
scattering and by ``3'' (``4'') after it, and the squaring is meant to be with
respect to the Lorentz-invariant scalar product $p^2=(p^0)^2-\vec{p}^2=
(p^0)^2 - (p^1)^2 - (p^2)^2 - (p^3)^2$. That the matrix element does not
depend on more than these two kinematic variables has the following reason:
The matrix element is a Lorentz-invariant quantity, \ie a scalar. It can
therefore be only a function of scalars. The only scalars that can be formed
out of $p_1$ and $p_2$ are $s$ and $t$ \cite[p.~168]{martin-spearman}. One can
convince oneself also by expressing explicitly any kinematical quantities in
terms of these two, similarly to what I have done for the kinematics of a
three-body decay in chapter~\ref{cha:kin}.

The momentum transfer $t$ and the scattering angle $\theta$ (in the
cms) are related ($m_a=m_b\equiv m$, $q$: cms momentum):
\begin{equation}\label{eq:8}
  \begin{split}
    t &= 2m_a^2 - 2E_a^2 + \vec{p}_1\vec{p}_3\\
    &= 2p^2(\cos\theta - 1).
\end{split}
\end{equation}
The $\mathcal{M}$-matrix elements may thus be written as a function of $s$ and
$z\equiv\cos\theta$,
\begin{equation}
  \bra{ab}\mathcal{M}\ket{ab}=\mathcal{M}(s,z).
\end{equation}

The differential cross section for elastic scattering, $d\sigma(ab\to ab)$,
per solid angle $d\Omega\equiv dz d\phi$ is given in terms of the
$\mathcal{M}$-matrix element,
\begin{align}
    \frac{d\sigma}{d\Omega} &= \frac{1}{64\pi^2 s}
    |\mathcal{M}(s,z)|^2,\\ 
    \frac{d\sigma}{dz} &= \frac{1}{32\pi s}
    |\mathcal{M}(s,z)|^2.
\end{align}
It is sometimes useful to work with the amplitude
\begin{equation}\label{eq:54}
  f(s,z) = \frac{1}{8\pi\sqrt{s}} \mathcal{M}(s,z) ,
\end{equation}
such that
\begin{equation}
  \frac{d\sigma}{d\Omega} = |f(s,z)|^2.
\end{equation}

\section{Unitarity}\label{sec:unit}

Collecting all quantities characterizing a free-particle state in one single
symbol $\alpha$ (or $\beta$ or the like), cf.\ equation~\eqref{eq:43}, we have
the following short-hand notation for the normalization,
\begin{equation}
  \braket{\alpha}{\beta} = \delta(\alpha-\beta).
\end{equation}
Using the completeness relation~\eqref{eq:43} we find
\begin{multline}
  \delta(\alpha-\gamma) = {_\text{out}\braket{\alpha}{\gamma}_\text{out}} =
  \int d\beta\,
  {_\text{out}\braket{\alpha}{\beta}_\text{in}}{_\text{in}\braket{\beta}{\gamma}_\text{out}}
  \\
=\int d\beta\,
  S_{\alpha\beta}S^{\dagger}_{\beta\gamma} = [SS^{\dagger}]_{\alpha\gamma}.
\end{multline}
Applying in a similar vein the completeness relation using the out-states one
obtains
\begin{equation}
  [S^{\dagger}S]_{\alpha\gamma} = \delta(\alpha-\gamma).
\end{equation}
These two relations mean that $S$ is an \emph{unitary} matrix.

In section~\ref{sec:event-distr} I showed in some detail the two methods to
deal with state vectors normalized to delta functions. I will often assume
that one can anyhow discretize the states and normalize them to unity for
instance as done in section~\ref{sec:box}. Then, the probability that a system
in an initial state $\ket{i}$ evolves in \emph{any} final state is one. This
postulate is sometimes called \emph{probability conservation}. The probability
for the transition $\ket{\alpha}_\text{in}\to \ket{\alpha}_\text{out}$ is
given by the absolute value squared of the corresponding $S$-matrix element.
The condition for probability conservation thus reads
\begin{equation}\label{eq:24}
  \sum_f |\bra{f}S\ket{i}|^2 = \sum_f \bra{i}S^\dagger\ket{f} \bra{f}S\ket{i}
  = \bra{i}S^\dagger S\ket{i} = 1. 
\end{equation}
This means that the diagonal elements of the operator $S^\dagger S$ are 1. The
off-diagonal elements are zero as can be shown by the following argument.

Take two orthonormal state vectors $\ket{m}$ and $\ket{n}$, and set the
initial state $\ket{i}$ to $(|\alpha|^2 + |\beta|^2)^{-1/2}(\alpha\ket{m} +
\beta\ket{n})$ with arbitrary $\alpha$ and $\beta$. Eq.~\eqref{eq:24} then
implies
\begin{multline}
\frac{1}{|\alpha|^2 + |\beta|^2} (|\alpha|^2 \bra{m}S^\dagger S\ket{m} +
|\beta|^2 \bra{n}S^\dagger S\ket{n} \\
+ \alpha^*\beta\bra{m}S^\dagger S\ket{n} +S^\dagger S\ket{n} +
\alpha\beta^*\bra{n}S^\dagger S\ket{m})  = 1. 
\end{multline}
Since according to eq.~\eqref{eq:24} the diagonal elements are 1 we get
\begin{equation}
  \alpha^*\beta\bra{m}S^\dagger S\ket{n} + \alpha\beta^*\bra{n}S^\dagger
  S\ket{m} = 0,
\end{equation}
and because $\alpha$ and $\beta$ are arbitrary it must be that
\begin{equation}
  \bra{m}S^\dagger S\ket{n} = \bra{n}S^\dagger
  S\ket{m} = 0.
\end{equation}
All in all we have derived that $S^\dagger S$ is the unit operator,
\begin{equation}\label{eq:25}
 S^\dagger S = \mathbbm{1}. 
\end{equation}
Multiplying eq.~\eqref{eq:25} from the left with $S$ and from the right with
$S^{-1}$ we get that also $SS^\dagger$ is the unit operator. This means that
$S$ is a \emph{unitary} operator.

Plugging $S=\mathbbm{1}+iT$ into eq.~\eqref{eq:25} yields
\begin{equation}\label{eq:26}
  T^\dagger T = i(T^\dagger -T).
\end{equation}

\section{Partial wave amplitudes (without spin)}

The mathematical objects that are interpreted as elementary particles are
irreducible representations of the Lorentz group.  For the question whether in
scattering of two particles resonances or new particles are formed by these,
it is therefore important to decompose the two-body states $\ket{\psi}$, which
are in general \emph{not} an irreducible representation of the rotation group,
into states of an irreducible representation. In other words we have to expand
it in eigenstates of angular momentum.
\begin{equation}
 \ket{\psi} = \sum_{J,M} C_{JM}\ket{JM,\Lambda} 
\end{equation}
Such an expansion is also useful inasmuch
as angular momentum is a conserved quantity, such that the matrix element can
only be non-zero between states of the same total angular momentum and third
component, \ie we can write
\begin{equation}\label{eq:55}
  \bra{J'M',\Lambda'}\mathcal{M}\ket{JM,\Lambda} =
  \delta_{J'J}\delta_{M'M}\mathcal{M}^J_{\Lambda'\Lambda}. 
\end{equation}
As a consequence of the Wigner-Eckart theorem
$\mathcal{M}^J_{\Lambda'\Lambda}$ does, as the notation suggests, not depend
on the third component $M$ of angular momentum. If we expand in- and
out-states in eigenstates of angular momentum we thus obtain\footnote{For
  details see \eg \cite[p.~177ff.]{martin-spearman}. $P_J$: Legendre
  polynomials.} 
\begin{equation}
  \begin{split}
    \bra{\psi_f}\mathcal{M}\ket{\psi_i} &= \sum_{J,M,J',M'} C_{JM}C^*_{J'M'}
    \bra{J'M',f}T\ket{JM,i} \\
    &= \sum_{J,M,J',M'} C_{JM}C^*_{J'M'}\delta_{J'J}\delta_{M'M}
    \mathcal{M}^J_{\Lambda'\Lambda}\\ 
    &= \sum_{J,M}|C_{JM}|^2 \mathcal{M}^J_{\Lambda'\Lambda} \\
    &= \frac{1}{4\pi} \sum_J (2J+1) P_J(z) \mathcal{M}^J_{\Lambda'\Lambda},
\end{split}
\end{equation}
Note that $\bra{\psi_f}\mathcal{M}\ket{\psi_i}$ is a function of $s$ and
$z\equiv\cos\theta$, while $\mathcal{M}^J_{\Lambda'\Lambda}$ is only a
function of $s$,
\begin{equation}
\mathcal{M}(s,z) = \frac{1}{4\pi} \sum_J (2J+1) P_J(z)
  \mathcal{M}^J(s),
\end{equation}
\ie the $\theta$ dependence is entirely contained in the Legendre polynomials
$P_J$, and the dependence on the cms energy is contained in the partial wave
$\mathcal{M}$-matrix element $\mathcal{M}^J(s)$.  Again (cf.\ 
equation~\eqref{eq:54}, we may define a sometimes more convenient quantity,
the partial wave amplitude
\begin{equation}
  f^J(s) = \frac{1}{32\pi^2 \sqrt{s}} \mathcal{M}^J(s)
\end{equation}
such that\footnote{I have dropped the indices ``$\Lambda$'' for brevity.}
\begin{equation}\label{eq:23}
f(s,z) = \sum_J (2J+1)P_J(z)f^J(s).
\end{equation}
The inverse of equation~\eqref{eq:23} is
\begin{equation}
  f^J(s) = \frac{1}{2} \int_{-1}^1 P_J(z) f(s,z) dz.
\end{equation}

\section{Phase shifts}

The postulate of probability conservation says that an initial state is to be
found with certainty in \emph{any} of the final states. By definition of the
concept of a conserved quantity, the probability of finding the system in a
final state that differs in a conserved quantity is zero. Two conserved
quantities that characterize hadronic states are the total angular momentum
and isospin.\footnote{Isospin is only \emph{approximately} conserved in the
  strong interactions.} In hadron spectroscopy it is therefore adequate to
decompose in- and out-states in eigenstates of these two conserved quantities.
As a consequence of the conservation laws unitarity relations hold for each
component.

As in the discussion of the partial wave expansion we can define a new
scattering matrix element by factoring out Kronecker deltas representing the
conservation of total angular momentum and isospin and their respective third
components (cf.\ eq.~\eqref{eq:55}).
\begin{equation}
  \bra{J'M'I'I_3',\Lambda'}\mathcal{M}\ket{JMII_3,\Lambda} =
  \delta_{J'J}\delta_{M'M}\delta_{I'I}\delta_{I'_3I_3}\mathcal{M}^J_I
\end{equation}
where $\Lambda$ symbolizes all quantum numbers other than angular momentum and
isospin. $\mathcal{M}^J_I$ depends in general on $\Lambda$; I dropped the
indices for brevity. As consequence of the Wigner-Eckart theorem it does not
depend on the third components of angular momentum and isospin.

It is possible (cf.~\cite[p.~152]{weinberg}) and useful to define an $S$
operator in the little Hilbert space where sofar we only have defined the
elements of $\mathcal{M}$, such that
\begin{equation}
  \bra{J'M'I'I_3',\Lambda'}S\ket{JMII_3,\Lambda} =
  \delta_{J'J}\delta_{M'M}\delta_{I'I}\delta_{I'_3I_3}S^J_I,
\end{equation}
and
\begin{equation}
 S^J_I(\Lambda'\Lambda) =
 \delta_{J'J}\delta_{M'M}\delta_{I'I}\delta_{I'_3I_3}\delta_{\Lambda'\Lambda}
 - i\mathcal{M}^J_I(\Lambda'\Lambda).
\end{equation}

The condition of probability conservation (eq.~\eqref{eq:24}) then takes the
form
\begin{equation}
  1 =
  \sum_{J'M'I'I_3'\Lambda'}|\bra{J'M'I'I_3',\Lambda'}S\ket{JMII_3,\Lambda}|^2
  =  |S^J_I(\Lambda'\Lambda)|^2.
\end{equation}
From this one can derive in a similar vein as the unitarity of $S$ in general
has been derived (section~\ref{sec:unit}) that also each partial wave
$S$-operator, $S^J_I$, is unitary and that $\mathcal{M}^J_I$ satisfies the
same constraint as $\mathcal{M}$,
\begin{gather}
  (S^J_I)^\dagger S^J_I = \mathbbm{1}, \\
  (\mathcal{M}^J_I)^\dagger \mathcal{M}^J_I = i((\mathcal{M}^J_I)^\dagger - \mathcal{M}^J_I).\label{eq:3}
\end{gather}
If only the elastic channel is open, \ie $\Lambda'=\Lambda$, then $S^J_I$ is
just a complex number. Unitarity is then the condition that this complex
number be of modulus unity and so can be written in the general form
\begin{equation}
  S^J_I = e^{2i\delta^J_I},
\end{equation}
where we have introduced the \emph{phase shift} $\delta\in\mathbb{R}$. (The
factor~2 is convention.)

If also other than the elastic channels are allowed, it is useful to introduce
another (real) quantity that characterizes a certain scattering process, the
\emph{elasticity} $\eta_J$. This is in a given partial wave the square root of
the fraction of the probability of elastic scattering and the total
probability of any scattering, which is unity,
\begin{equation}
   (\eta^J_I)^2 =
   \frac{|S^J_I(\Lambda\Lambda)|^2}{\sum_{\Lambda'}|S^J_I(\Lambda'\Lambda)|^2}=|S^J_I(\Lambda\Lambda)|^2.   
\end{equation}
With this definition, the elastic partial wave amplitude
$S^J_I(\Lambda\Lambda)$ can be written in terms of the elasticity factor and
the phase shift $\delta^J_I$ in the respective partial wave characterized by
angular momentum and isospin,
\begin{equation}
 S^J_I(\Lambda\Lambda) = \eta^J_I e^{2i\delta^J_I}.
\end{equation}
In the case of purely elastic scattering, \ie when no other final channels are
available as the initial one, we have $\eta_J=1$. Using this form of
$S^J_I(\Lambda\Lambda)$ the partial wave $\mathcal{M}$-matrix element for elastic
scattering takes the form
\begin{multline}\label{eq:57}
  \mathcal{M}^J_I(\Lambda\Lambda) = \frac{(S^J_I(\Lambda\Lambda)-1)}{i} = \frac{\eta^J_I
  e^{2i\delta^J_I} -1}{i} \\ 
  = 2\eta^J_I\frac{e^{i\delta^J_I}(e^{i\delta^J_I}-e^{-i\delta^J_I})}{2i} +
  \frac{\eta^J_I-1}{i} = 2\eta^J_I e^{i\delta^J_I}\sin\delta^J_I + \frac{\eta^J_I-1}{i}.
\end{multline}
For elastic scattering we have ($\eta^J_I=1$)
\begin{equation}\label{eq:5}
  \mathcal{M}^J_I(\Lambda\Lambda) = 2e^{i\delta^J_I}\sin\delta^J_I.
\end{equation}
If the elasticity tends to zero, $\eta^J_I=0$, the corresponding partial wave
$\mathcal{M}$-element tends to
\begin{equation}
  \mathcal{M}^J_I(\Lambda\Lambda) = i.
\end{equation}

\section{Resonances}

\subsection{Argand diagram}

The relation (see eq.~\eqref{eq:57})
\begin{equation}
  \mathcal{M}^J_I(\Lambda\Lambda) \equiv \mathcal{M}^J_I = \frac{\eta^J_I
  e^{2i\delta^J_I} -1}{i}
\end{equation}
can be transformed in
\begin{equation}
  \mathcal{M}^J_I = i - i \eta^J_I \cos2\delta^J_I + \eta^J_I \sin2\delta^J_I.
\end{equation}
From this we can see that the complex number $\mathcal{M}^J_I$ can be
represented as a point in the complex plane in the interior or on the boundary
of the \emph{unitarity circle}. 

\begin{figure}
  \centering
  \begin{pspicture}(-5,-1.5)(5,10)
\pscircle(0,4){4}
\psline[arrowsize=5pt]{->}(-5,0)(5,0)
\psline[arrowsize=5pt]{->}(0,0)(0,10)
\psline[arrowsize=5pt]{->}(0,0)(2.5,5.5)
\psline[arrowsize=5pt]{->}(0,4)(2.5,5.5)
\psarc{->}(0,4){.5}{-90}{30.96}
\uput[180](0,4){$i$}
\psline(4,.1)(4,-.1)
\rput(4,-.5){1}
\rput(.8,3.6){$2\delta^J_I$}
\rput(1,5){$\eta^J_I$}
\rput(1.8,2.5){$\mathcal{M}^J_I$}
\rput(5,-1){Re$\mathcal{M}^J_I$}
\rput(-1,9){Im$\mathcal{M}^J_I$}
\psarc[linewidth=.1](0,4){4}{-90}{-30}
\psbezier[linewidth=.1]{->}(3.46,2)(3.7,2.3)(3.9,3.2)(2.5,5.5)
\psline[arrowsize=5pt]{->}(3.5,7.5)(2.9,6.9)
\rput[l](3.7,7.7){Unitarity circle}
\end{pspicture}

  \caption{Argand diagram.}
  \label{fig:argand}
\end{figure}
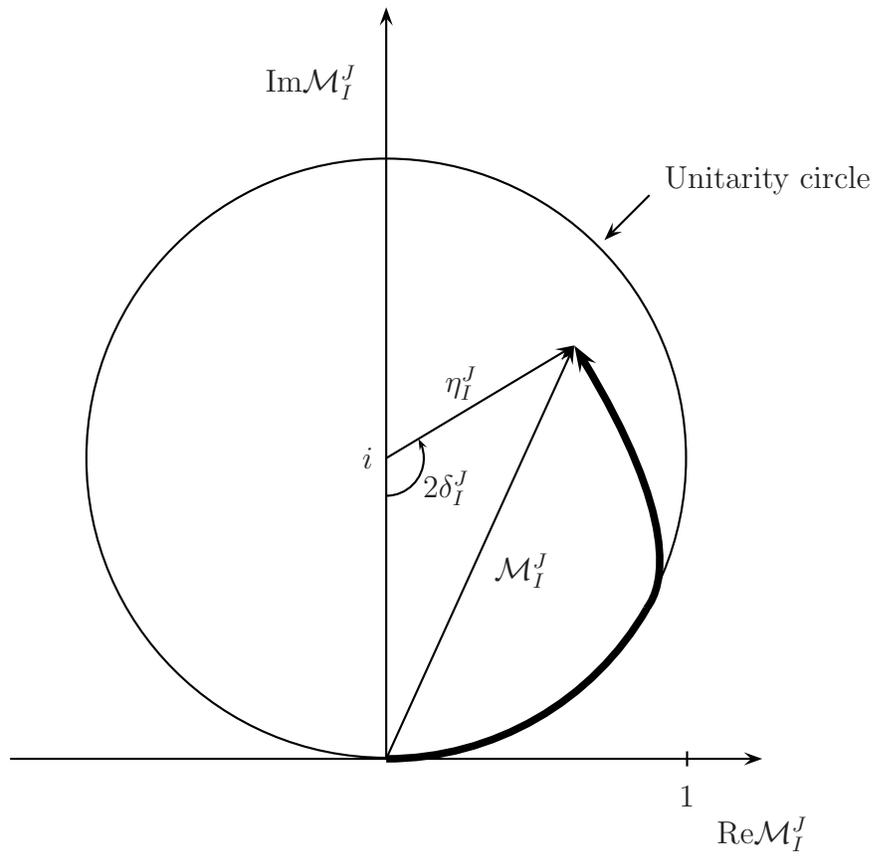

\subsection{Breit-Wigner resonance}

\begin{figure}
  \centering
  \subfigure[Breit-Wigner; the amplitude is
    maximal at $s=m_R^2$.]{\includegraphics[scale=.7]{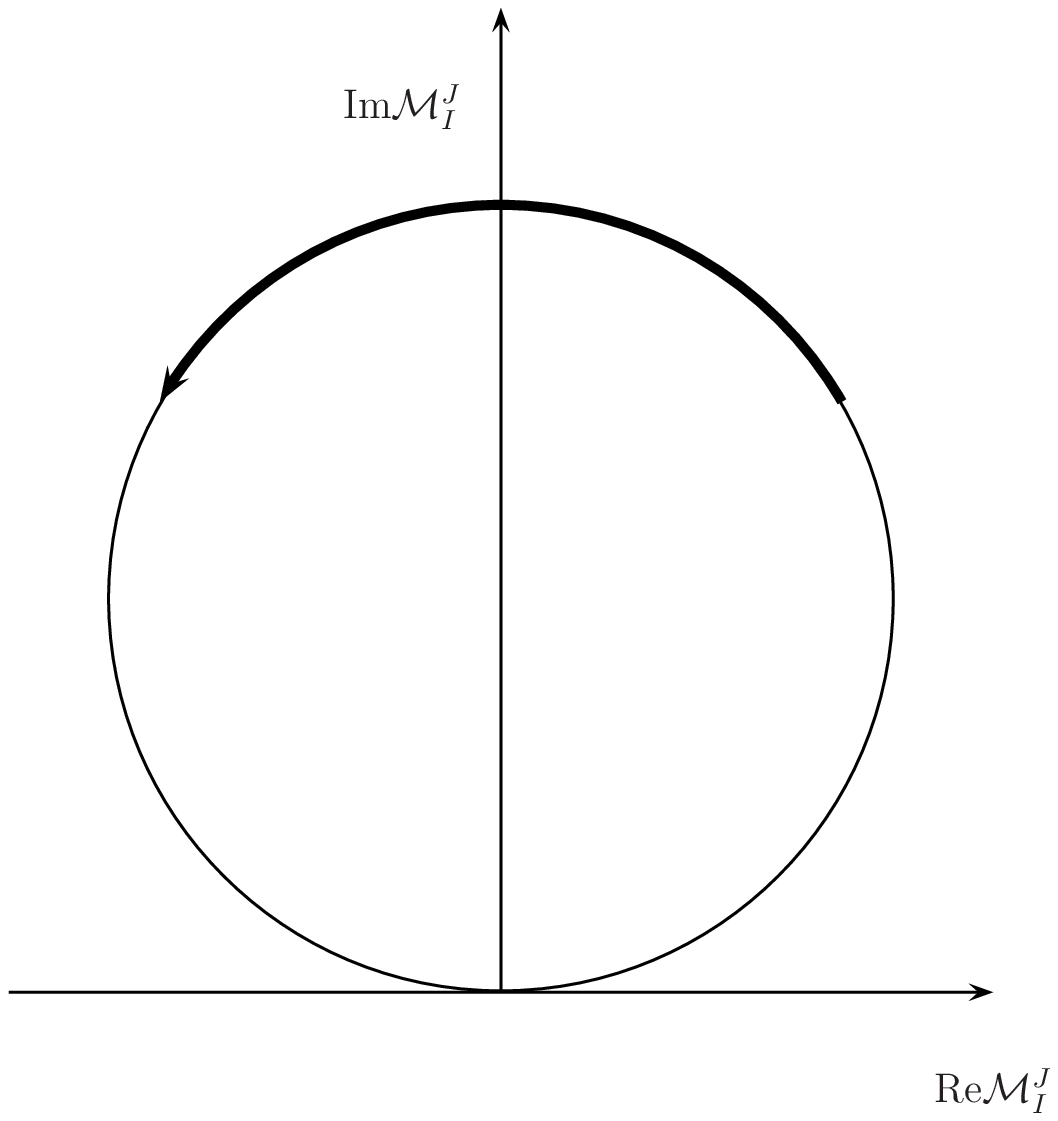}}\quad
  \subfigure[Resonance with background phase of about $\pi/2$.
    The amplitude is zero at $s=m_R^2$.]{\includegraphics[scale=.7]{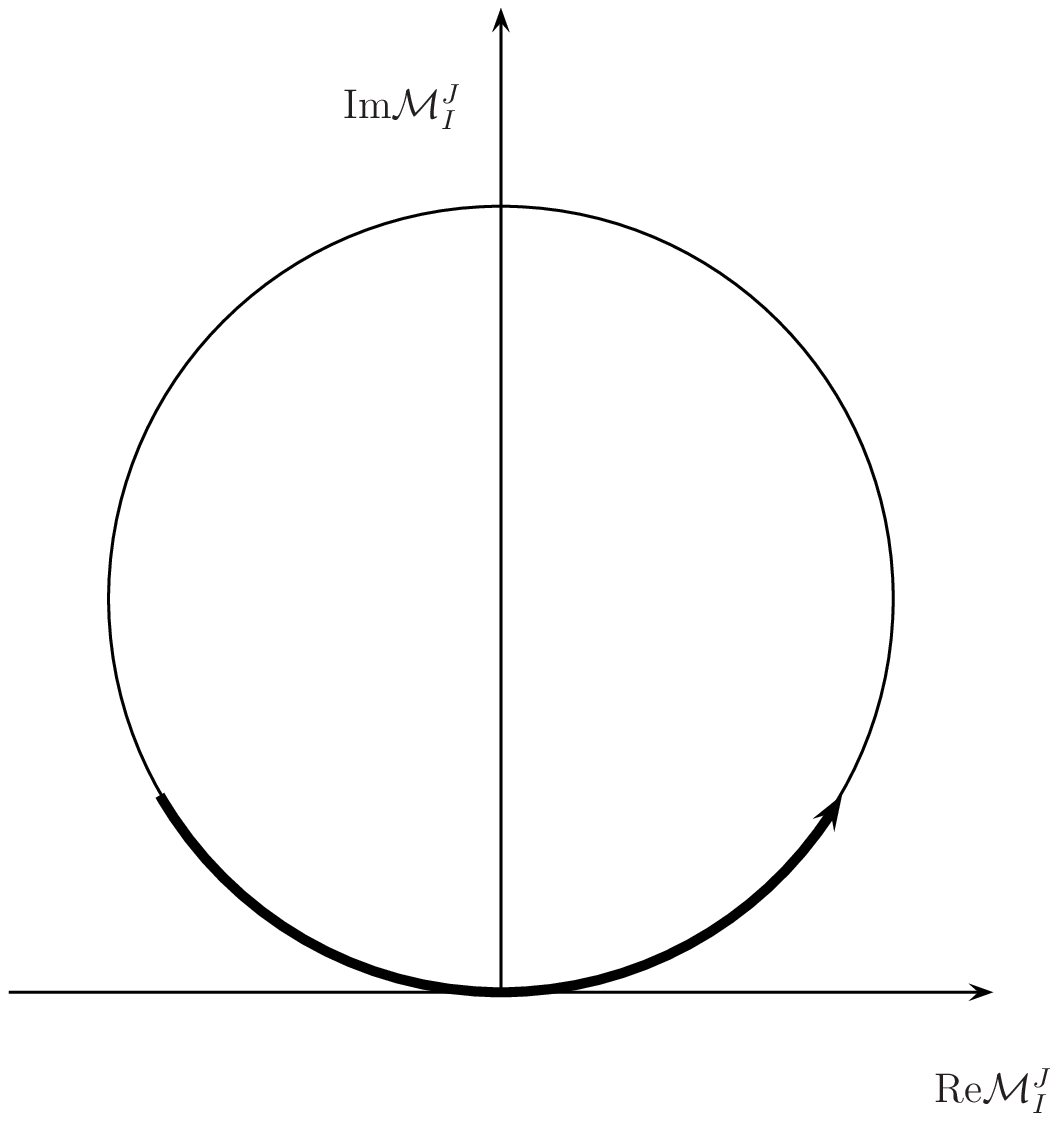}}
  \caption{Resonances in an Argand diagram.}
  \label{fig:argand-res}
\end{figure}

The epitome of a resonance is a peak in a partial cross section with definite
total spin; a peak that is caused by an amplitude with elasticity $\eta^J=1$
and a rapid shift of the phase with energy through $\delta^J=\pi/2$. In the
Argand diagram this is represented by a rapid movement through the top of the
unitarity circle, see figure~\ref{fig:argand-res}~(a).

Such a resonance is called \emph{Breit-Wigner} after the inventors of the
formula for the cross section describing such a resonance. For the scattering
amplitude we have in the Breit-Wigner case (dropping isospin labels) in the
neighborhood of $m_R^2$, the mass of the resonance squared,
\begin{equation}\label{eq:9}
  \mathcal{M}^J_I = \frac{\Gamma}{(m_R^2-s) - i\Gamma/2},
\end{equation}
with $\Gamma$ the width of the resonance. The phase shift is then given by
\begin{equation}
\tan \delta =  \frac{\Gamma/2}{m_R^2-s}.
\end{equation}
If $s$ changes from values below $m_R^2$ to values above, $\delta$ changes
rapidly from zero through $\pi/2$ to $\pi$ (or odd multiples thereof).

\subsection{Background} 

In general resonances are as in the Breit-Wigner case characterized by the two
parameters $\Gamma$ and $m_R$ and a rapid phase shift. However, these
quantities are in general not directly related to the position and width of a
peak in cross section. They rather denote the position of the pole of
$\mathcal{M}$.\footnote{For a discussion of pole parameters and Breit-Wigner
  parameters see \cite{against-bw}.} Also the rapid phase shift does not
necessarily occur at $\pi/2$ (or a multiple thereof). It may happen that the
rapid phase shift begins only when the phase has already appreciable values
different from zero.

\begin{figure}
  \centering
  \includegraphics[width=\linewidth]{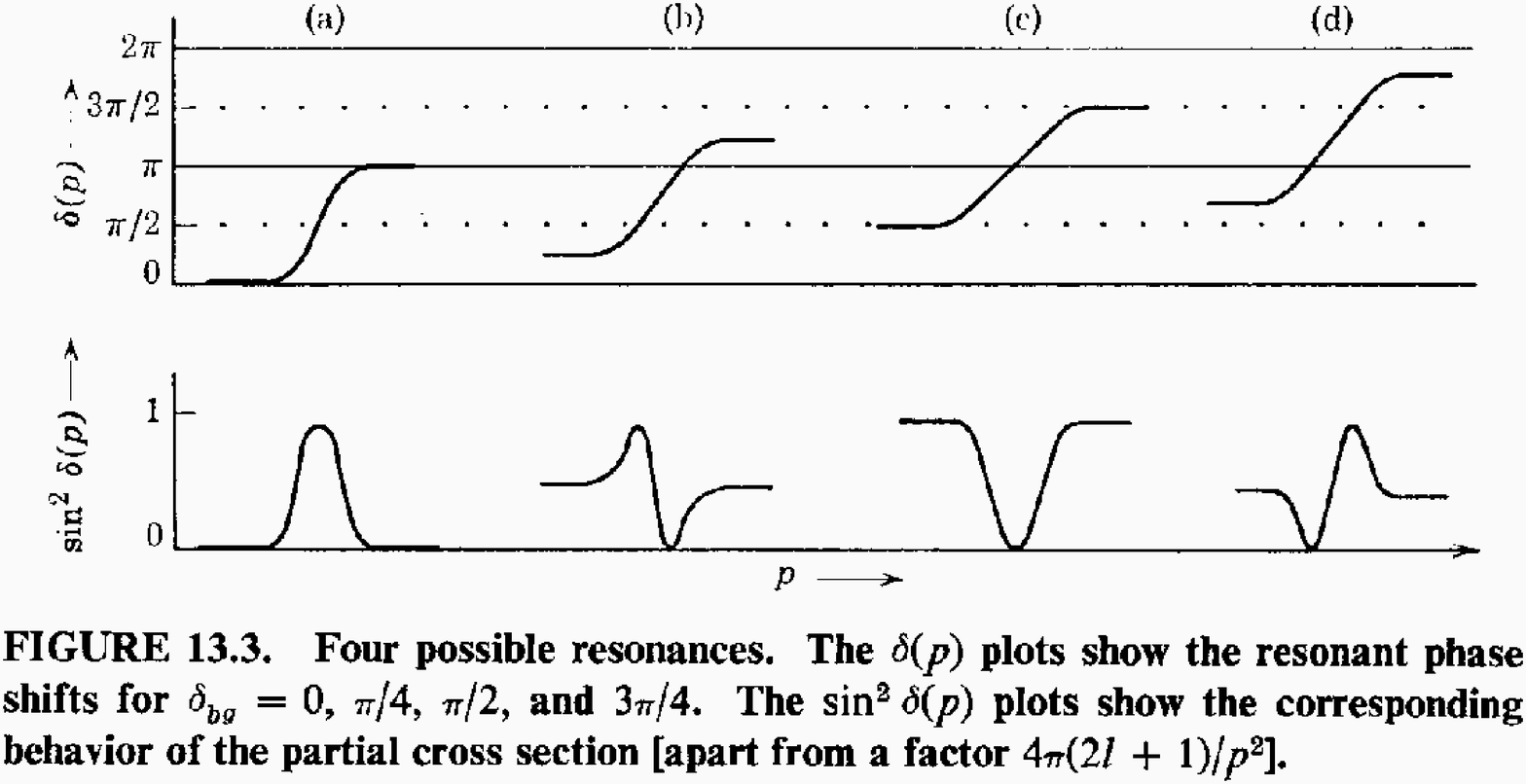}
  \caption[Phase shifts and cross sections.]{Phase shifts and corresponding
    characteristic feature of the cross section; from \cite[p.~242]{taylor}.}
  \label{fig:taylor}
\end{figure}

\begin{figure}
  \centering
  \includegraphics[width=.6\linewidth]{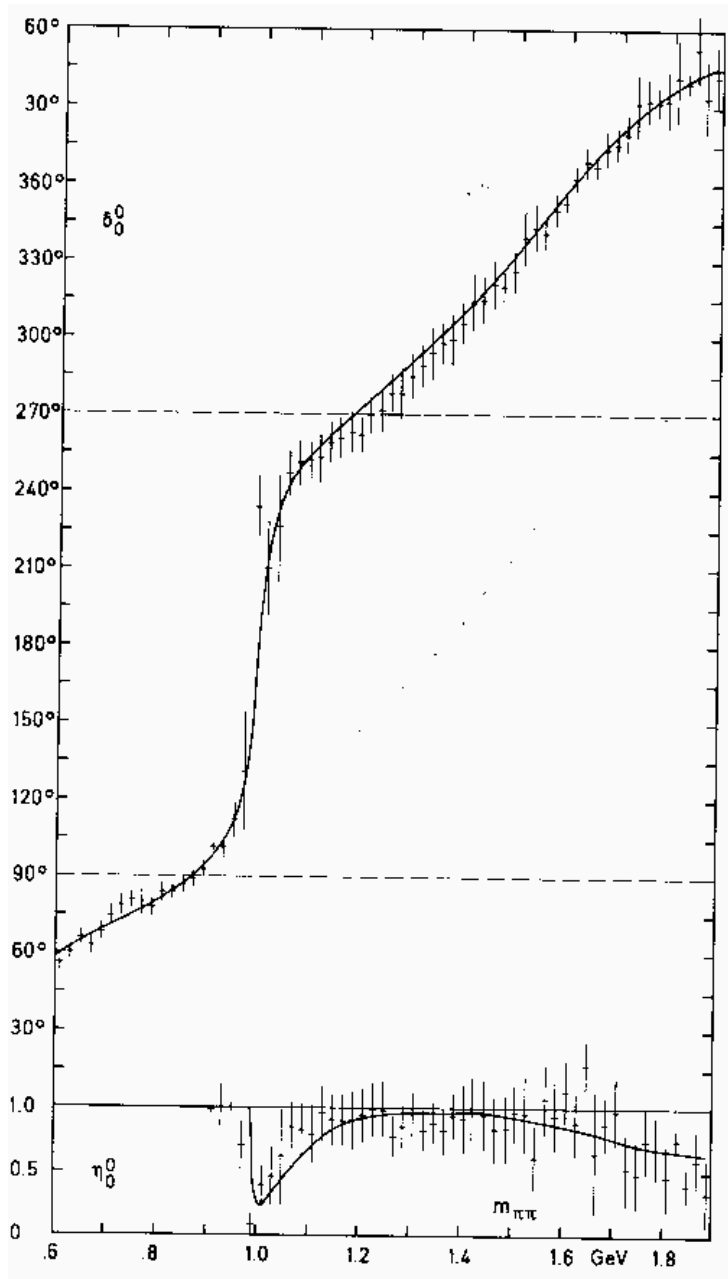}
  \caption[Resonant phase shift with background in the isoscalar S
  wave.]{Resonant phase shift starting from an appreciable background phase in
    the isoscalar S wave. In the lower panel the behavior of the inelasticity
    in this partial wave is shown; from \cite{hyams}.}
  \label{fig:hyams}
\end{figure}

\begin{figure}
  \centering
  \includegraphics[angle=180,width=.6\linewidth]{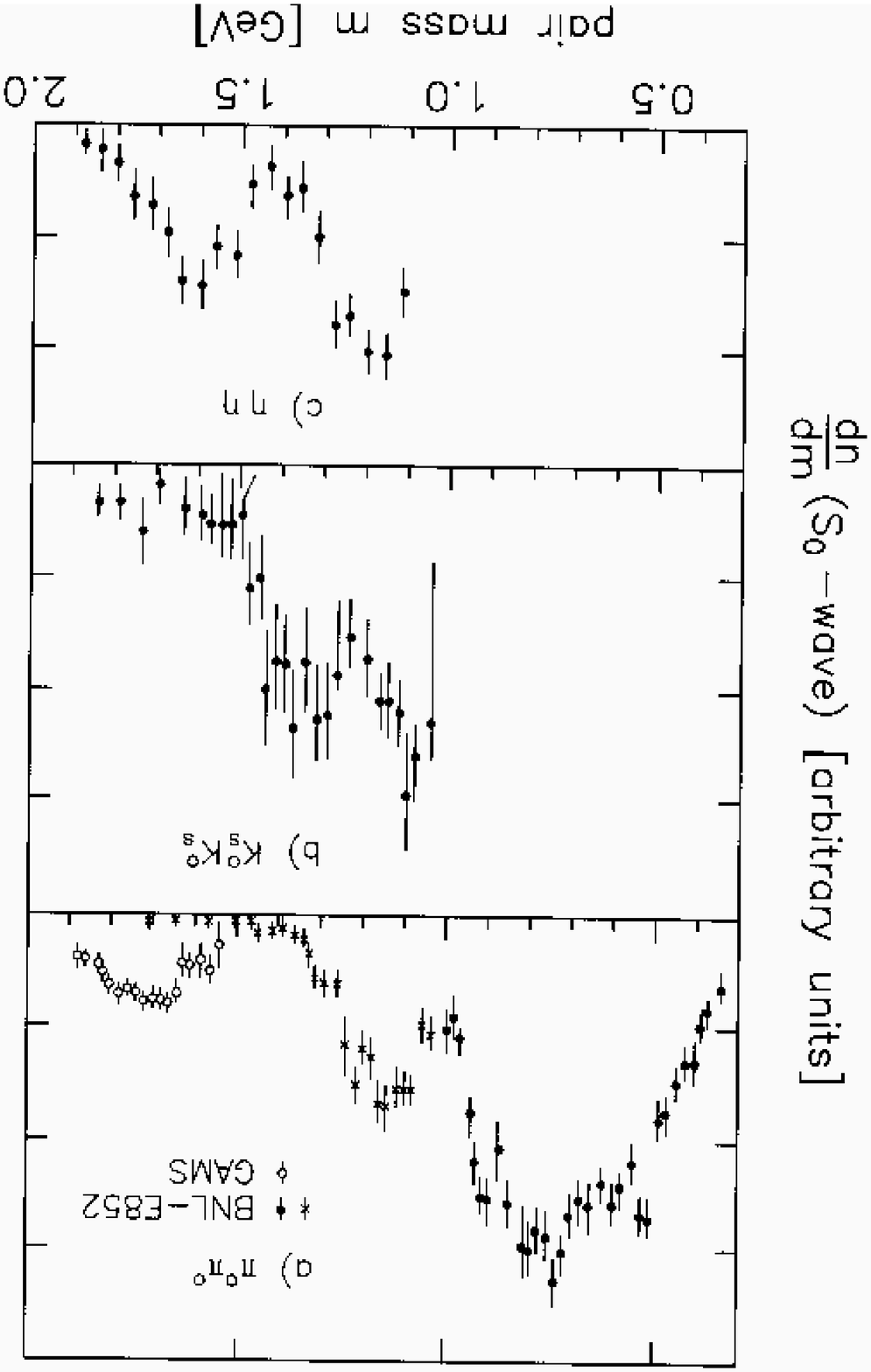}
  \caption[Compilation of the isoscalar S wave.]{Compilation of the isoscalar
    S wave. Three major peaks or alternatively two major dips can be seen.
    Copied from \cite{mink-eur}, see references therein.}
  \label{fig:dragon}
\end{figure}

\begin{figure}
\begin{center}
\subfigure[]{%
\begin{pspicture}(-1,-1)(10,8)
  \rput[bl](0,0){\includegraphics[width=.6\linewidth,height=.5\linewidth]{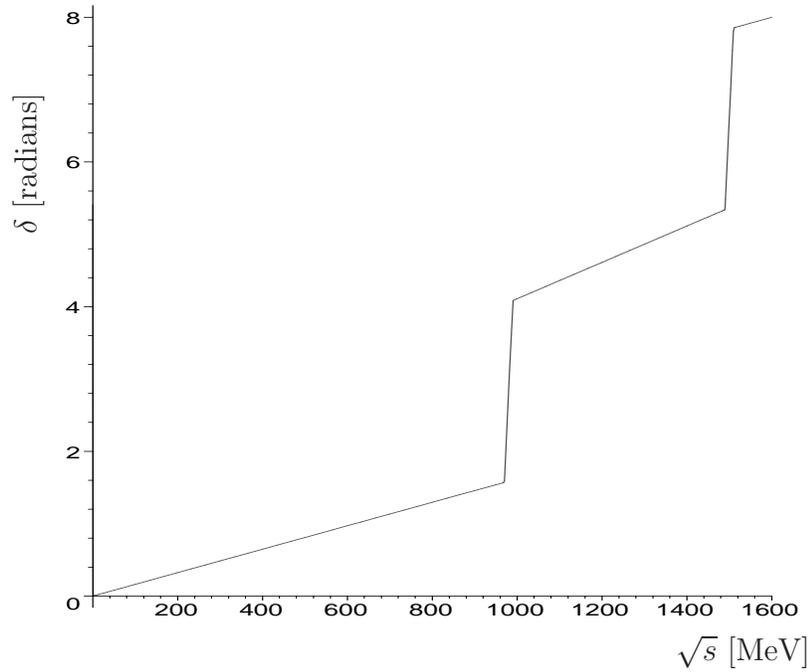}}
  \rput(9,-.5){$\sqrt{s}$ [MeV]}
  \rput{L}(-.5,6){$\delta$ [radians]}
\end{pspicture}}
\subfigure[]{%
\begin{pspicture}(-1,-1)(10,8)
  \rput[bl](0,0){\includegraphics[width=.6\linewidth,height=.5\linewidth]{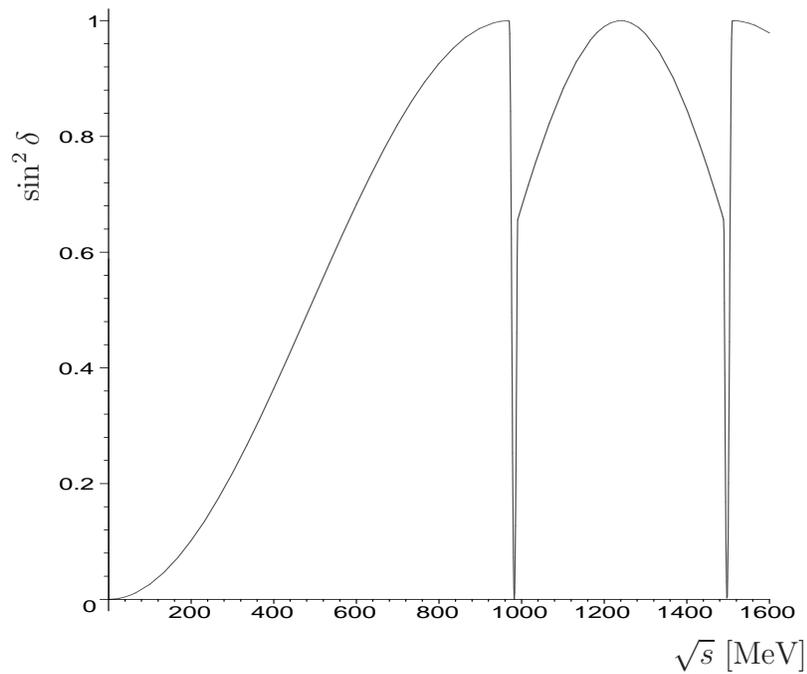}}
  \rput(9,-.5){$\sqrt{s}$ [MeV]}
  \rput{L}(-.5,6){$\sin^2\delta$}
\end{pspicture}}
\end{center}
  \caption[Rough parametrization of a phase shift.]{(a) Very rough
    parametrization of a phase shift that gives rise to a cross-section
    comparable to that of figure~\ref{fig:dragon}. (b) $\sin^2$ of the rough
    parametrization of the phase shift, which gives the characteristic form of
    the corresponding partial cross section.}
  \label{fig:phase-maple}
\end{figure}

Figure~\ref{fig:taylor} shows four possibilities of a resonant phase shift and
its corresponding characteristic behavior of the cross section.
Appreciable background phases are not just a mathematical possibility but seem
to occur in actual physical reactions. One prominent example is the $f_0(980)$
which appears as a dip in the $\pi\pi$ spectrum because the corresponding
phase shift increases rapidly through $\pi$ and not through $\pi/2$ as for a
pure resonance. In the Argand diagram the amplitude moves rapidly through the
bottom of the unitarity circle, see figure~\ref{fig:argand-res}~(b). The
experimental data and a parametrization of the relevant phase shift is shown
in figure~\ref{fig:hyams} together with the inelasticity.  In
figure~\ref{fig:dragon} the cross-section of the isoscalar S wave is compiled.
Such a structure can be explained by two rapid phase shifts through $\pi$ and
$2\pi$, as I roughly simulated in figure~\ref{fig:phase-maple}.

\section{Optical theorem}
\label{sec:opt-th}

In the special case where the initial an the final state are the same,
so-called \emph{forward scattering}, the unitarity condition for the $T$
matrix (equation~\eqref{eq:26}) reads
\begin{equation}\label{eq:45}
  \begin{split}
    i(T_{\alpha\alpha}^* - T_{\alpha\alpha}) &= \int d\beta\,
    T^{\dagger}_{\alpha\beta} T_{\beta\alpha}, \\
    2\text{Im}T_{\alpha\alpha} &= \int d\beta\, |T_{\beta\alpha}|^2, \\
    \text{Im}T_{\alpha\alpha} &= 8\pi^2 p \sqrt{s} \sigma_\text{tot}.
\end{split}
\end{equation}

To obtain a form of the optical theorem in terms of $\mathcal{M}$ we have to
take care again about delta functions. Therefore, we cannot substitute simply
$T = \delta^4(p_i-p_f)\mathcal{M}$ in the above equation. In
ref.~\cite[p.~147]{weinberg} this is done by dividing through delta function,
which I want to avoid. The alternative that I prefer is again by means of
wave-packets, cf.\ \cite[p.~183ff]{martin-spearman}. 
I consider states of the form
\begin{equation}
  \ket{i} = \sum_{m_f} \int d^4p_i \Phi^{(i)}(p_i)
  \ket{p_i}\otimes\ket{\psi_i}. 
\end{equation}
Sandwiched between two such states $\ket{i}$ and $\ket{f}$ the operator
equation~\eqref{eq:26} takes the form
\begin{multline}
  \sum_{m_f} \int d^4p_i \Phi^{(i)}(p_i)\Phi^{(f)*}(p_i) (\mathcal{M}_{fi} -
  \mathcal{M}^{\dagger}_{fi}) \\
  = i \int d^4p_i \Phi^{(i)}(p_i)\Phi^{(f)*}(p_i) \sum_{m_f=2}^{M} 
  \sum_\Lambda \int dQ_{m_f}\, \mathcal{M}^{\dagger}_{(f_\Lambda)\psi}
  \mathcal{M}_{\psi_i}.   
\end{multline}
Since the momentum distribution functions $\Phi$ are arbitrary the integrands
have to be equal. Therefore, by equation~\eqref{eq:44}, we obtain for forward
scattering, \ie $i=f$,
\begin{equation}\label{eq:56}
  \text{Im}\mathcal{M}_{ii} = 8\pi^2 p \sqrt{s} \sigma_\text{tot},
\end{equation}
the right-hand side of which is the same as in equation~\eqref{eq:45}, in this
sense the delta-function relating $T$ and $\mathcal{M}$ has canceled. The
result of equation~\eqref{eq:56} could have been found also by first showing
that $\mathcal{M}$ satisfies the same form of unitarity condition as $T$, \ie 
\begin{equation}\label{eq:m-unit}
  i(\mathcal{M}^\dagger - \mathcal{M}) = \mathcal{M}^\dagger \mathcal{M},
\end{equation}
see, for instance \cite[p. 147]{weinberg}.

Because $\mathcal{M}$ and $T$ satisfy the same form of unitarity condition
(equations~\eqref{eq:26} and \eqref{eq:m-unit}) I will often use $T$ and
$\mathcal{M}$ interchangeable. In particular when discussing unitarity
constraints (section~\ref{sec:el-unit}) I will use $T$ elements without
worrying much about the relation to the $\mathcal{M}$ elements in my
conventions. None of the results obtained later hinge on the difference
between $T$ and $\mathcal{M}$. Thus I adopt the notation of
\cite{morgan93,amp87} for formulating unitarity constraints.

\section{Diffraction peak}\label{sec:diff-peak}

The element of phase space of a two-body, \ie $m_{f}=2$, final state is
\begin{equation}
  dQ_2 = \frac{d^3p_1 }{2E_1} \frac{d^3p_2 }{2E_2} \delta^4(p_1+p_2-p_a-p_b),
\end{equation}
where $p_1$ and $p_2$ are the four-momenta of the two final state particles,
and $p_a$ and $p_b$ the four-momenta of the two initial particles. We now
discuss this expression in the center-of-mass frame where we have, by
definition, that $\vec{p_a}+\vec{p_b}=0$. Thus, cf.\ 
\cite[p.~139ff.]{weinberg},
\begin{multline}
  dQ_2 = \frac{d^3p_1 }{2E_1} \frac{d^3p_2
  }{2E_2}\delta^3(\vec{p_1}+\vec{p_2}) \delta(E_1+E_2-E_a-E_b) \\
  = \frac{1}{2\sqrt{m_1^2+\vec{p_2}^2}} \frac{d^3p_2
  }{2E_2} \delta(\sqrt{m_1^2+\vec{p_2}^2}+E_2-E_a-E_b) \\
= \frac{1}{2\sqrt{m_1^2+\vec{p_2}^2}} \frac{|\vec{p_2}|^2 d|\vec{p_2}| d\phi\,
  d\theta\, \sin\theta}{2\sqrt{m_2^2+\vec{p_2}^2}}
  \delta(\sqrt{m_1^2+\vec{p_2}^2}+\sqrt{m_2^2+\vec{p_2}^2}-E_a-E_b).
\end{multline}
For the delta function we can use
\begin{equation}
  \delta(f(|\vec{p_2}|)) = \frac{\delta(|\vec{p_2}|- q )}{|f'(q)|}
\end{equation}
with $q$ the zero of $f(|\vec{p_2}|)$ and find
\begin{equation}
  \delta(\sqrt{m_1^2+\vec{p_2}^2}+\sqrt{m_2^2+\vec{p_2}^2}-E_a-E_b) \\
  = \delta(|\vec{p_2}|- q ) \frac{E_1 E_2 q}{(E_a + E_b)}, 
\end{equation}
with 
\begin{equation}
  q = \frac{\sqrt{[(E_a + E_b)^2 - m_1^2 - m_2^2]^2 - 4 m_1^2 m_2^2}}{2(E_a +
  E_b)}. 
\end{equation}
Thus we obtain the relation between the two-body phase space and the
scattering angles $\theta$ and $\phi$ in the cms frame:
\begin{equation}
  dQ_2 = \frac{q}{4(E_a + E_b)}d\phi\,
  d\theta\, \sin\theta \equiv \frac{q}{4(E_a + E_b)} d\phi\, d(\cos\theta)
  \equiv  \frac{q}{4(E_a + E_b)} d\Omega.
\end{equation}

We can now make an estimation that reveals a certain behavior of the total
cross section for two-particle to two-body scattering, cf.\ 
equation~\eqref{eq:46}. We assume that the matrix elements of $\mathcal{M}$
are smooth functions of the scattering solid angle $\Omega$ and that they fall
off to zero for large angles. Then we can find a solid angle
$\Delta\Omega$ such that 
\begin{equation}
  \int  d\Omega\, |\mathcal{M}_{(f_\Lambda)i}|^2 \geq \frac{1}{2}\Delta\Omega
  |\mathcal{M}_{ii}|^2. 
\end{equation}
Then we obtain
\begin{multline}
    \sigma_{\text{tot}} \geq \sigma_\Lambda(2\to 2) = \frac{1}{16\pi^2
  p\sqrt{s}} 
\int dQ_{m_f}\, |\mathcal{M}_{(f_\Lambda)i}|^2 \\
= \frac{1}{16\pi^2
  p\sqrt{s}} 
\frac{q}{4(E_a + E_b)} \int  d\Omega\, |\mathcal{M}_{(f_\Lambda)i}|^2 \\
\geq \frac{1}{16\pi^2
  p\sqrt{s}} 
\frac{q}{4(E_a + E_b)} \Delta\Omega |\mathcal{M}_{ii}|^2 \\
\geq \frac{1}{16\pi^2
  p\sqrt{s}} 
\frac{q}{4(E_a + E_b)} \Delta\Omega |\text{Im}\mathcal{M}_{ii}|^2 \\
= \frac{1}{16\pi^2
  p\sqrt{s}} 
\frac{q}{4(E_a + E_b)} \Delta\Omega 64\pi^4 p^2 s \sigma^2_\text{tot}.
\end{multline}
In the cms $p=q$, $E_a + E_b=\sqrt{s}$.  So we obtain an upper bound for
$\Delta\Omega$,
\begin{equation}\label{eq:47}
  \Delta\Omega \leq \frac{1}{\pi^2 p^2 \sigma_{\text{tot}}}. 
\end{equation}

\begin{figure}
  \centering
  \includegraphics[angle=180,width=\linewidth]{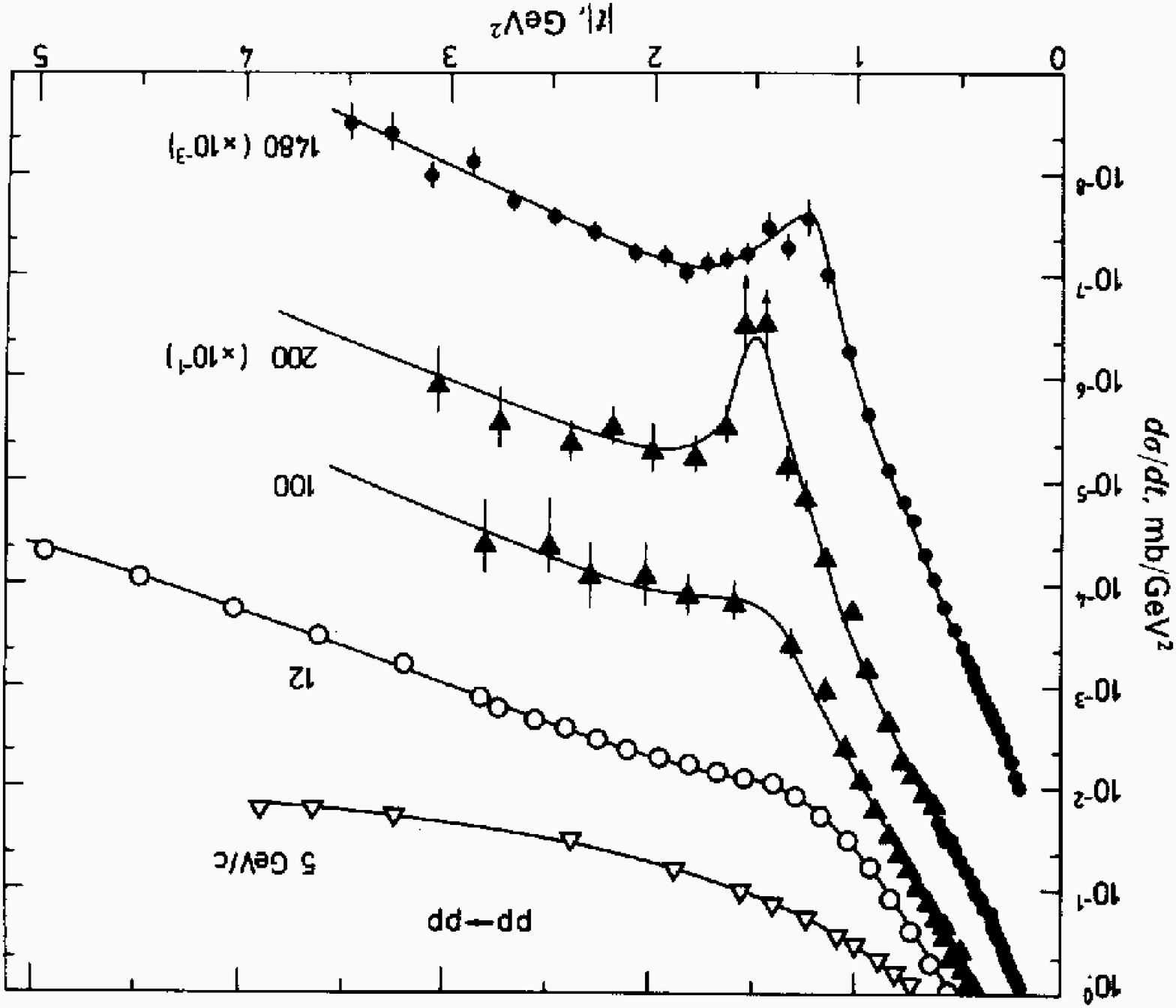}
  \caption[Differential cross section for elastic $pp$
  scattering]{Differential cross section for elastic $pp$ scattering as a
    function of the square of the momentum transfer, $|t|$; from
    \cite[p.~137]{perkins}.}
  \label{fig:diff-peak}
\end{figure}

Hadronic cross sections show a \emph{diffraction peak}, \ie the differential
cross section falls off exponentially with $t=2p^2(\cos\theta -1)$ (cf.\ 
eq.~\eqref{eq:8}). In the special case of a totally absorbing disk of radius
$R$ (see \eg \cite[p.~136f.]{perkins} and \cite[p.~210f.]{martin-spearman}) we
have approximately
\begin{equation}
  \frac{d\sigma}{dt} \propto \exp\left( -\frac{R^2|t|}{4} \right).
\end{equation}
In figure~\ref{fig:diff-peak} the differential cross section for elastic
proton-proton scattering is shown. There we see that already for incident
momenta of 5 GeV (\ie $\sqrt{s}\approx 3.4\ \text{GeV}$) the fall-off is
visible but that the diffraction peak is more and more pronounced for higher
incident momenta (in the figure up to 1480 GeV, \ie $\sqrt{s}\approx 53\ 
\text{GeV}$).

\begin{figure}
  \centering
  \includegraphics[width=\linewidth]{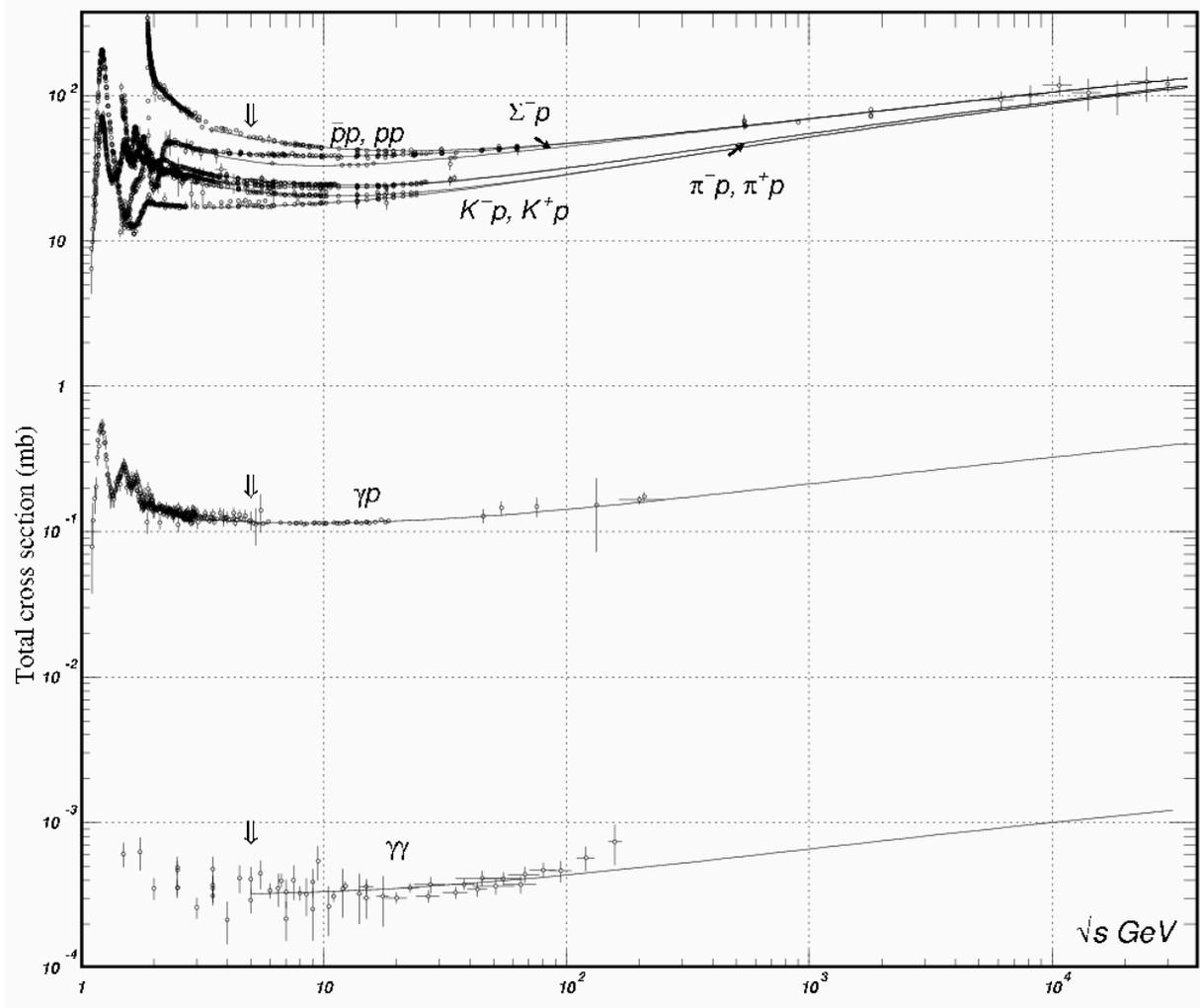}
  \caption[High energy behavior of typical hadronic cross sections.]{High
    energy behavior of typical hadronic cross sections; from \cite{PDBook}.}
  \label{fig:pp}
\end{figure}

As shown in figure~\ref{fig:pp}, typical hadronic cross sections rise with
energies for energies greater than roughly 20 GeV. Thus also
equation~\eqref{eq:47} shows that the forward peak in the two-body to two-body
total hadronic cross section becomes more and more narrow at these high
energies.

\chapter{Kinematics of a three body decay}\label{cha:kin}

In this chapter I collect kinematical relations that could eventually be used
for a partial wave analysis. As a numeric example I take the decay $B^+\to
\pi^-\pi^+K^+$.

\section{Rest system of decaying particle}

First I consider the kinematical relations that hold in the rest system of the
$B$ meson. By definition its three momentum is zero, $\vec{p}_B=0$, in this
frame of reference. Its energy is given by its mass $m_B$. The squares of the
respective two-body invariant masses are defined by
\begin{equation}
  \begin{split}
    s_{\pi^-\pi^+} &= (p_{\pi^-} + p_{\pi^+})^2 =
    (E_{\pi^-}+E_{\pi^+})^2 - (\vec{p}_{\pi^-} + \vec{p}_{\pi^+} )^2, \\
    s_{\pi^-K^+} &= (p_{\pi^-} + p_{K^+})^2 =
    (E_{\pi^-}+E_{K^+})^2 - (\vec{p}_{\pi^-} + \vec{p}_{K^+} )^2, \\
    s_{\pi^+K^+} &= (p_{\pi^+} + p_{K^+})^2 =
    (E_{\pi^+}+E_{K^+})^2 - (\vec{p}_{\pi^+} + \vec{p}_{K^+} )^2.
    \end{split}
\end{equation}
Recall that these quantities are Lorentz invariant; the energies and the
three-momenta for one particle pair can be taken with respect to \emph{any}
reference frame.  (But it has to be the same reference frame for the energy
and the three-momentum.) For the sake of brevity I will sometimes refer to
these quantities just as ``two-body masses'', ``pair masses'' or the like.
These masses (squared) of one pair of particles are related to the energy of
the remaining particle, \eg
\begin{equation}\label{eq:15}
  \begin{split}
   s_{\pi^-\pi^+} &=  (E_{\pi^-}+E_{\pi^+})^2 - (\vec{p}_{\pi^-} +
   \vec{p}_{\pi^+} )^2 \\
   &= (m_B - E_{K^+} )^2 - \vec{p}_{K^+}^2 \\
   &= m_B^2 + m_{K^+}^2 - 2m_B E_{K^+},
  \end{split}
\end{equation}
and therefore also to the modulus of the three-momentum of the remaining
particle:
\begin{equation}\label{eq:16}
  \begin{split}
    \vec{p}_{K^+}^2 &=  E_{K^+}^2 - m_{K^+}^2 \\
                    &=  \left( \frac{ m_B^2 + m_{K^+}^2 -  s_{\pi^-\pi^+}
                    }{2m_B} \right)^2 - m_{K^+}^2 \\
                    &= \frac{m_B^4 + m_{K^+}^4 + s_{\pi^-\pi^+}^2 - 2 ( m_B^2
                    m_{K^+}^2 + m_B^2 s_{\pi^-\pi^+} + m_{K^+}^2
                    s_{\pi^-\pi^+}  )  } { 4m_B^2 }.
  \end{split}
\end{equation}
Similarly, one obtains expressions for $E_{\pi^-}$ and $\vec{p}_{\pi^-}^2$,
and $E_{\pi^+}$ and $\vec{p}_{\pi^+}^2$ in terms of $s_{\pi^+K^+}$ and
$s_{\pi^-K^+}$ respectively, see table~\ref{tab:coll}.

\section{Two-body system}\label{sec:two-body}

We now make a Lorentz transformation into the center-of-mass system (cms) of
the two pions, say. The kinematic situation in this reference system is shown
in figure~\ref{fig:kin-pipi}. Compare it to the kinematic situation in
scattering in the cms shown in figure~\ref{fig:cms-scatt}. 
\begin{figure}
  \centering
  \begin{pspicture}(-7,-1)(7,7)
\pscircle[fillstyle=solid,fillcolor=black](0,6){.1}
\pscircle[fillstyle=solid,fillcolor=black](2,2){.1}
\pscircle[fillstyle=solid,fillcolor=black](-3.5,3.5){.1}
\pscircle[fillstyle=solid,fillcolor=black](-.5,.5){.1}
\psline[linestyle=dashed](-6,6)(6,6)
\psline[linestyle=dashed](-6,2)(6,2)
\psline[linestyle=dotted](-3.5,3.5)(-.5,.5)
\psarc(-2,2){.5}{0}{135}
\psline[arrowsize=5pt]{->}(0,6)(2,6)
\psline[arrowsize=5pt]{->}(2,2)(4,2)
\psline[arrowsize=5pt]{->}(-3.5,3.5)(-5,5)
\psline[arrowsize=5pt]{->}(-.5,.5)(1,-1)
\uput[90](0,6){$B^+$}
\uput[90](2,2){$K^+$}
\uput[ur](-3.5,3.5){$\pi^-$}
\uput[dl](-.5,.5){$\pi^+$}
\rput(-1.45,2.6){$\theta$}
\uput[-90](1,6){$l$}
\uput[-90](3,2){$l$}
\uput[dl](-4.25,4.25){$k$}
\uput[45](.25,-.25){$k$}
\end{pspicture}

  \caption[Kinematic situation in the $\pi\pi$ rest system.]{Kinematic
    situation for the three-body decay $B^+\to \pi^-\pi^+ K^+$ in the $\pi\pi$
    rest system with the two-body angle $\theta$ as defined in
    equation~\eqref{eq:10}, $z\equiv\cos\theta$. Top of the figure: before the
    decay of the $B$. Bottom of the figure: after the decay and the strong
    final state interactions.}
  \label{fig:kin-pipi}
\end{figure}
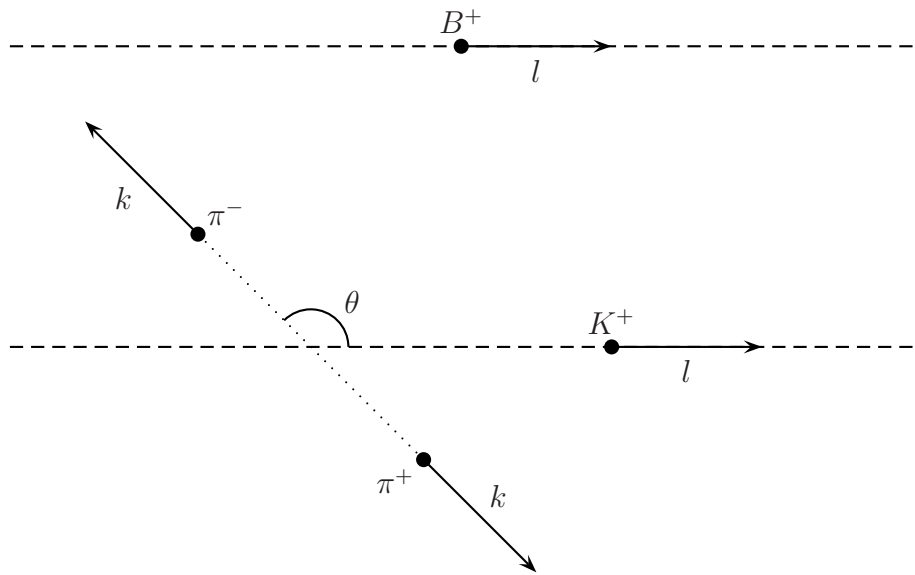
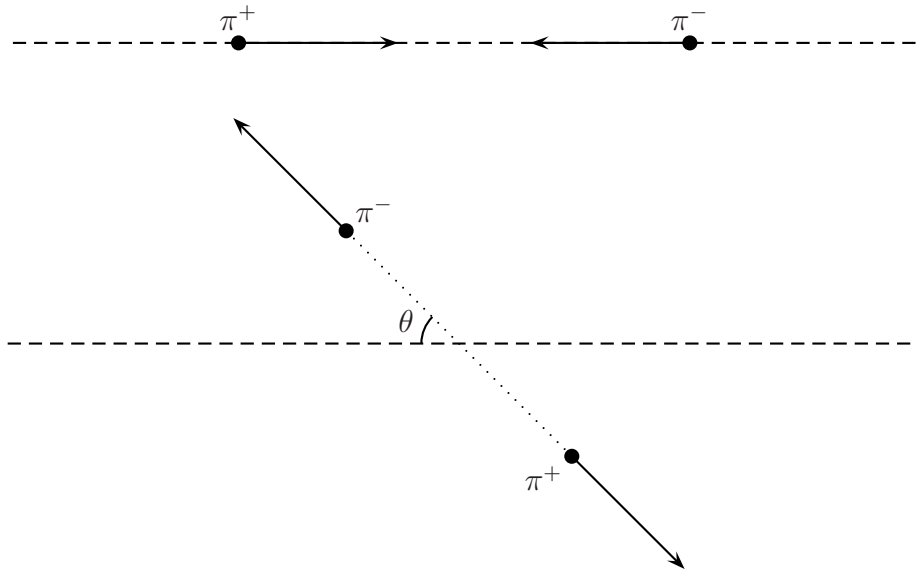
\begin{figure}
  \centering
  \newcommand{\sub}{%
\begin{pspicture}(-3.5,-1)(1,5)
\pscircle[fillstyle=solid,fillcolor=black](-3.5,3.5){.1}
\pscircle[fillstyle=solid,fillcolor=black](-.5,.5){.1}
\psline[linestyle=dashed](-8,2)(4,2)
\psline[linestyle=dotted](-3.5,3.5)(-.5,.5)
\psarc(-2,2){.5}{135}{180}
\psline[arrowsize=5pt]{->}(-3.5,3.5)(-5,5)
\psline[arrowsize=5pt]{->}(-.5,.5)(1,-1)
\uput[45](-3.5,3.5){$\pi^-$}
\uput[dl](-.5,.5){$\pi^+$}
\rput(-2.7,2.3){$\theta$}
\end{pspicture}
}
\begin{pspicture}(-5,-1)(5,7)
\pscircle[fillstyle=solid,fillcolor=black](3,6){.1}
\pscircle[fillstyle=solid,fillcolor=black](-3,6){.1}
\psline[linestyle=dashed](-6,6)(6,6)
\psline[arrowsize=5pt]{->}(3,6)(.878,6)
\psline[arrowsize=5pt]{->}(-3,6)(-.878,6)
\uput[90](3,6){$\pi^-$}
\uput[90](-3,6){$\pi^+$}
\rput(.75,2){\sub}
\end{pspicture}

  \caption[Kinematic situation for scattering.]{Kinematic situation for
    scattering in the center-of-mass system. Top of the figure: before the
    strong interactions. Bottom of the figure: after the strong interactions.}
  \label{fig:cms-scatt}
\end{figure}
The appropriate transformation from the $B$ rest system to the $\pi\pi$ system
is a Lorentz boost along the direction opposite to $\vec{p}_{K^+}$. The
relative velocity between the $B$ rest system and the two-body system
considered now is (in unities of $c$ and with respect to the direction of
$\vec{p}_{K^+}$)
\begin{equation}
  v = - \frac{ |\vec{p}_{\pi^-} + \vec{p}_{\pi^+}| }{ E_{\pi^-} + E_{\pi^+}  }
  = - \frac{ |\vec{p}_{K^+}| }{ m_B - E_{K^+}  }.
\end{equation}
This velocity defines the hyperbolic angle $\alpha$ by which a Lorentz boost
along a given axis (here the axis given by $\vec{p}_{K^+}$) can be specified:
$\tanh\alpha = v$. The matrix that represents the Lorentz boost then reads
\begin{equation}
  \begin{pmatrix}
    \cosh\alpha & -\sinh\alpha & 0 & 0 \\
    -\sinh\alpha & \cosh\alpha & 0 & 0 \\
    0 & 0 & 1 & 0 \\
    0 & 0 & 0 & 1
      \end{pmatrix}.
\end{equation}
Accordingly, the energy of the $B$ in the new reference frame is
\begin{equation}
  \begin{split}
    E_B' = m_B \cosh\alpha &= m_B \frac{m_B - E_{K^+}}{\sqrt{s_{\pi^-\pi^+}}}
    \\
    &=
    \frac{m_B}{\sqrt{s_{\pi^-\pi^+}}} \left( m_B - \frac{ m_B^2 + m_{K^+}^2 -
        s_{\pi^-\pi^+}}{2m_B} \right)\\
    &= \frac{m_B^2 - m_{K^+}^2 + s_{\pi^-\pi^+} }{2 \sqrt{s_{\pi^-\pi^+}}}.
\end{split}
\end{equation}
In the center-of-mass system of the two pions we have $E_B' =
\sqrt{s_{\pi^-\pi^+}} + E_{K^+}'$. The energy of the $K^+$ in this system is
therefore given by
\begin{equation}\label{eq:12}
  E_{K^+}' = E_B' - \sqrt{s_{\pi^-\pi^+}} = \frac{m_B^2 - m_{K^+}^2 -
  s_{\pi^-\pi^+} }{2 \sqrt{s_{\pi^-\pi^+}}}. 
\end{equation}
By definition of this reference frame the three-momenta of the two pions are
equal, $\vec{p}_{\pi^-}' = \vec{p}_{\pi^+}'$. So are, by momentum
conservation, the three-momenta of the $B$ and the $K$, $\vec{p}_B' =
\vec{p}_{K^+}'$. I denote the respective moduli of the three-momenta of the
pions and the $B$ (or $K$) as $k$ and $l$. $l$ is determined by energy
conservation in the new reference frame,
\begin{equation}
  \sqrt{m_B^2 + l^2} = \sqrt{m_{K^+} + l^2} + \sqrt{s_{\pi^-\pi^+}},
\end{equation}
which equation can be solved for $l$, yielding
\begin{equation}\label{eq:14}
  l^2 = \frac{ m_B^4 + m_{K^+}^4 + s_{\pi^-\pi^+}^2 - 2( m_B^2 m_{K^+}^2 +
  m_B^2s_{\pi^-\pi^+} + m_{K^+}^2 s_{\pi^-\pi^+}) }{ 4 s_{\pi^-\pi^+}  }.
\end{equation}
As to the three-momentum $k$ of the pions, it can be obtained by evaluating
the Lorentz invariant quantity $s_{\pi^-\pi^+}$ in the new reference system,
\ie
\begin{equation}
  \sqrt{s_{\pi^-\pi^+}} = \sqrt{m_{\pi^-}^2 + k^2} + \sqrt{m_{\pi^+}^2 + k^2},
\end{equation}
which upon solving yields
\begin{equation}\label{eq:13}
  k^2 = \frac{ m_{\pi^-}^4 + m_{\pi^+}^4 + s_{\pi^-\pi^+}^2 - 2( m_{\pi^-}^2
  m_{\pi^+}^2 + 
  m_{\pi^-}^2s_{\pi^-\pi^+} + m_{\pi^+}^2 s_{\pi^-\pi^+}) }{ 4 s_{\pi^-\pi^+}
  }. 
\end{equation}
The respective energies of the pions in the primed reference system are then 
\begin{equation}\label{eq:11}
  E_{\pi^-}' = \sqrt{m_{\pi^-}^2 + k^2}, \qquad  E_{\pi^+}' =
  \sqrt{m_{\pi^+}^2 + k^2}. 
\end{equation}

One relevant quantity for a partial-wave analysis is the angle $\theta$
between the respective directions of the $\pi ^-$ and the $K^+$ in the new
reference frame, \ie the center-of-mass system of the two pions. We are now
going to derive an expression for the cosine of this angle, \ie
$z=\cos\theta$, in terms of the two variables of the Dalitz plot,
$s_{\pi^-\pi^+}$ and $s_{\pi^-K^+}$, and the masses of the four particles
concerned. By definition we have
\begin{equation}\label{eq:10}
  z = \frac{\vec{p}_{\pi^-}' \vec{p}_{K^+}'}{|\vec{p}_{\pi^-}'|
  |\vec{p}_{K^+}'|} = \frac{ -p_{\pi^-} p_{K^+} + E_{\pi^-}' E_{K^+}'
  }{k l}  . 
\end{equation}
Suitable expressions for $E_{\pi^-}'$, $E_{K^+}'$, $k$ and $l$ are
given, respectively, in the equations~\eqref{eq:11}, \eqref{eq:12},
\eqref{eq:13} and \eqref{eq:14}.  For $p_{\pi^-} p_{K^+}$ we have
\begin{equation}
    p_{\pi^-} p_{K^+} = E_{\pi^-}E_{K^+} - \vec{p}_{\pi^-}\vec{p}_{K^+}.
\end{equation}
For $E_{K^+}$ we have already derived a relation between it and the Dalitz
plot variable $s_{\pi^-\pi^+}$ (equation~\eqref{eq:15}).  The scalar product
between the two three-momenta can be expressed by the moduli squared of the
three-momenta of all three decay products, by transforming the relation for
momentum conservation first by putting it in an appropriate form then squaring
it and finally solving it for $\vec{p}_{\pi^-}\vec{p}_{K^+}$:
\begin{equation}
  \begin{split}
    \vec{p}_{\pi^-} + \vec{p}_{K^+} + \vec{p}_{\pi^+} &= \vec{p}_{B} = 0, \\
    \vec{p}_{\pi^-} + \vec{p}_{K^+} &= - \vec{p}_{\pi^+}, \\
    \vec{p}_{\pi^-}^2 + \vec{p}_{K^+}^2 + 2\vec{p}_{\pi^-}\vec{p}_{K^+} &=
    \vec{p}_{\pi^+}^2, \\ 
    \vec{p}_{\pi^-}\vec{p}_{K^+} &= \frac{\vec{p}_{\pi^+}^2 -
    \vec{p}_{\pi^-}^2 - \vec{p}_{K^+}^2}{2}. 
  \end{split}
\end{equation}
The squares of three momenta in terms of the three two-body masses are given
by equation~\eqref{eq:16} and the respective entries in table~\ref{tab:coll}.
To express all quantities in the two Dalitz plot variables $s_{\pi^-\pi^+}$
and $s_{\pi^-K^+}$, it remains to find an expression for $s_{\pi^+K^+}$ in
terms of these. Such an expression can be obtained as follows.
\begin{multline}\label{eq:17}
   s_{\pi^-\pi^+} +  s_{\pi^-K^+} + s_{\pi^+K^+} = (p_{\pi^-} + p_{\pi^+})^2 +
   (p_{\pi^-} + p_{K^+})^2 + (p_{\pi^+} + p_{K^+})^2 \\
   = p_{\pi^-}^2 + p_{\pi^+}^2 + p_{K^+}^2 \\
   + p_{\pi^-}^2 + p_{\pi^+}^2 +
   p_{K^+}^2 + 2( p_{\pi^-}p_{\pi^+} + p_{\pi^-}p_{K^+} + p_{\pi^+}p_{K^+}  ).
\end{multline}
Because of energy-momentum conservation we have
\begin{equation}
  \begin{split}
    m_B^2 = p_B^2 &= ( p_{\pi^-} + p_{\pi^+} + p_{K^+} )^2 \\
    &= p_{\pi^-}^2 + p_{\pi^+}^2 +
   p_{K^+}^2 + 2( p_{\pi^-}p_{\pi^+} + p_{\pi^-}p_{K^+} + p_{\pi^+}p_{K^+}  ).
\end{split}
\end{equation}
Using this result and since $p_{\pi^-}^2 = m_{\pi^-}^2$ etc.,
equation~\eqref{eq:17} reads
\begin{equation}\label{eq:27}
  s_{\pi^-\pi^+} +  s_{\pi^-K^+} + s_{\pi^+K^+} = m_{\pi^-}^2 + m_{\pi^+}^2 +
  m_{K^+}^2 + m_B^2. 
\end{equation}
So to obtain expressions in the two Dalitz plot variables $s_{\pi^-\pi^+}$ and
$s_{\pi^-K^+}$ one can replace every occurrence of $s_{\pi^+K^+}$ by $m_B^2 +
m_{\pi^-}^2 + m_{\pi^+}^2 + m_{K^+}^2 - s_{\pi^-\pi^+} - s_{\pi^-K^+}$.

The corresponding formulae for the other two two-body systems, \ie the cms of
the negative or positive pion and the kaon, can be obtained by making
appropriate substitutions. In the case considered so far, \ie the cms of the
two pions, the kaon played the role of the spectator, the negative pion the
role of the particle with respect to which the angle $\theta$ is measured. In
the cms of the negative pion and the kaon, the positive pion is the spectator,
the reference particle for the angle I take to be the negative pion. (It could
equally well be the kaon, this is just my convention. With the kaon as
reference for the angle, $z$ would change sign.) In the cms of the positive
pion and the kaon, the negative pion is the spectator and the kaon (or the
positive pion) is the reference particle for the angle.

\begin{table}[h]
  \centering
\includegraphics[angle=90]{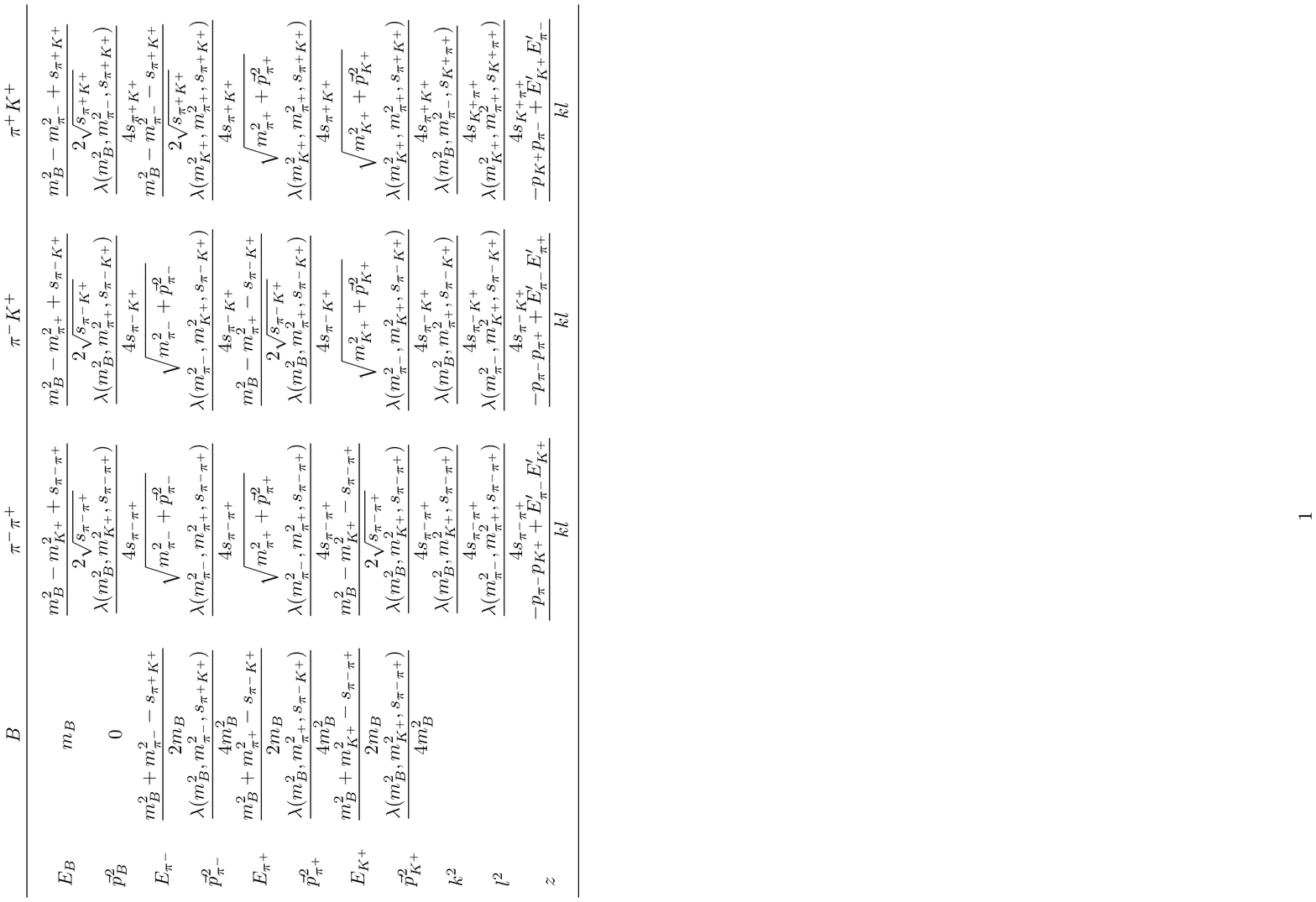}
  \caption{Expressions for energies, three-momenta and cosine of the
    scattering angle in the $B$ rest system (2nd column), the center-of-mass
    system of the two pions (3rd column),  the center-of-mass
    system of the negative pion and the kaon (4th column), and the
    center-of-mass 
    system of the positive pion and the kaon (5th column). $\lambda(a,b,c) =
    a^2 + b^2 + c^2 - 2( ab + ac + bc )$.} 
  \label{tab:coll}
\end{table}

\clearpage

\section{Pair masses and two-body angles}

\subsection{Resonance band or angular peak?}

\begin{figure}
  \centering
  \includegraphics[scale=.5]{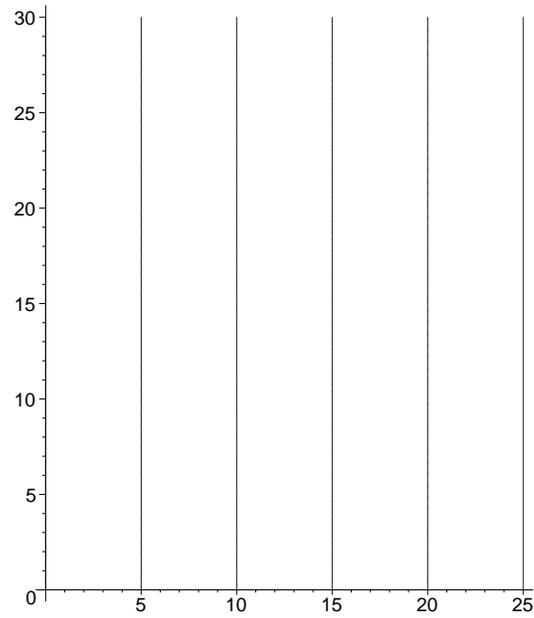}
  \caption[Lines of constant $s_{\pi^-\pi^+}$.]{Lines where $s_{\pi^-\pi^+}=
    (5, 10, 15, 20, 25)\, \text{GeV}^2$. Abscissa: $s_{\pi^-\pi^+}\ 
    [\text{GeV}^2]$. Ordinate: $s_{\pi^-K^+}\ [\text{GeV}^2]$.}
  \label{fig:s12}
\end{figure}

\begin{figure}
  \centering
  \includegraphics[scale=.5]{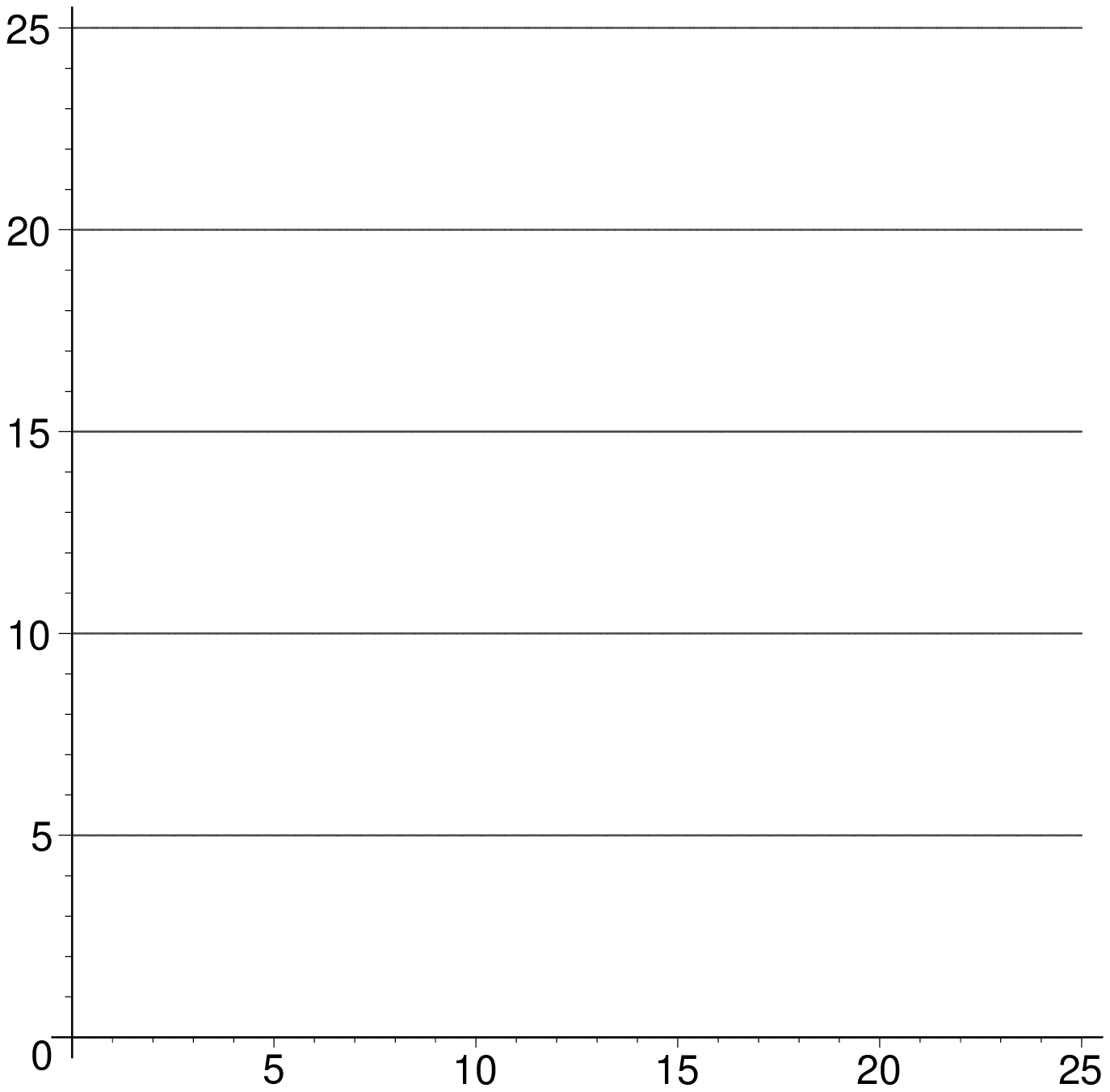}
  \caption[Lines of constant $s_{\pi^-K^+}$.]{Lines where $s_{\pi^-K^+}= (5,
    10, 15, 20, 25)\, \text{GeV}^2$. Abscissa: $s_{\pi^-\pi^+}\ 
    [\text{GeV}^2]$. Ordinate: $s_{\pi^-K^+}\ [\text{GeV}^2]$.}
  \label{fig:s13}
\end{figure}

\begin{figure}
  \centering
\begin{pspicture}(-1,-1)(7,8)
\psfrag{s[12]}{$s_{\pi^-\pi^+}\ [\text{GeV}^2]$}
\psfrag{s[13]}{\rotatebox{90}{$s_{\pi^-K^+}\ [\text{GeV}^2]$}}
    \rput[bl](-1,-1){\includegraphics[scale=.5]{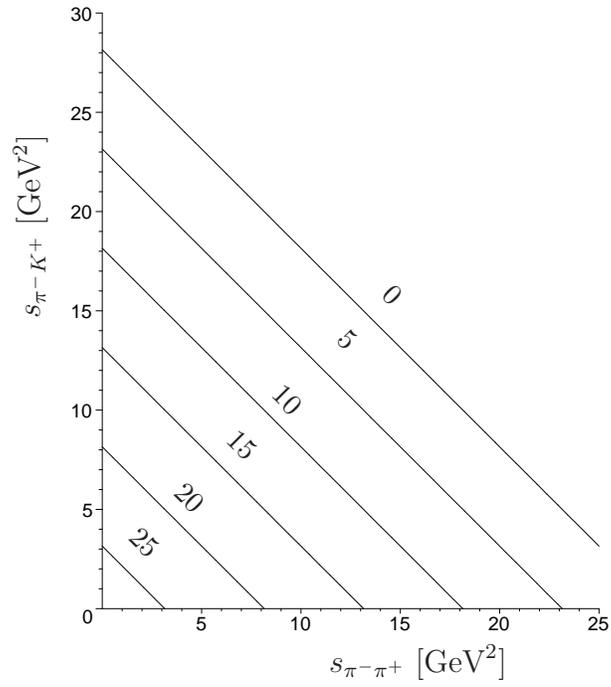}}
    \rput{-45}(4,4){{\small 0}}
    \rput{-45}(3.4,3.4){{\small 5}}
    \rput{-45}(2.6,2.6){{\small 10}}
    \rput{-45}(2,2){{\small 15}}
    \rput{-45}(1.3,1.3){{\small 20}}
    \rput{-45}(.7,.7){{\small 25}}
\end{pspicture}
  \caption[Lines of constant $s_{K^+\pi^+}$.]{Lines where $s_{K^+\pi^+}= (0,5,
    10, 15, 20, 25)\, \text{GeV}^2$.}
  \label{fig:s23}
\end{figure}

\begin{figure}
  \centering
\begin{pspicture}(0,0)(7,8)
  \rput[bl](0,0){\includegraphics[scale=.5]{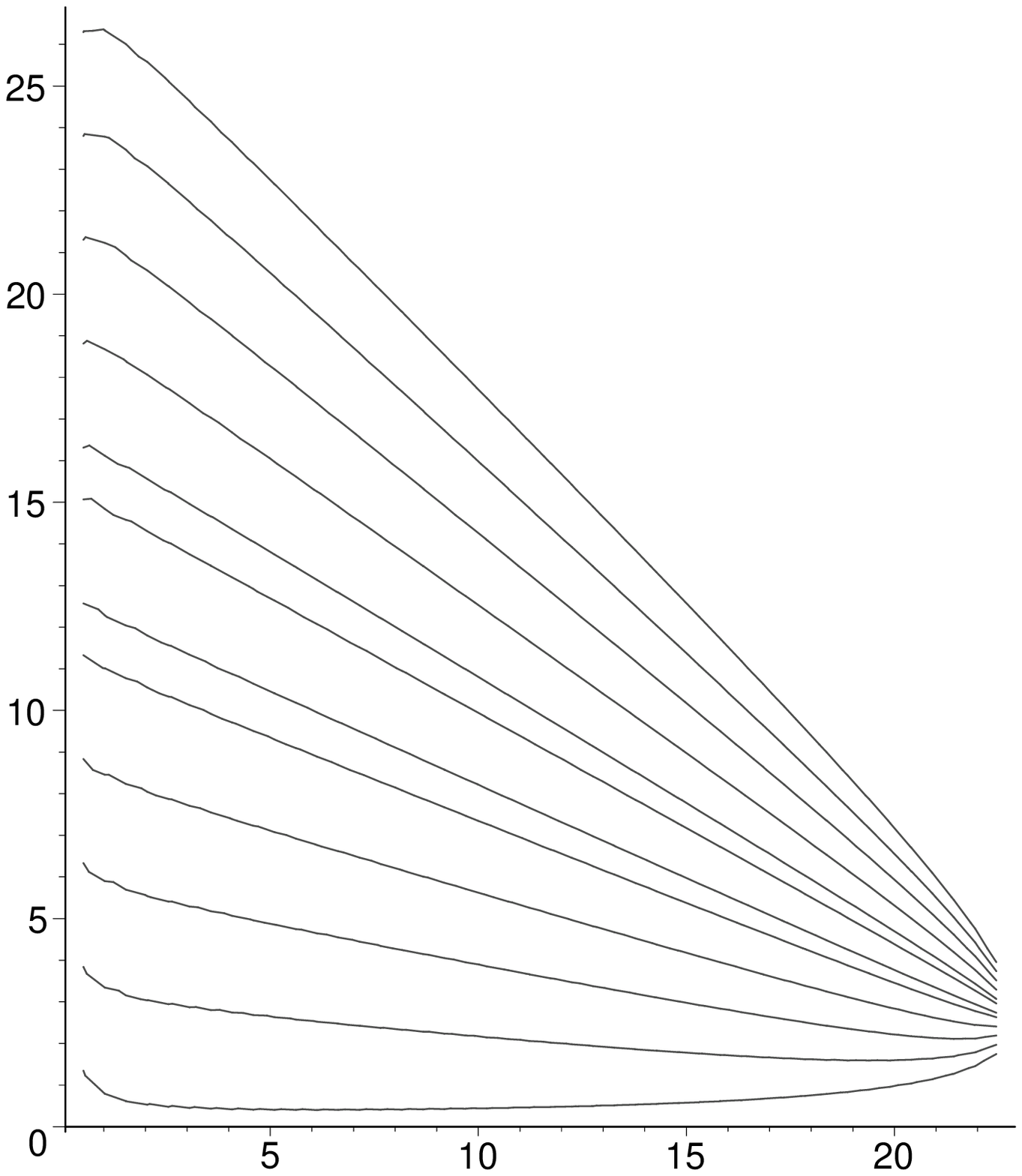}}
  \psline{->}(-.4,7.9)(.4,7.9)
  \rput[r](-.4,7.9){{\small $z_{\pi^-\pi^+}=-1$}}
  \psline{->}(-.4,4.3)(1,4.3)
  \rput[r](-.4,4.3){{\small $z_{\pi^-\pi^+}=-0.1$}}
  \psline{->}(-.4,3.9)(.4,3.9)
  \rput[r](-.4,3.9){{\small $z_{\pi^-\pi^+}=0.1$}}
  \psline{->}(-.4,.6)(.4,.6)
  \rput[r](-.4,.6){{\small $z_{\pi^-\pi^+}=1$}}
\end{pspicture}
  \caption[Lines of constant $z_{\pi^-\pi^+}$.]{Lines where $z_{\pi^-\pi^+}=
    \pm 1, \pm 0.8, \pm 0.6, \pm 0.4, \pm 0.2, \pm 0.1$. Abscissa:
    $s_{\pi^-\pi^+}\ [\text{GeV}^2]$.  Ordinate: $s_{\pi^-K^+}\ 
    [\text{GeV}^2]$.}
  \label{fig:z12}
\end{figure}

\begin{figure}
   \centering
\begin{pspicture}(0,0)(7,8)
  \rput[bl](0,0){\includegraphics[scale=.5]{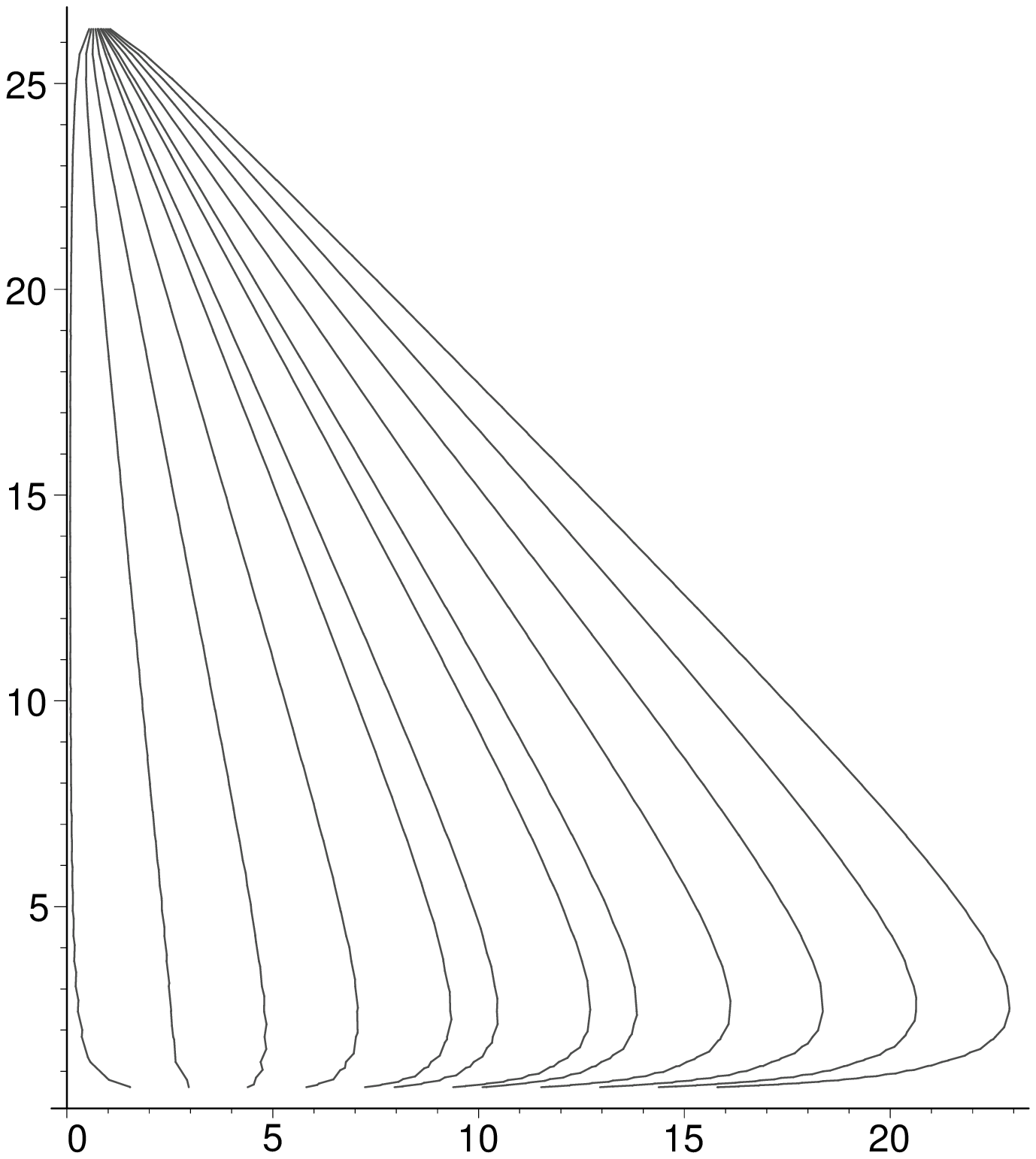}}
   \psline{->}(7.2,2.2)(6.6,1.6)
   \rput[bl](7.2,2.2){{\small $z_{\pi^-K^+}=-1$}}
\end{pspicture}
   \caption[Lines of constant $z_{\pi^-K^+}$.]{Lines where $z_{\pi^-K^+}= \pm
     1, \pm 0.8, \pm 0.6, \pm 0.4, \pm 0.2, \pm 0.1$. Abscissa:
     $s_{\pi^-\pi^+}\ [\text{GeV}^2]$. Ordinate: $s_{\pi^-K^+}\ 
     [\text{GeV}^2]$.}
   \label{fig:z13}
 \end{figure}

\begin{figure}
   \centering
\begin{pspicture}(0,0)(7,8)
  \rput[bl](0,0){\includegraphics[scale=.5]{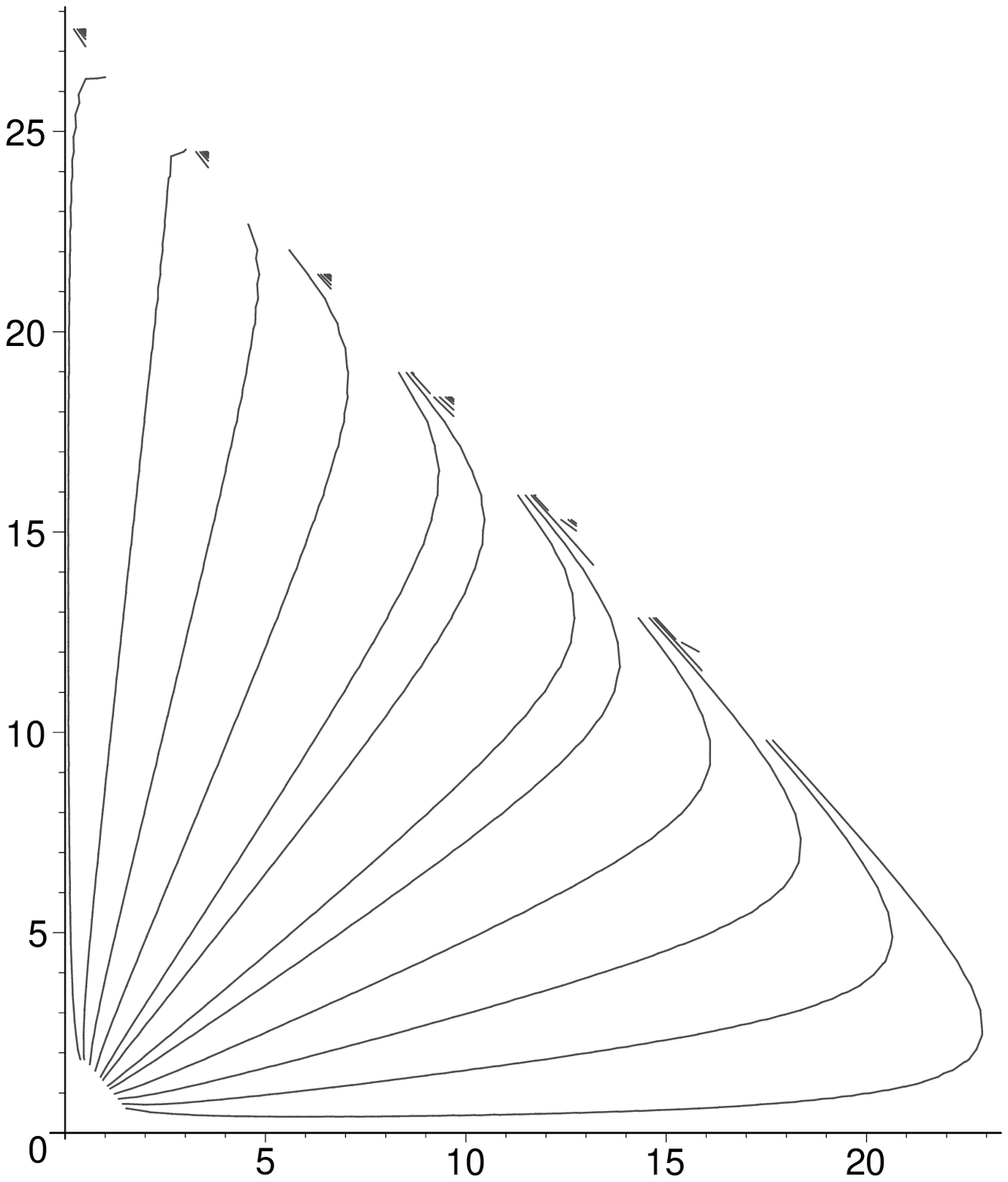}}
   \psline{->}(7.2,2.2)(6.6,1.6)
   \rput[bl](7.2,2.2){{\small $z_{K^+\pi^+}=1$}}
\end{pspicture}
    \caption[Lines of constant $z_{K^+\pi^+}$.]{Lines where $z_{K^+\pi^+}= \pm
      1, \pm 0.8, \pm 0.6, \pm 0.4, \pm 0.2, \pm 0.1$. Abscissa:
      $s_{\pi^-\pi^+}\ [\text{GeV}^2]$. Ordinate: $s_{\pi^-K^+}\ 
      [\text{GeV}^2]$.}
    \label{fig:z23}
 \end{figure}

In a typical Dalitz plot for $B^+\to\pi^-\pi^+ K^+$ each decay event is
assigned a coordinate pair $(s_{\pi^-\pi^+},s_{\pi^-K^+})$. With the kinematic
constraints discussed in this chapter the decay events can be
identified\footnote{Actually, the decay events are not uniquely determined by
  pairs of two appropriate kinematic quantities like
  $(s_{\pi^-\pi^+},s_{\pi^-K^+})$. The mirror image of a kinematic situation
  in a three-body decay is not distinguished from its original by such
  coordinates.} equivalently by other pairs of kinematic quantities. For the
task of identifying resonance masses and spins one is interested in the
invariant pair masses $s_{\pi^-\pi^+}$, $s_{\pi^-K^+}$ and $s_{K^+\pi^+}$, and
in the two-body angles $\theta_{\pi^-\pi^+}$, $\theta_{\pi^-K^+}$ and
$\theta_{K^+\pi^+}$. In figures~\ref{fig:s12} to~\ref{fig:z23} lines of
constant $s$ and $z\equiv\cos\theta$ are drawn using the Maple command
\texttt{implicitplot}. With any two out of these six patterns one can
construct one of ${6\choose 2}=15$ coordinate systems for the three-body decay
events.

The possibility of using coordinate systems consisting of angles and pair
masses shows that the interpretation of structures in the Dalitz plot event
distribution like bands (see figures~\ref{fig:bands} and \ref{fig:overlap}) is
a priori not unique. A concentration of events in a certain region of the
Dalitz plot can obtain because certain pair masses are more likely to be
produced in the decay or because certain angular distributions in a two-body
subsystem are preferred.

\subsection{Partial waves}

The coordinate lines of the two-body angle can eventually be used for a
partial wave analysis. In the decay $B^+\to\pi^-\pi^+ K^+$ the line of
constant $z_{\pi^-K^+}=0$ could be of particular importance. In \cite{belle04}
the reference amplitude and phase is taken to be the signal from
\begin{equation}
  B^+\to K^*(892)\pi^+ \to  K^+\pi^-\pi^+.
\end{equation}
$K^*(892)$ has spin $J=1$ and should thus show its particular angular
dependence given by the Legendre polynom
\begin{equation}
  P_1(z_{\pi^-K^+}) = z_{\pi^-K^+} \equiv \cos\theta_{\pi^-K^+}.
\end{equation}
The partial wave amplitude representing a decay via the $K^*(892)$ is
therefore expected to vanish where $z_{\pi^-K^+}=0$.

\begin{figure}
\centering
\subfigure[]{\includegraphics[scale=.4]{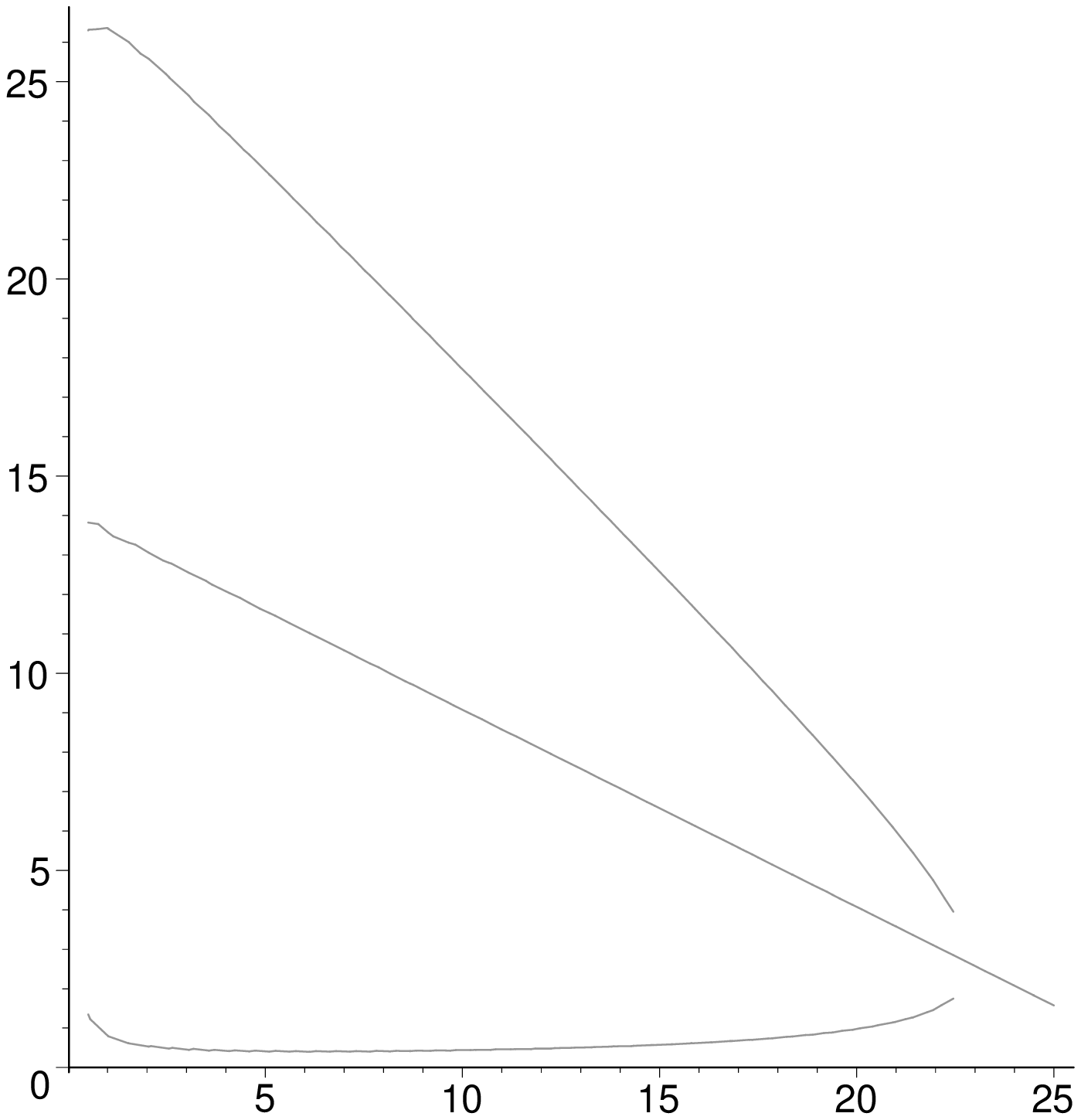}}\quad
\subfigure[]{\includegraphics[scale=.4]{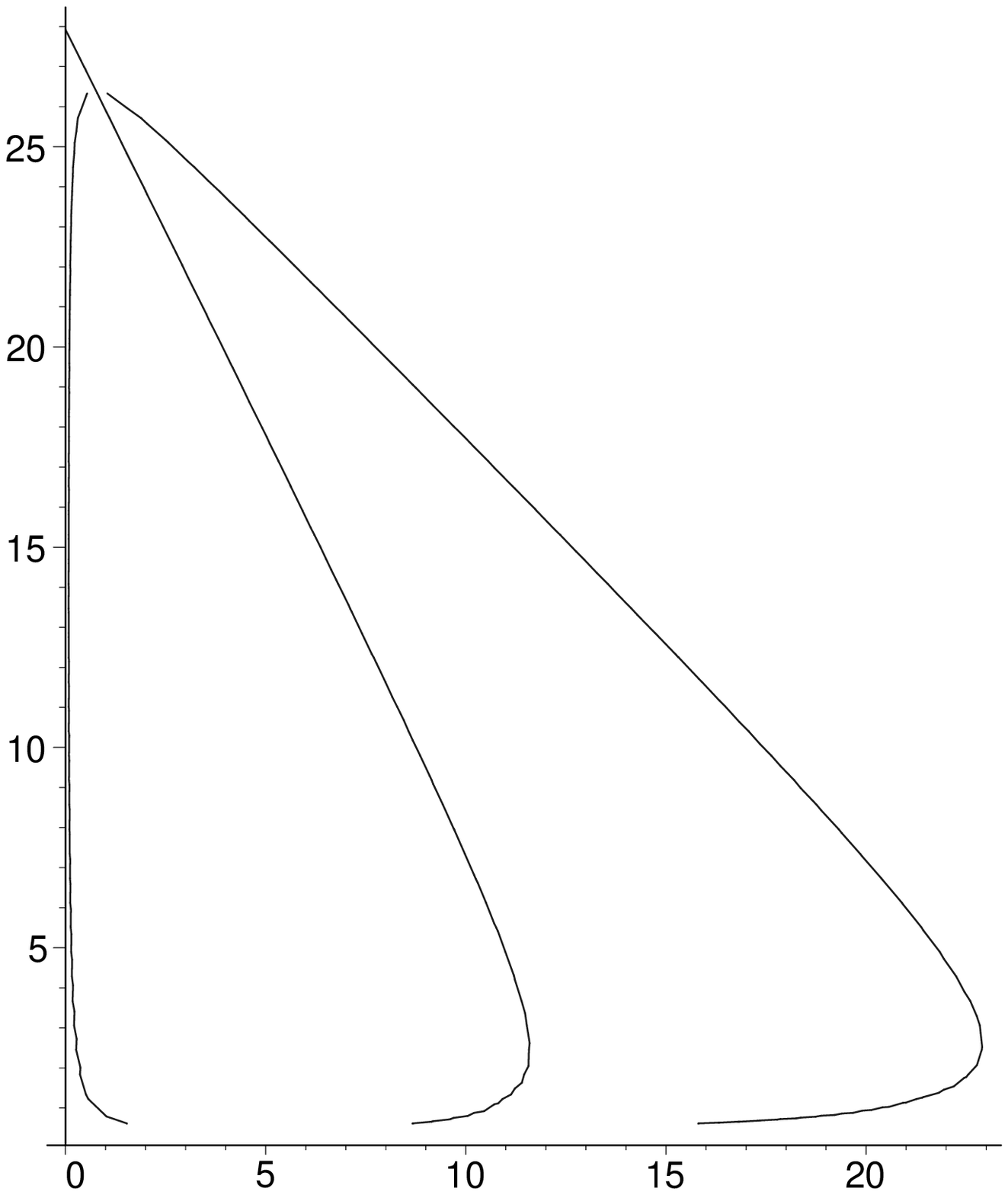}}\quad
\subfigure[]{\includegraphics[scale=.4]{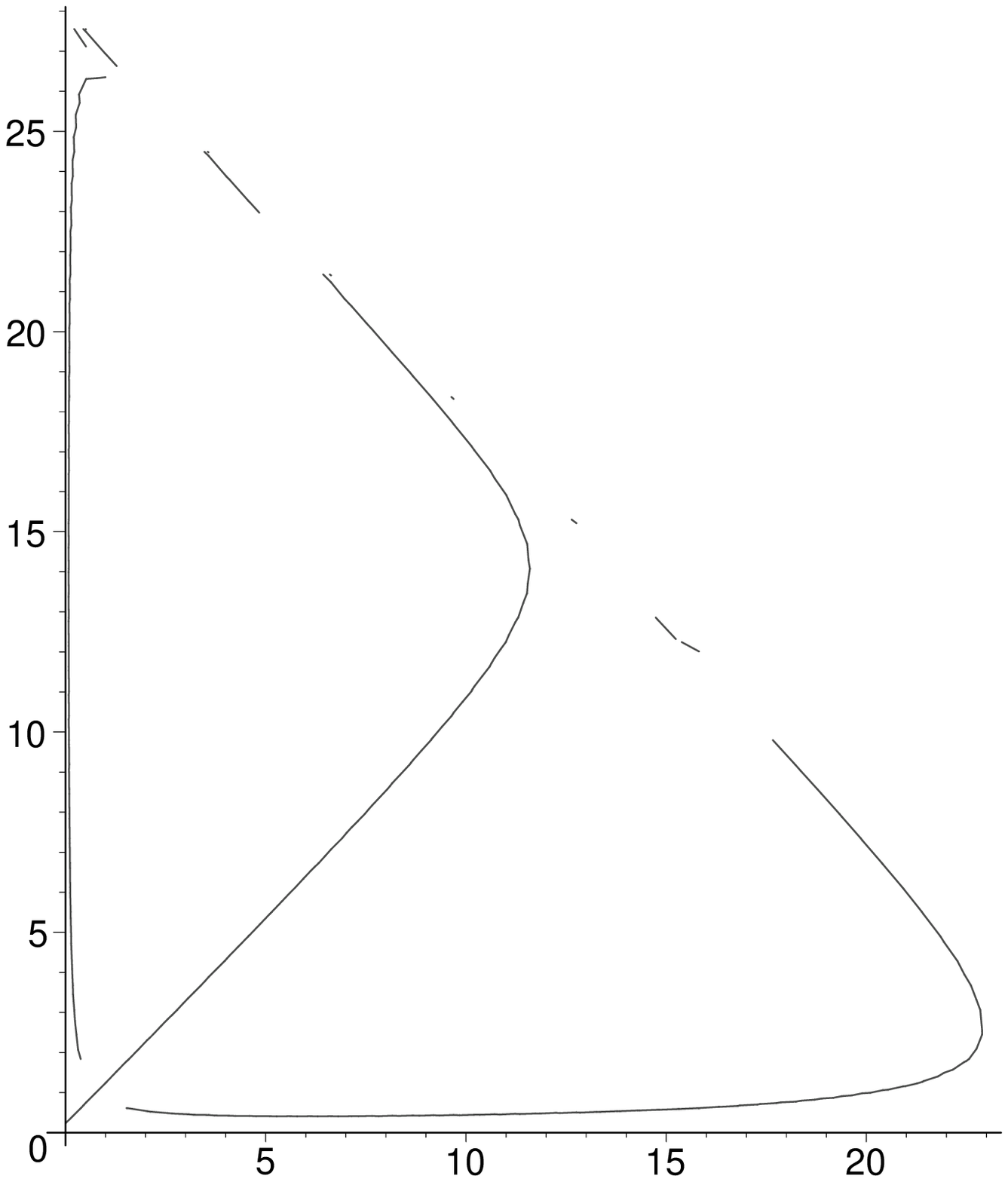}}
  \caption[Lines with $z=0$.]{Lines with (a) $z_{\pi^-\pi^+}=0$, (b)
    $z_{\pi^-K^+}=0$, (c) $z_{\pi^+K^+}=0$ and parts of the Dalitz-plot
    boundary, \ie with $|z|=1$ for the respective $z$'s.  Abscissa:
    $s_{\pi^-\pi^+}\ [\text{GeV}^2]$. Ordinate: $s_{\pi^-K^+}\ 
    [\text{GeV}^2]$.}
  \label{fig:z=0}
\end{figure}

In figure~\ref{fig:z=0} the three lines $z_{\pi^-\pi^+}=0$, $z_{\pi^-K^+}=0$,
$z_{\pi^+K^+}=0$ are plotted.

\chapter{Guidelines for Dalitz plot analysis}\label{cha:guide}

\section{Surrogate scattering laboratories}\label{sec:vic-scatt}

Experimentally, collisions with an unstable target (and beam) particle like
$\pi\pi\to\pi\pi$ and $K\pi\to K\pi$ are not feasible. Chew and Low
\cite{chew} proposed that one can study such reactions in scattering of the
beam particle off a stable target whereby the beam particle actually scatters
with a virtual particle that is exchanged between the beam and the target
particle. Two typical reactions of this type which serve as surrogate
laboratory for scattering of unstable particles are $\pi^- p \to \pi^-\pi^+ n$
and $K^- p \to K^-\pi^+ n$, see also section~\ref{sec:kpi}. Another
possibility in a similar vein is to study \eg the scattering of two pions in
the final state of a three-body decay such as $J/\psi\to\phi\pi\pi$ where as a
good approximation the $\phi$ can be assumed not to interact with the pions
\cite[p.~1188, p.~1193]{morgan93}. As a by-product of the extensive
CP-violation studies in $B$ decays like $B^+ \to K^+\pi^+\pi^-$ there is now a
lot of new data available from a similar type of surrogate laboratory for
final-state scattering of pions and kaons. One great advantage of using this
new data is the huge amount of events such that it should be possible to
interpret structures of the corresponding Dalitz plots that in earlier Dalitz
plots would not be distinguishable from statistical fluctuations or not
visible at all.

Another feature of the reaction $B^+ \to K^+\pi^+\pi^-$ is that contrary to
the case of $J/\psi\to\phi\pi\pi$ the approximation that only two of the three
final state particles interact is not applicable. Rather the pairwise
interactions between all three final state particles are of the same order in
strength.  However, as an approximation one can assume that the interactions
can be regarded as a superposition of three pairwise interactions and neglect
the \emph{simultaneous} interaction of all three particles in the final state,
see equation~\eqref{eq:59}.

I will propose some guidelines for a Dalitz plot analysis for the three-body
decay $B^+ \to K^+\pi^+\pi^-$. I concentrate on this example because it allows
to show that an analysis in terms of a sum of Breit-Wigner terms and a
non-resonant contribution à la Belle and Babar \cite{belle03,belle04,babar04}
may be at odds with what is known about the involved two-body systems and
insufficient to extract new or confirm old information about those.

The goal of investigating the strong interactions of the final state particles
of the three-body decay defines a region of interest in the Dalitz plot.
Controversial issues in the hadron spectroscopy of pions and kaons include the
$\sigma$, the $\kappa$, the $f_0(980)$, the $f_0(1370)$, the $f_0(1500)$ and
the $K^*(1430)$.  These are structures in the $\pi\pi$ and $K\pi$ mass-region
from zero to about 1500 MeV. These mass intervals corresponds to the lower
left corner of the Dalitz plot of $B^+ \to K^+\pi^+\pi^-$ in the usual
representations of \eg Belle.\footnote{Note that Belle has $s_{\pi^-K^+}$ as
  $x$-axis and $s_{\pi^-\pi^+}$ as $y$-axis while in my plots I have
  (unintendedly) reversed the assignment.}

\section{Three-body decay and two-body amplitudes}

As an approximation one can think of a three-body decay like
$B^+\to\pi^-\pi^+K^+$ as the superposition of three amplitudes for the
production of a two-body state with an accompanying \emph{spectator}, \ie a
particle which apart from its mere presence does not make any difference for
the interactions taking place,
\begin{multline}\label{eq:59}
    T(B^+\to \pi^-\pi^+ K^+) = T(B^+\to (\pi^-\pi^+) K^+) +
    T(B^+\to (\pi^- K^+)\pi^+) \\ + T(B^+\to \pi^-(\pi^+
    K^+)).
\end{multline}
Each of the production amplitudes on the right-hand side can be related to the
scattering amplitude of the respective two-body system. The relation that I
will give is distinguished by that it is by construction consistent with
unitarity constraints as discussed in the following, cf.~\cite{morgan93,amp87}.

Here and in the following I use with refs.~\cite{morgan93,amp87} $T$ elements
for discussing unitarity constraints. In my conventions the relevant quantity
would rather be $\mathcal{M}$. $\mathcal{M}$ and $T$ do, however, satisfy the
same form of unitarity constraint, see equations~\eqref{eq:26} and
\eqref{eq:m-unit}. Therefore, the difference between $T$ and $\mathcal{M}$ is
irrelevant here.

\subsection{Unitarity and production amplitudes}
\label{sec:el-unit}

To see how the production amplitudes of the right-hand side of
equation~\eqref{eq:59} relates to the respective two-body scattering
amplitudes, let us take as an example the amplitude $T(B^+\to \pi^-(\pi^+
K^+))$. The unitarity constraint takes into consideration the channels $B^+$,
$\pi^-(\pi^+ K^+)$ and inelastic channels of the scattering of $\pi^+ K^+$
(with $\pi^-$ as spectator) like $\pi^-(\pi^+ K^+\pi^-\pi^+)$.  I enumerate
the channels as
\begin{equation}
  1 = B^+,\qquad 2 = \pi^-(\pi^+ K^+),\qquad 3 = \pi^-(\pi^+
K^+\pi^-\pi^+),\quad \dots, 
\end{equation}
where the dots stand for the prescription to enumerate all further inelastic
$\pi^+ K^+$-scattering channels. (There are no other channels with only
two-particles to which  $\pi^+ K^+$ couples; no other two-particle state has
the same quantum numbers as $\pi^+ K^+$.)

I use the unitarity constraint $T^\dagger T = i(T^\dagger -T)$ in the form
\begin{equation}\label{eq:18}
  \text{Im} T_{ji} = \frac{1}{2} \sum_k T^*_{jk}T_{ki}.
\end{equation}
This means that I assume that the states can be properly discretized (cf.\ 
section~\ref{sec:box}) and that time reversal invariance is
given,\footnote{Because of CPT invariance the assumption of time reversal
  invariance is incompatible with the possibility of CP violation. For the
  issues discussed here the assumption of $T$ invariance is simplifying but
  not necessary. CP violating phases can be neglected in the present context
  as overall phases. They can be absorbed by a redefinition of the state
  vectors.} which I take to imply that the $T$ matrix is symmetric, in other
words
\begin{equation}
  T^\dagger = T^*.
\end{equation}

The $B^+$ meson consists of an up- and an anti-bottom quark. It has thus
bottomness~1. Strong interactions conserve the flavors. Therefore, since the
$B$-meson is the lightest state with bottomness~1 it can only decay by
electro-weak interactions. The elements $T_{1k}$ ($k\neq 1$) representing a
transition from channel 1 (the $B^+$) to some other channel are therefore of
order of magnitude of the electro-weak coupling constants, which in the energy
region to be discussed here are small compared to the coupling constant of the
strong interaction. In the following formulation of unitarity constraints I
will neglect all terms quadratic in electro-weak couplings. To keep track of
the different orders of magnitude of the electro-weak couplings I use a
superscript `$w$' for electro-weak amplitudes and a superscript `$s$' for
strong amplitudes. 

The $S$ element for interactions in channel~1 and transitions from channel~1
to any other channel is purely weak,
\begin{equation}
  S_{1i} = S_{1i}^w, \quad i \in \{1,2,3,\dots\}.
\end{equation}
On the other hand, interactions in and transitions between channels~2, 3,
\dots\ are described by a Hamiltonian that is a sum of an electro-weak and a
strong Hamiltonian,
\begin{equation}
  H_{lm} = H_{lm}^s + H_{lm}^w, \quad l,m \in \{2,3,\ldots\}.
\end{equation}
The corresponding $S$ operator is therefore to leading order in $H_{lm}^w$
given by (see \cite[p.~108]{peskin}).
\begin{equation}
  S = \lim_{t\to\infty} \exp(-iH_{lm}(2t)) \approx \lim_{t\to\infty}
  (\mathbbm{1}-iH_{lm}^wt)\exp(-iH_{lm}^s(2t)). 
\end{equation}
For the $T_{lm}$ ($l,m \in \{2,3,\ldots\}$) elements we obtain
\begin{multline}
  T_{lm} = -i(S_{lm} - \delta_{lm}) \approx \lim_{t\to\infty} -i
  [(\delta_{lm}-iH_{lm}^w(2t))(\delta_{lm}+iT_{lm}^s) -
  \delta_{lm}] \\
  = T_{lm}^sS_{lm}^w + T_{lm}^w,
\end{multline}
with
\begin{equation}
  T_{lm}^w = \lim_{t\to\infty} (-H_{lm}^w(2t)) \quad \text{and}\quad
  S_{lm}^w=\lim_{t\to\infty} \exp(-iH_{lm}^w(2t)). 
\end{equation}
The factor $S^w$ gives a weak phase to the strong
amplitude. This phase can be neglected in the present context as an overall
phase. It can be absorbed by a redefinition of the state vectors. Not keeping
track of this overall phase and neglecting quadratic and higher order terms of
the weak $T$ elements we then have in and between channels 2, 3, \dots
\begin{equation}
T_{ji} = T_{ji}^w + T_{ji}^s, \quad j,i \in \{2,3,\ldots\}.
\end{equation}
The unitarity condition for the production amplitudes, $T^w_{1k}$ ($k\neq 1$),
reads
  \begin{align}
    2\text{Im}T^w_{1k} &= \sum_{r=1}^{\dots} T^*_{1r}T_{rk}\\
    &= T^{w*}_{11}T^w_{1k} + \sum_{r=2}^{\dots}
    T^{w*}_{1r}(T^{s}_{rk}+T^{w}_{rk})\\  
    &= \sum_{r=2}^{\dots} T^{w*}_{1r}T^{s}_{rk},\label{eq:28}
  \end{align}
  where in the last step terms quadratic in the weak coupling have again been
  dropped.
  
If the production amplitudes are written as,
\begin{equation}\label{eq:prod-ampl}
  T_{1k} = T^w_{1k} = \sum_{n=2}^{\dots} \alpha_n T_{nk}^s, \quad
  k\neq 1,
\end{equation}
where, importantly, the $\alpha$'s are real, the constraint of
equation~\eqref{eq:28} is satisfied by construction; provided that the strong
amplitudes $T_{nk}^s$ satisfy the unitartity condition among themselves, \ie
\begin{equation}
  2\text{Im}T^s_{lm} = \sum_k T^{s*}_{lk}T^{s}_{km}, \quad k,l,m \in
  \{2,3,\ldots\}. 
\end{equation}
Indeed we then have:
\begin{multline}
  2\text{Im}T^w_{1k} = 2\text{Im} \sum_{n=2}^{\dots} \alpha_n T_{nk}^s =
  \sum_{n=2}^{\dots} \alpha_n 
  2\text{Im} T_{nk}^s = \sum_{n=2}^{\dots} \alpha_n \sum_{r=2}^{\dots}
  T^{s*}_{nr}T^{s}_{rk} \\  
  = \sum_{r=2}^{\dots} \sum_{n=2}^{\dots} \alpha_n T^{s*}_{nr} T^{s}_{rk} =
  \sum_{r=2}^{\dots} 
  T^{w*}_{1r}T^{s}_{rk}.
\end{multline}

\subsection[Elastic region for $\pi^-\pi^+$ and $\pi^-K^+$]{Elastic region for
  $\boldsymbol{\pi^-\pi^+}$ and $\boldsymbol{\pi^-K^+}$}\label{sec:watson}

Because of equation~\eqref{eq:17} not all three two-body cms energies can be
low. The region of the Dalitz plot we are interested in is \eg characterized
by low values of $s_{\pi^-\pi^+}$ and $s_{\pi^-K^+}$ and high values of
$s_{\pi^+K^+}$, roughly
\begin{alignat}{2}\label{eq:60}
  0\ \text{GeV} &\leq \sqrt{s_{\pi^-\pi^+}} \leq &1.6\ \text{GeV}, \\
  0\ \text{GeV} &\leq \sqrt{s_{\pi^-K^+}} \leq &1.6\ \text{GeV}, \\
  3\ \text{GeV} &\leq \sqrt{s_{\pi^+K^+}} \leq &5\ \text{GeV}.  
\end{alignat}

From analyses of $\pi\pi$ scattering (\eg CERN-Munich \cite{hyams}, see
figure~\ref{fig:hyams}, lower panel) it is known that indeed the inelasticity
is almost 1 up to 1.5 GeV, except for a small energy region around 980 MeV.
This dip in the inelasticity may be due to the decay of $f_0(980)$ into
$\pi\pi$ as well as into $K\bar{K}$. I neglect this inelasticity in the hope
that the resulting ansatz is while simpler still able to represent the most
relevant features of the process.

For the production amplitudes $T(B^+\to (\pi^-\pi^+) K^+)$ and $T(B^+\to (\pi^-
K^+)\pi^+)$ we can therefore use an \emph{elastic} unitarity condition that
takes only elastic strong amplitudes to be non-zero. The respective production
amplitudes have then the simple form $T_{1k} = T^w_{1k} = \alpha_k T_{kk}^s$,
\ie
\begin{equation}\label{eq:4}
  \begin{split}
    T(B^+\to (\pi^-\pi^+) K^+) &= \alpha_{\pi^-\pi^+}
    T^s(\pi^-\pi^+\to\pi^-\pi^+) \\
    T(B^+\to (\pi^- K^+)\pi^+) &= \alpha_{\pi^- K^+} T^s(\pi^- K^+\to\pi^-
    K^+).
\end{split}
\end{equation}
This form of the production amplitude shows that, since $\alpha_k$ is real,
the production amplitude and the strong amplitude of the interaction of the
final state particles have the same phase. This result is known as
\emph{Watson's theorem}, \emph{final state interaction theorem} or just
\emph{elastic unitarity}. The name ``final state interaction theorem'' is
justified by that the particular form of the production amplitude
(eq.~\eqref{eq:prod-ampl}), which is a solution for the unitarity constraint
can be interpreted to the effect that the decay from channel~1 into a channel
$k\neq 1$ is a sum of amplitudes representing the decay from channel~1 into
channel $n\neq 1$ with coupling strength $\alpha$ followed by a strong final
state interaction represented by $T_{nk}^s$ \cite{morgan93,amp87}. In the
original Watson theorem \cite{watson52} it is explicitly two potentials that
are considered: The potential responsible for the production of the hadrons,
and the potential of the interaction of these hadrons in the final state.

\subsubsection{Singular couplings}\label{sec:sing-coup}

Since the partial wave amplitudes satisfy a unitarity condition of the same
form as the elements of the $T$ matrix, see equation~\eqref{eq:3} unitarity
for production amplitudes requires also each partial wave and isospin
component to be of the form of equation~\eqref{eq:prod-ampl}. Thus we obtain
unitarity constraints for the elastic amplitude in terms of phase shifts (cf.\
equation~\eqref{eq:5}), for example:
\begin{equation}
  \begin{split}
    \mathcal{M}^J_I(B^+\to (\pi^-\pi^+) K^+) &= \alpha^J_I
    (\mathcal{M}^J_I)^s(\pi^-\pi^+\to\pi^-\pi^+) \\
    &= 2e^{i\delta^J_I}\sin\delta^J_I.
  \end{split}
\end{equation}

As discussed and emphasized in \cite[p.~1188]{morgan93} and
\cite{pennington97} the $\alpha$'s are not necessarily regular functions. They
may be such that zeros of the elastic amplitudes are removed. This is the case
if $\alpha$ behaves like $1/\sin\delta$ in the region where $\delta$ is a
multiple of $\pi$. Also the $\alpha$'s may introduce new zeros. It is adequate
to redefine the real coupling constants
\begin{equation}
  2\alpha^J_I\sin\delta^J_I \equiv \tilde{\alpha}^J_I,
\end{equation}
such that
\begin{equation}
 \mathcal{M}^J_I(B^+\to (\pi^-\pi^+) K^+) =  \tilde{\alpha}^J_I
 e^{i\delta^J_I}.
\end{equation} 

In ref.~\cite[p.~1189]{morgan93} it is emphasized that, while the production
and the scattering amplitude do not necessarily have the same zeros, they do
indeed have the same resonance poles. Resonance poles are in this sense
universal: ``A further \emph{consequence} of unitarity, vital for our
discussion, is that it requires that resonance poles be universal, \ie, a
given resonance pole occurs at the same complex energy $E_R$ in all processes
to which it couples. This is automatically built in our solution, [\dots]''

I share the view that the universality of resonance poles is indeed an
important feature of the solution given in ref.~\cite{morgan93}. The
universality gives the resonances an identity. However, since poles are
complex quantities they are not observable and their determination therefore
requires an extrapolation from real to complex values of energy. This in turn
presupposes a lot of theoretical input.

More specifically, I do not see how the universality of the
resonance poles is \emph{implied} by unitarity. With the solution of
ref.~\cite{morgan93} the scattering and the production amplitude do have the
same poles; it respects the universality of the resonance poles. But I cannot
see an argument why this is the \emph{only} way to satisfy the unitarity
constraint.

\subsection[High energy amplitude for $K^+\pi^+$]{High energy amplitude for
  $\boldsymbol{K^+\pi^+}$}

In the region of interest defined in equation~\eqref{eq:60} we have to
consider the high energy behavior for the amplitude $T(B^+\to \pi^-(\pi^+
K^+))$ in contrast to the amplitudes $T(B^+\to (\pi^-\pi^+) K^+)$ and
$T(B^+\to (\pi^- K^+)\pi^+)$, where we are concerned with low cms energies.
Unitarity as formulated in section~\ref{sec:el-unit} is satisfied by the
following ansatz (see eq.~\eqref{eq:prod-ampl}),
\begin{multline}
  T(B^+\to \pi^-(\pi^+K^+)) = \alpha_{\pi^+K^+}T^s(\pi^+K^+\to \pi^+K^+) \\
  + \alpha_{\pi^+K^+\pi^-\pi^+}T^s(\pi^+K^+\pi^-\pi^+\to \pi^+K^+) + \ldots 
\end{multline}

\subsubsection{No resonances}

The two particle state $\pi^+K^+$ has isospin
$I=3/2$ and has electric charge 2. A $q\bar{q}$ resonance cannot have this
electric charge and this isospin. Baryon resonances can, \eg $\Delta^{++}$,
but because of baryon number conservation $\pi^+K^+\to \text{baryon}$ is not
possible. To conserve baryon number, baryonic resonances should be produced as
pairs of baryon and antibaryon. Also there is the a priori possibility of
resonance of four quarks. In any case, no resonances in the $\pi^+K^+$ channel
are known as of today.

The S-wave phase shift with $I=3/2$ is at least in the interval 0.8 GeV to 2
GeV negative \cite{Buettiker:2003pp}. This supports the claim that there are
no resonances in this channel; a repulsive force cannot lead to resonances.

\subsubsection[Forward peak in $K\pi$ scattering?]{Forward peak in
  $\boldsymbol{K\pi}$ scattering?}\label{sec:kpi}

I do not see any reason why the $\pi^+K^+$ total cross section should vanish.
By the optical theorem (see section~\ref{sec:opt-th}) then the imaginary part
of the elastic amplitude in the forward direction is non-zero and with it the
absolute value of the elastic amplitude. So although there are no resonances
in the $\pi^+K^+$ channel we have reasons to expect that the elastic amplitude
is not zero in the region of the Dalitz plot under consideration. Contrary to
this expectation the amplitude in the $\pi^+K^+$ channel is set to zero in the
default-model of ref.~\cite{belle04}, see section~\ref{sec:belle}.
\label{non-vanish} There is in principle the possibility that
$\alpha_{\pi^-(\pi^+ K^+)}$ tends to zero in such a manner that the product
$\alpha_{\pi^-(\pi^+ K^+)} T (\pi^+ K^+\to\pi^+ K^+)$ is zero at values for
the $\pi^+ K^+$ invariant mass of about the $B$ mass.  However, since I know
of no reason why this should happen to be so I assume that also the product is
different from zero.

Not only we have reasons to expect that the $K^+\pi^+$ amplitude is not zero.
There is also experimental indication that the $K^+\pi^+$ amplitude as defined
in chapter~\ref{cha:guide} is forward peaked at energies relevant for the
Dalitz plot analysis of $B^+\to \pi^-\pi^+K^+$: The Lass collaboration obtains
in ref.~\cite{aston} results about $K^-\pi^+$ scattering by studying the
reaction
\begin{equation}
  K^-p\to K^-\pi^+n.
\end{equation}
Basically, the method to extract $K\pi$ scattering data from such reactions is
the one of Chew and Low \cite{chew}: One selects the events with low momentum
transfer $t$. In these events the exchange of a virtual pion in the $t$
channel is the dominant contribution to the interaction of the proton and the
kaon. By an extrapolation from virtual $t$'s to the pion pole one can then
obtain information about the scattering of pions and kaons. The scattering
angle $\theta$ for $K\pi\to K\pi$ is essentially the \emph{Gottfried-Jackson}
angle $\theta_{\text{GJ}}$ of the reaction $K^-p\to K^-\pi^+n$.

\begin{figure}
  \centering \includegraphics[width=.7\linewidth,angle=180]{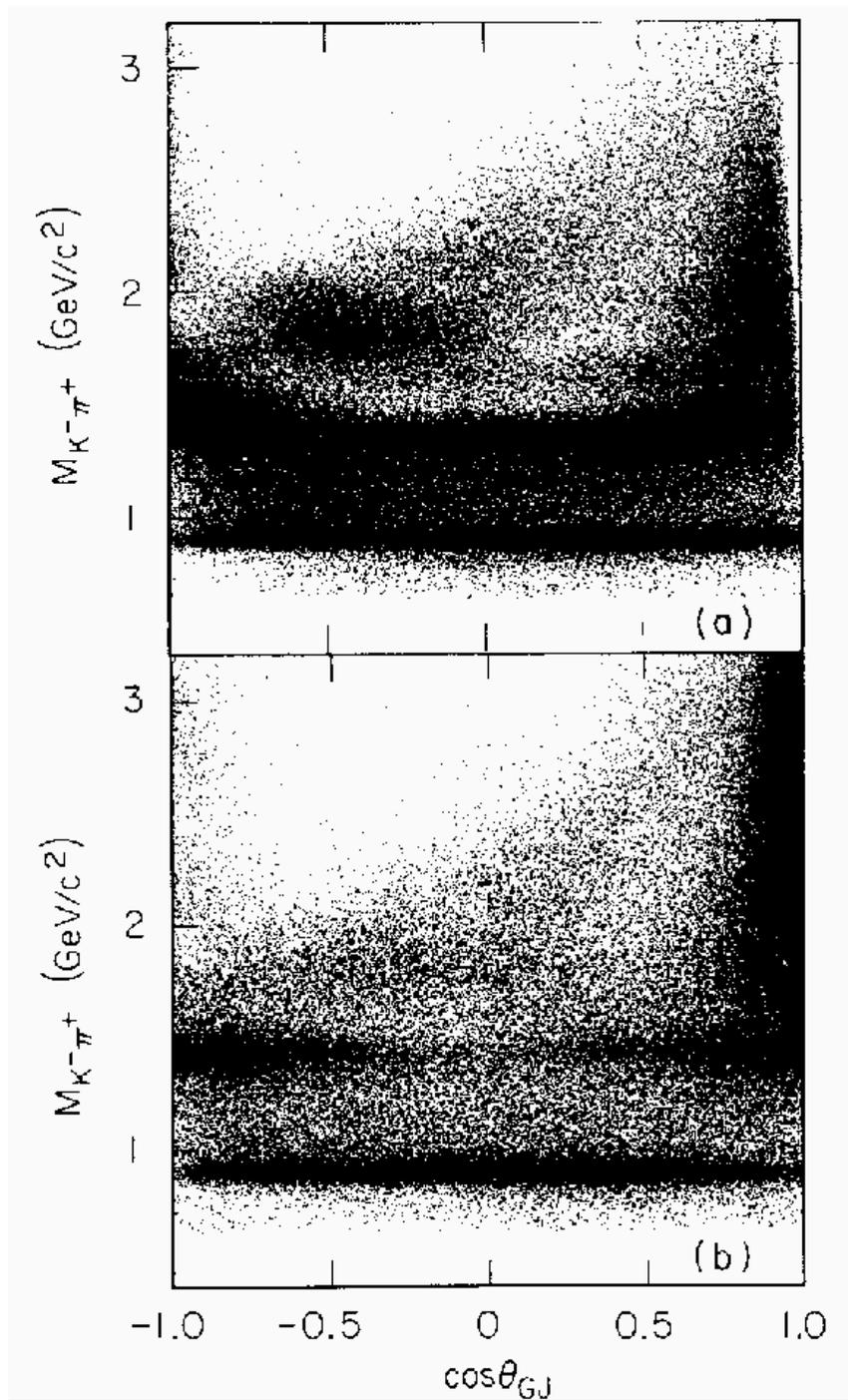}
  \caption[The cosine of the \emph{Gottfried-Jackson} angle of $K^-p\to
  K^-\pi^+n$.]{The cosine of the \emph{Gottfried-Jackson} angle of $K^-p\to
    K^-\pi^+n$ versus the pair mass of $K^-\pi^+$. Top panel: $|t|\leq
    0.2\,\text{GeV}^2$. Bottom panel: $0.2\leq |t|\leq 1.0\,\text{GeV}^2$;
    from \cite{aston}.}
  \label{fig:peak}
\end{figure}

Figure~\ref{fig:peak} shows a distinct enhancement of the $K^-\pi^+$ cross
sections in the forward direction
($\cos\theta_{\text{GJ}}=1$)\footnote{\label{fn:1}I am not sure if
  $\theta_{\text{GJ}}$ corresponds indeed to the \emph{forward} scattering
  angle or rather to the one \emph{backwards}. However, what is of importance
  here is only whether the $K^+\pi^+$ amplitude can generate an enhancement
  near the boundary of the Dalitz plot for $B^+\to \pi^-\pi^+K^+$; and
  \emph{both}, $\theta=0$ and $\theta=180$ degrees are lines on the boundary,
  see figures~\ref{fig:z12}--\ref{fig:z23}.} and energies larger than 2 GeV.
This is particularly so for events with high $t$ (bottom panel) where the
exchange of heavier particles as the $\rho$ dominate over the pion
contribution. But also in the case of relevance here---low $t$'s, pion
exchange---the events at energies around 2.5 GeV and higher are concentrated
at values of $0.5\leq \cos\theta_{\text{GJ}}\leq 1.$

The $K^+\pi^+$ scattering is purely isospin $I=3/2$. The $K^-\pi^+$ amplitude,
studied in \cite{aston}, also contains an $I=3/2$ component. It is not clear
to me whether the forward peak comes only from the isospin components with
$I\neq 3/2$. \emph{As a hypothesis} I nevertheless assume the amplitude
$T^s(\pi^+K^+\to \pi^+K^+)$ at energies of about 4--5 GeV, which are of
interest here, to share the feature of being forward peaked (or backwards, see
footnote~\ref{fn:1}).  From these rough arguments I dare conclude that as a
hypothesis we may write for the $\pi^+ K^+$ production amplitude a
non-resonant amplitude with an enhancement of its absolute value in the region
where $|z_{\pi^+ K^+}|\approx 1$. This is near the boundary of the Dalitz
plot. In particular for pair masses of about 4--5 GeV for $\pi^+K^+$ the
enhancement would be near the lower left corner of the Dalitz plot.  I denote
the amplitude by
\begin{equation}\label{eq:6}
  T(B^+\to \pi^-(\pi^+K^+)) = \mathcal{A}_{\pi^+ K^+}^{\text{boundary}}.
\end{equation}

\subsection{Pronounced S-wave?}\label{sec:s-wave}

\begin{figure}
  \centering
  \input{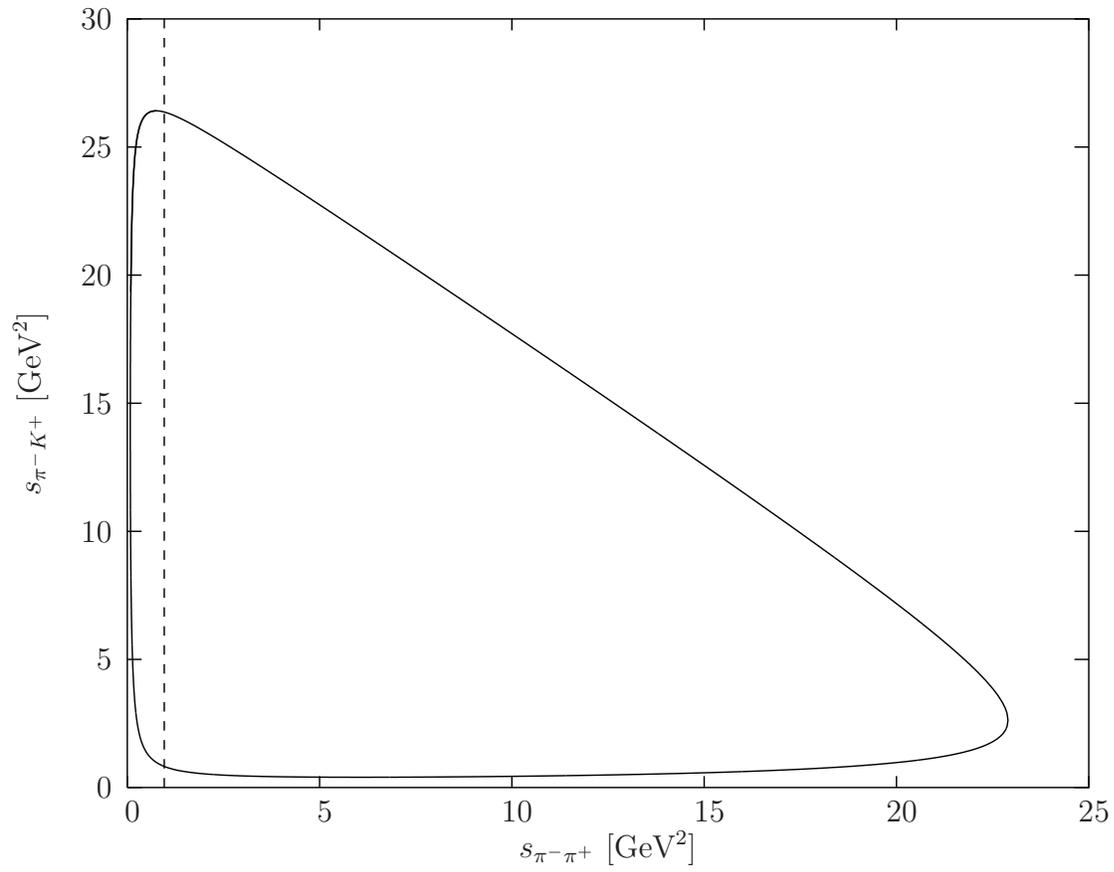}
  \caption[Line with $s_{\pi^-\pi^+}=(.980 \text{GeV})^2$.]{Line with
    $s_{\pi^-\pi^+}=(0.980\ \text{GeV})^2$ and the boundary of the Dalitz plot
    for $B^+ \to \pi^-\pi^+K^+$.}
  \label{fig:f980}
\end{figure}

Compared to the $K$ and the $D$ the $B$ has a much higher mass ($m_K\approx
495\text{MeV}$, $m_D\approx 1900\text{MeV}$, $m_B\approx 5300\text{MeV}$). The
phase space of the three-body final state is therefore much larger in
three-body decays of the latter. The region of interest (see
equations~\eqref{eq:60}ff.) for present problems of hadron spectroscopy is
given by relatively low masses of the $\pi^-\pi^+$ and $\pi^- K^+$ pair. By
equation~\eqref{eq:27} the pair mass of $\pi^+ K^+$ is therefore about 5 GeV.
If the $\pi^+ K^+$ amplitude is indeed forward (or backward) peaked as
suggested in section~\ref{sec:kpi}, this could lead to an enhancement near the
boundary of the Dalitz plot in the lower left corner. I would be interested in
seeing whether such an enhancement explains at least in part the purported
observation of an important S wave signal. The risk of mistaking the $\pi^+
K^+$ forward peak for a strong scalar signal of \eg $f_0(980)$ is given
because the bands or dips expected for such a state is very close to the left
boundary, see figure~\ref{fig:f980}.

\subsection{Resulting ansatz}

Putting together equations~\eqref{eq:59}, \eqref{eq:4} and \eqref{eq:6} we
obtain for the three-body decay amplitude
\begin{multline}
 T(B^+\to\pi^-\pi^+ K^+) = \alpha_{\pi^-\pi^+}
 T^s(\pi^-\pi^+\to\pi^-\pi^+) \\
 + \alpha_{\pi^- K^+}
    T^s(\pi^- K^+\to\pi^- K^+) + \mathcal{A}_{\pi^+
 K^+}^{\text{boundary}}. 
\end{multline}
Performing isospin decomposition (with conventions of the Particle Data Group)
we get
\begin{equation}
  \begin{split}
    T(B^+\to\pi^-\pi^+ K^+) &= \alpha_{\pi^-\pi^+} [ 1/6
    T_2(\pi\pi) + 1/2 T_1(\pi\pi) + 1/3
    T_0(\pi\pi) ] \\ 
    &\quad + \alpha_{\pi^- K^+}
    [ 1/3 T_{3/2}(\pi^- K^+) + 2/3 T_{1/2}(\pi^- K^+) ] \\
    &\quad + \mathcal{A}_{3/2}^{\text{boundary}}(\pi^+
 K^+). 
\end{split}
\end{equation}

Partial-wave decomposition for the $\pi^-\pi^+$ and the $\pi^- K^+$ amplitudes
(taking into account only S and P wave, \ie $J=0$ and $J=1$) further
yields\footnote{Because of Bose symmetry there is no component with $J=0$ and
  $I=1$ or $J=1$ and $I=0$ in the $\pi\pi$ channel.}
\begin{equation}\label{eq:7}
  \begin{split}
    T(B^+\to\pi^-\pi^+ K^+) &= \frac{1}{4\pi} \big(
    \alpha_{\pi^-\pi^+} [ 1/6 T^0_2(\pi\pi) + 1/2
    \cos\theta_{\pi^-\pi^+ }T^1_1(\pi\pi)  + 1/3
    T^0_0(\pi\pi) ] \\
    &\quad + \alpha_{\pi^- K^+} [ 1/3 \{ T^0_{3/2}(\pi^- K^+) +
    \cos\theta_{\pi^- K^+}
    T^1_{3/2}(\pi^- K^+) \} \\
    &\quad + 2/3 \{ T^0_{1/2}(\pi^- K^+) + \cos\theta_{\pi^- K^+}
    T^1_{1/2}(\pi^- K^+) \}
    ] \big)\\
    &\quad + \mathcal{A}_{3/2}^{\text{boundary}}(\pi^+ K^+).
\end{split}
\end{equation}

Because the (real) coupling constants $\alpha$ may be singular (see
section~\ref{sec:sing-coup}) it is appropriate to define the following (still
real) coupling constants:
\begin{equation}
  \begin{split}
    (\alpha_{\pi^-\pi^+})^J_I &\equiv 2 \alpha_{\pi^-\pi^+}
    \sin\delta^J_I(\pi^-\pi^+), \\ 
    (\alpha_{\pi^- K^+})^J_I &\equiv 2 \alpha_{\pi^- K^+} \sin\delta^J_I(\pi^-
    K^+).
  \end{split}
\end{equation}

Then equation~\eqref{eq:7} reads
\begin{equation}
  \begin{split}
    T(B^+\to\pi^-\pi^+ K^+) &= \frac{1}{4\pi} \big( 1/6
    (\alpha_{\pi^-\pi^+})^0_2 e^{i\delta^0_2 (\pi\pi)} + 1/2
    \cos\theta_{\pi^-\pi^+ }(\alpha_{\pi^-\pi^+})^1_1 e^{i\delta^1_1 (\pi\pi)}
    \\
    &\quad + 1/3 (\alpha_{\pi^-\pi^+})^0_0 e^{i\delta^0_0 (\pi\pi)} \\
    &\quad + 1/3 \{ (\alpha_{\pi^- K^+})^0_{3/2} e^{i\delta^0_{3/2}(\pi^-
      K^+)} + \cos\theta_{\pi^- K^+} (\alpha_{\pi^- K^+})^1_{3/2}
    e^{i\delta^1_{3/2}(\pi^- K^+)}\} \\
    &\quad + 2/3 \{ (\alpha_{\pi^- K^+})^0_{1/2}e^{i\delta^0_{1/2}(\pi^- K^+)}
    + \cos\theta_{\pi^- K^+} (\alpha_{\pi^-
      K^+})^1_{1/2}e^{i\delta^1_{1/2}(\pi^- K^+)} \} \big)\\
    &\quad + \mathcal{A}_{3/2}^{\text{boundary}}(\pi^+ K^+).
\end{split}
\end{equation}

In this ansatz the phases of the $\pi^-\pi^+$ and $\pi^- K^+$ amplitudes are
not free fit parameters but are the strong phases of the respective scattering
amplitude. The phase $\delta^0_0 (\pi\pi)$ for example is shown in
figure~\ref{fig:hyams} (top panel). In the Dalitz plot event distribution,
which is essentially given by $|T|^2$ the phases show up in interference terms
of type
\begin{equation}
  \cos(\delta^0_2 (\pi\pi) - \delta^0_0 (\pi\pi)).
\end{equation}

\section{Two examples}

In the following two subsections I briefly report two analyses of $B^+\to
K^+\pi^+\pi^-$ and $B^+\to K^+K^+K^-$. The idea is to give an impression of
how different approaches lead to different conclusions.

\subsection{Minkowski and Ochs}

In ref.~\cite[table~VII]{belle04} the branching fraction for the decay $B^+\to
K^+\pi^+\pi^-$ via the resonance $f_0(980)$ is found to be $7.55\times
10^{-6}$ while in \cite[Table~6]{Minkowski:2004xf} it is concluded to be
$19.3\times 10^{-6}$. One of the reasons for such discrepancies in the
analysis of the three body decays $B^+\to K^+\pi^+\pi^-$ and $B^+\to
K^+K^+K^-$ is the role played by the object called ``non-resonant background''
or ``gb'' respectively. In \cite{Minkowski:2004xf}, the gb interferes in the
$\pi^+\pi^-$ channel destructively with both $f_0(980)$ and $f_0(1500)$ such
that the peaks in the corresponding mass projection of the Dalitz plot appears
smaller than might be expected or even as a dip, see \cite{Minkowski:2004xf}
for details, in particular figure~4. In the channel $K^+K^-$, on the other
hand, the gb interferes \emph{constructively} with the $f_0(1500)$ such that,
contrary to the expectations, the number of events in the region of the
$f_0(1500)$ in the $K^+K^-$ mass projection is actually much higher than in
the $\pi^+\pi^-$ projection, see \cite{Minkowski:2004xf}, in particular
figure~5.

\subsubsection[Large background phase, no $f_0(1370)$]{Large background phase,
  no $\boldsymbol{f_0(1370)}$}

Minkowski and Ochs fit only the mass projections on $\pi^+\pi^-$ and $K^+K^-$.
According to them a fit that takes into account the two-dimensional Dalitz
plot distribution would be preferable but as a first exploration a fit of the
one-dimensional mass projection should give a rough picture of the dominant
aspects. The ansatz for the rate per pair mass interval is roughly the
following ($m=\sqrt{s}$), see \cite[eq.~12]{Minkowski:2004xf}
\begin{multline}
  \frac{d\Gamma}{dm_{hh}} \propto | |T_{gb}|e^{i\delta_{bg}} +
  c_2T_{f_0(980)}e^{2i\delta_{bg}}
  + c_3T_{f_0(1500)}e^{2i\delta_{bg}}|^2, \\
  hh=\pi^+\pi^-, K^+K^-.
\end{multline}
This is a superposition of an amplitude representing the broad resonant
glueball, $|T_{gb}|e^{i\delta_{bg}}$, and two amplitudes representing the
resonances $f_0(980)$ and $f_0(1500)$. The two resonances are represented by
Breit-Wigner terms\footnote{The $T$ of ref.~\cite{Minkowski:2004xf} is
  normalized differently than my $\mathcal{M}$. Cf.\ \eg the expression for a
  Breit-Wigner resonance, equation~\eqref{eq:9}.} with an energy dependent
width which, moreover, is characterized by a particular shape function $G$,
\begin{equation}
  T_a = \frac{m_a\Gamma_a}{m_a^2-m^2-im_a\Gamma_a(1+G_a(m))}, \quad
  a=gb, f_0(980), f_0(1500). 
\end{equation}

The relative phases between the three resonant contribution are not free fit
parameters but share the common factor of $e^{i\delta_{bg}}$. This has as a
consequence that the $f_0(980)$ and $f_0(1500)$ are not typical Breit-Wigner
resonances the phases of which move rapidly through $\pi/2$ but are rotated in
the Argand diagram by two times the phase of the glueball amplitude at
resonance ($\pi/2$). This ansatz is motivated by the interpretation of the
isospin S wave as the \emph{red dragon}, the broad $0^{++}$ glueball
interfering destructively with the narrower resonances $f_0(980)$ and
$f_0(1500)$, see figures~\ref{fig:hyams}--\ref{fig:phase-maple}.

Interestingly enough, the superposition of all three amplitudes leads to an
enhancement in the region of $m=\sqrt{s_{hh}}=1300\text{MeV}$, which the Belle
Collaboration (see section~\ref{sec:belle}) would rather interpret as a signal
from the controversial $f_0(1370)$.

\subsection{Belle Collaboration}\label{sec:belle}

I do not discuss here the event reconstruction, background suppression, and
corrections for efficiency and resolution of ref.~\cite{belle04}.\footnote{An
  important point to be checked would be how the pairs of signal events are
  formed for different mass projections, as discussed in
  section~\ref{sec:bg-pw}.}  I only report briefly the ansatz for fitting of
the signal. The model that is used to determine the branching fraction in
question consists of the following \emph{ansatz} for the S-matrix element:
\begin{equation}\label{eq:1}
  \begin{split}
  S_{A_J}(K^+\pi^+\pi^-) &=%
  a_{K^*}e^{i\delta_{K^*}}\mathcal{A}_1(\pi^+K^+\pi^-|K^*(892)^0)\\
  & \quad +
  a_{K^*_0}e^{i\delta_{K^*_0}}\mathcal{A}_0(\pi^+K^+\pi^-|K^*_0(1430))\\
  & \quad +
  a_{\rho}e^{i\delta_{\rho}}\mathcal{A}_1(K^+\pi^+\pi^-|\rho(770)^0)\\
  & \quad + a_{f_0}e^{i\delta_{f_0}}\mathcal{A}_0(K^+\pi^+\pi^-|f_0(980))\\
  & \quad + a_{f_X}e^{i\delta_{f_X}}\mathcal{A}_J(K^+\pi^+\pi^-|f_X)\\
  & \quad +
  a_{\chi_{c0}}e^{i\delta_{\chi_{c0}}}\mathcal{A}_0(K^+\pi^+\pi^-|\chi_{c0})\\ 
  & \quad + \mathcal{A}_{\text{nr}}(K^+\pi^+\pi^-),
\end{split}
\end{equation}
with \cite[eq.~(9)]{belle04}
\begin{equation}
  \mathcal{A}_{\text{nr}}(K^+\pi^+\pi^-) = a_1^{\text{nr}}e^{-\alpha
  s_{13}}e^{i\delta_1^{\text{nr}}} + a_2^{\text{nr}}e^{-\alpha
  s_{23}}e^{i\delta_2^{\text{nr}}}.
\end{equation}

Here I adopt the notation of ref.~\cite{belle04}. In this notation the
order of the symbols `$K^+$', `$\pi^+$' and `$\pi^-$' in the brackets of
$\mathcal{A}$ indicates that the first two terms in equation~\eqref{eq:1}
denote an amplitude in the $K^+\pi^-$ channel whereas the second to sixth term
denote each an amplitude in the $\pi^+\pi^-$ channel. As to the non-resonant
amplitude $\mathcal{A_{\text{nr}}}$, the subscripts `13' and `23' show that
this amplitude is the (coherent) sum of one amplitude in the $K^+\pi^-$
channel and one amplitude in the $\pi^+\pi^-$ channel.

In order to estimate model dependent uncertainties a fit was also carried out
using a non-resonant amplitude with a contribution
\begin{equation}
a_3^{\text{nr}}e^{-\alpha s_{12}}e^{i\delta_3^{\text{nr}}} 
\end{equation}
from the $K^+\pi^+$ channel. In the default model, called $K\pi\pi-\text{C}_0$,
to determine the branching fractions, however such a contribution is not
present.

Let me spell out more explicitly what the $\mathcal{A}$'s are. I take as an
example $\mathcal{A}_1(\pi^+K^+\pi^-|K^*(892)^0)$.
\begin{equation}\label{eq:2}
  \begin{split}
    \mathcal{A}_1(\pi^+K^+\pi^-|K^*(892)^0) & \propto \frac{1}{M^2_R - s_{13}
      - iM_R\Gamma_R^1(s_{13})}\\
    & \quad \times \left( s_{12} - s_{23} +
      \frac{(M_R^2-m_2^2)(m_3^2-m_1^2)}{s_{13}} \right),
\end{split}
\end{equation}
with
\begin{equation}
  \Gamma_R^1(s_{13}) \propto \Gamma_R \left( \frac{p_{s13}}{p_0} \right)^2
  \left( \frac{M_R}{\sqrt{s_{13}}} \right) \frac{1+R^2p_0^2}{1+R^2p_{s13}^2}.
\end{equation}
The first factor on the right-hand side of equation~\eqref{eq:2} is a
relativistic Breit-Wigner function with energy-dependent width
$\Gamma_R^1(s_{13})$, the second factor describes the angular distribution of
the three decay products for a resonance with total spin $J=1$. 

For a $J=0$ resonance, like for example the $f_0(980)$, the amplitude
$\mathcal{A}$ takes a simpler form. The factor for the angular distribution is
1. For $M_R, \sqrt{s_{23}} \gg m_\pi$ the energy dependence of the width can be
neglected, $\Gamma_R^0(s_{23}) \approx \Gamma_R$. The corresponding density of
events in the Dalitz plot is proportional to $\mathcal{A}^2$, which gives in
case of no interference a band of width $\Gamma_R^2$ parallel to $s_{13}$ the
and perpendicular to the $s_{23}$ axis.

\subsubsection[Non-resonant background and $f_X(1300)$]{Non-resonant
  background and $\boldsymbol{f_X(1300)}$} 

In contrast to the ansatz of Minkowski and Ochs the analysis by the Belle
collaboration \emph{lacks a clear interpretation} for the introduction and
parametrization of their non-resonant amplitude $\mathcal{A}_{\text{nr}}$. As
to the controversial $f_0(1370)$, the Belle collaboration finds in their
default model $K\pi\pi-\text{C}_0$ that the ``the mass and width of the
$f_X(1300)$ state obtained from the fit with the model $K\pi\pi-\text{C}_0$
are consistent with those for the $f_0(1370)$'' \cite[p.~15]{belle04}.

\section{Conclusions}

In conclusion, I propose the following guidelines to begin with a Dalitz plot
analysis of the decay $B^+\to\pi^-\pi^+ K^+$ or comparable. It goes without
saying that I do not mean to have derived strict rules that are necessary and
sufficient for a successful analysis. I only try to stress some points that
might be neglected in common fit procedures and want to suggest some steps
that seem appropriate in the light of the issues discussed in precedent
sections and chapters.

\subsection{Assumptions to be checked}

The guidelines I propose rest in particular on the following assumptions,
which remain to be checked.
\begin{itemize}
\item The two-body angles defined in section~\ref{sec:two-body}, the
  \emph{Gottfried-Jackson} angle and the angle $\theta$ in two-body cms
  scattering are directly related.
\item For the $\pi^-\pi^+$ and $\pi^-K^+$ channel one can use as a constraint
  \emph{elastic} unitarity. Inelastic channels can be neglected.
\item The $\pi^+K^+$ amplitude has a forward (or backward or both) peak.
\item The coupling of of the $B$ to the $\pi^+K^+$ does not vanish, see
  page~\pageref{non-vanish}.
\end{itemize}

\subsection{Guidelines}

\paragraph{Forming triplets of signal events.}

One must not only held fix the signal to background ratio but form triplets of
signal events for the entire analysis including in particular different mass
projections, as discussed in section~\ref{sec:bg-pw}.

\paragraph{Two-dimensional binning.}

Dalitz plot analysis of a decay like $B^+\to\pi^-\pi^+ K^+$ differ from older
Dalitz plot analysis in that they have not the principal aim to determine spin
and parity of the decaying particle from characteristic patterns in the plot,
see figures~\ref{fig:omega} and \ref{fig:patterns}. Rather the aim is to fit
the ``landscape'' over the plot. To this end two dimensional binning of the
data and representation as a \emph{lego plot} is more appropriate than a
distribution of points.

\paragraph{Do not fit whole Dalitz plot.}

The interior of the Dalitz plot boundary is approximately 3 times larger than
the actual region of interest in the lower left corner with $s_{\pi^-\pi^+}$
and $s_{\pi^-K^+}$ between zero and 1.6 GeV, roughly.  A fit of the hadronic
structures under consideration should only be done in this region of interest.
If one fits the whole Dalitz-plot, a certain set of parameters may score
better in a $\chi^2$ test, but not because it represents better the structures
in question but the huge rest of the plot, which is irrelevant for the present
problems of hadron spectroscopy.

\paragraph{Elastic phases consistent with scattering data.}

Because of the at least approximate validity of elastic unitarity constraints
(Watson's theorem, see section~\ref{sec:watson}) the phases of the amplitudes
$T(B^+\to (\pi^-\pi^+) K^+)$ and $T(B^+\to (\pi^- K^+)\pi^+)$ should be
(approximately) equal to the phase shifts established in $\pi\pi$ and $K\pi$
scattering analyses.

\paragraph{Forward (or backward) peak from $\boldsymbol{K^+\pi^+}$?}

Without good reasons the amplitude $T(B^+\to \pi^-(\pi^+K^+))$ should not be
set to zero. Quite the contrary may be true: This amplitude may be
non-vanishing and generate an appreciable enhancement at the lower left
boundary of the Dalitz plot. If such a signal is indeed present, care should
be taken to separate it from searched for S wave signals; see
section~\ref{sec:s-wave}.

\paragraph{$\boldsymbol{K^*(892)^0}$ and $\boldsymbol{z_{\pi^-K^+}=0}$.}

The uncontroversial $K^*(892)^0$ resonance can serve as reference for fixing
overall coupling strength and phases. As a spin~1 particle the contribution
from this resonance should vanish with $z_{\pi^-K^+}$.

\paragraph{Peak from $\boldsymbol{K^+\pi^+}$ and zero from
  $\boldsymbol{K^*(892)^0}$.}

I could not establish in detail where the line with $z_{\pi^-K^+}$ passes
through the lower left corner of the Dalitz plot, see figure~\ref{fig:z13}.
Since the signal from $K^*(892)^0$ should vanish there the non-vanishing of the
$K^+\pi^+$ amplitude could most be noted.

\paragraph{Charge and flavor conjugated reactions.}

Analyses of charge and flavor conjugated reactions to $B^+\to\pi^-\pi^+ K^+$
like
\begin{equation}
    B^+,\ B^-,\ B^0,\ \bar{B^0} \to \pi \pi K,\ \eta\eta K,\ \eta\pi K
\end{equation}
may give additional input to decide between alternative fits, and
contributions from different isospin components.


\bibliographystyle{utcaps}
\bibliography{dip} \addcontentsline{toc}{chapter}{Bibliography}

\end{document}